\documentclass[twocolumn,showpacs,prb]{revtex4-1}

\usepackage{amsmath}
\usepackage{amssymb}

\usepackage{graphicx}
\usepackage{epstopdf}
\usepackage{subfigure}

\begin{document}

\title{Geometric and topological properties of \\the canonical grain growth microstructure}

\author{Jeremy K. Mason$^1$, Emanuel A. Lazar$^2$, Robert D. MacPherson$^3$, David J. Srolovitz$^2$}
\affiliation
{$^1$Bo\u{g}azi\c{c}i University, Bebek, Istanbul 34342 T\"urkiye.\\
 $^2$Materials Science and Engineering, University of Pennsylvania, Philadelphia, PA 19104.\\
 $^3$School of Mathematics, Institute for Advanced Study, Princeton, New Jersey 08540.
}
\date{\today}

\begin{abstract}

Many physical systems can be modeled as large sets of domains ``glued'' together along boundaries -- biological cells meet along cell membranes, soap bubbles meet along thin films, countries meet along geopolitical boundaries, and metallic crystals meet along grain interfaces.  Each class of microstructures results from a complex interplay of initial conditions and particular evolutionary dynamics.  The statistical steady-state microstructure resulting from isotropic grain growth of a polycrystalline material is canonical in that it is the simplest example of a cellular microstructure resulting from a gradient flow of a simple energy, directly proportional to the total length or area of all cell boundaries.  As many properties of polycrystalline materials  depend on their underlying microstructure, a more complete understanding of the grain growth steady-state can provide insight into the physics of a broad range of everyday materials.  In this paper we report geometric and topological features of these canonical two- and three-dimensional steady-state microstructures obtained through large, accurate simulations of isotropic grain growth.
\end{abstract}


\maketitle

\section{Introduction}
\label{sec:introduction}

\begin{figure*}
\begin{center}$
\begin{tabular}{ccc}
\fbox{\includegraphics[height=0.24\linewidth]{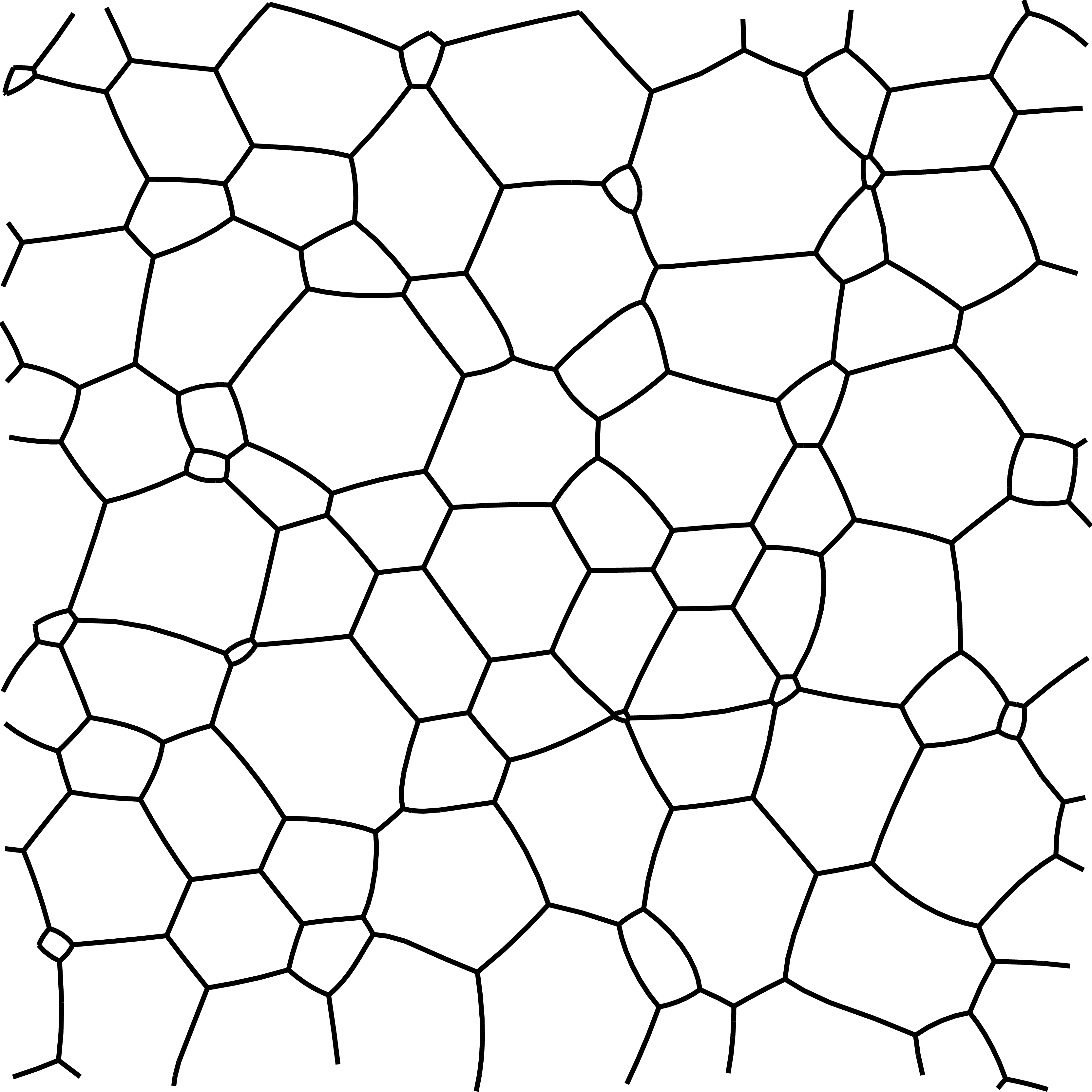}} \hspace{5mm} &
\fbox{\includegraphics[height=0.24\linewidth]{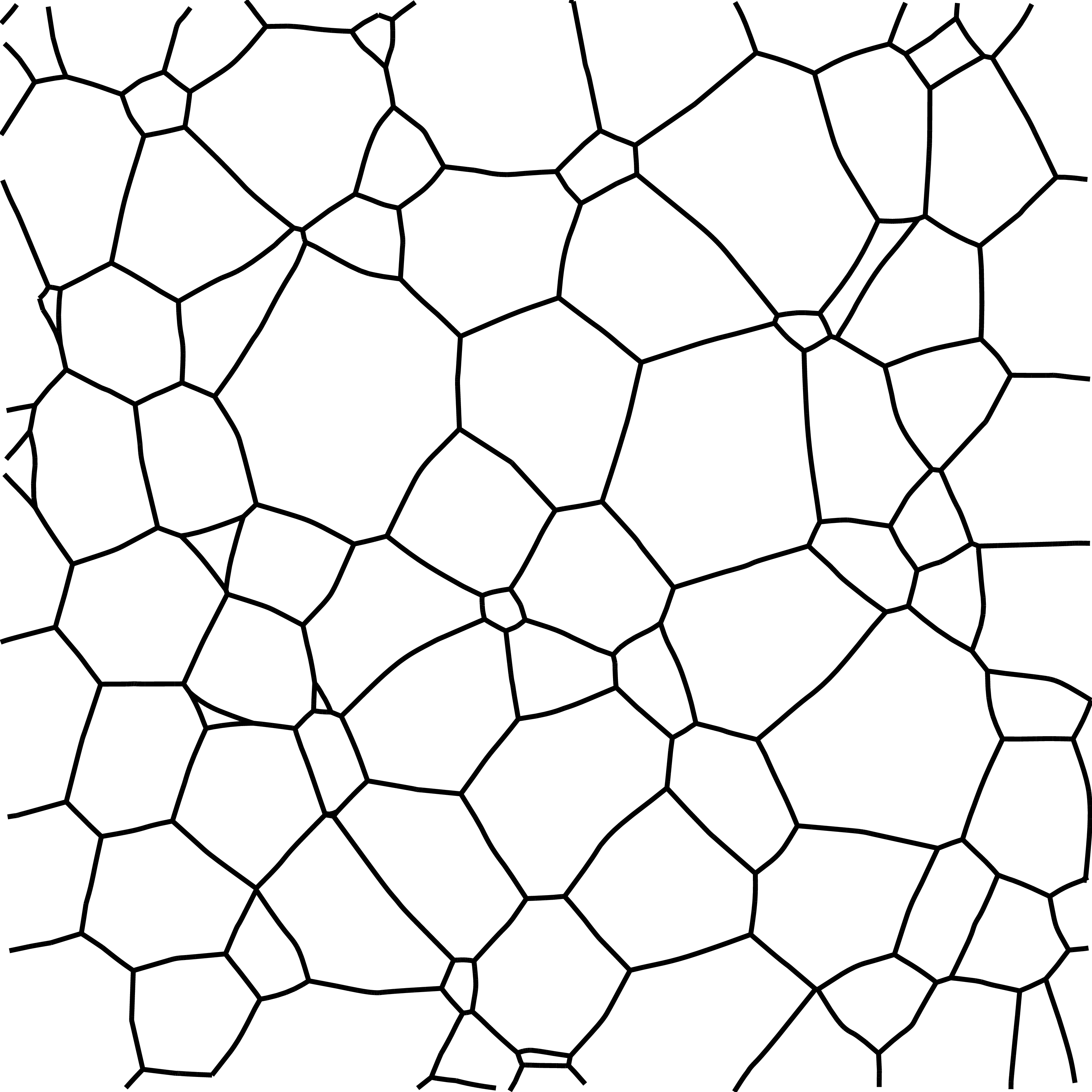}} \hspace{5mm} &
\fbox{\includegraphics[height=0.24\linewidth]{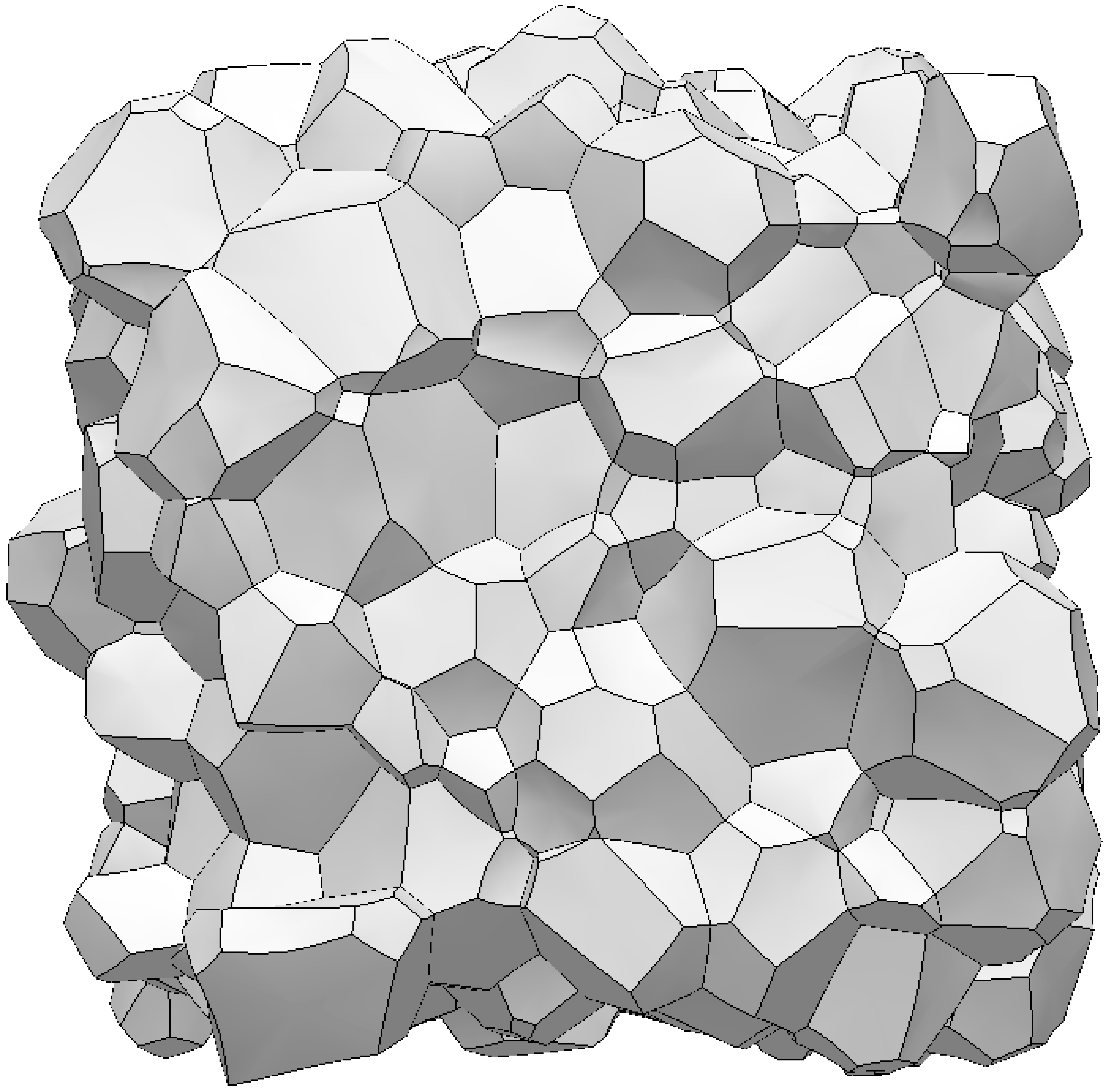}}\\
(a)&(b)&(c)
\end{tabular}$
\end{center}
\vspace{-10pt}
\caption{(a) A two-dimensional grain-growth microstructure, (b) a cross-section from a three-dimensional grain-growth microstructure, and (c) a three-dimensional grain-growth microstructure.}  
\label{cell_systems}
\end{figure*}

A polycrystalline material is an ensemble of individual crystallites or grains, joined along grain boundaries.  Polycrystalline microstructures are often formed during material synthesis and continue to evolve through processes such as grain growth and recrystallization.  Most metals and ceramics, and many semiconductors, ubiquitous in every day use and technology, are polycrystalline. One of the most common class of polycrystalline microstructures is that resulting from normal grain growth. Normal grain growth is the process in which grain boundaries migrate with a velocity proportional to their mean curvature, leading to a monotonic decrease of grain boundary area and excess energy, and a monotonic increase in mean grain size. In its ideal form, the grain boundary energy and mobility are isotropic and constant.  It has been observed that normal grain growth leads to the evolution of microstructures that exhibit a statistical steady-state, during which geometrical and topological properties of the microstructure remain statistically constant as the mean grain size increases \cite{mullins1986statistical}.  In many real experimental cases this picture is complicated by anisotropy and the presence of solutes or additional phases. Nonetheless, isotropic normal grain growth microstructures are the most generic of all polycrystalline microstructures.  Because of the simplicity of the grain growth process in the isotropic limit, and the ubiquity of grain growth microstructures it produces, normal grain growth microstructures can be thought of as the ``harmonic oscillators'' of polycrystalline microstructures or ``the canonical polycrystalline microstructure.'' 

We note that other classes of generic polycrystalline microstructures exist. For example, random grain nucleation and constant grain boundary velocities, appropriate for modeling certain solidification and phase transformation processes, leads to microstructures which are Voronoi tessellations of Poisson point processes \cite{1992okabe}. These Voronoi microstructures have flat grain boundaries, and distributions of their geometric and topological features are vastly different from those commonly observed in real polycrystalline materials; we return to this comparison at the end of this report. 

Characterizing steady-state microstructures experimentally in three dimensions is challenging due to the inherent difficulty in exploring the geometry and topology of grains buried deep within a three-dimensional microstructure.  Several such studies have been performed, including by embrittling the polycrystal grain boundaries and subsequently separating individual grains \cite{rhines1982effect}, by serial sectioning \cite{alkemper2001quantitative, uchic20063d, ullah2014three}, and more recently by synchrotron-based diffraction techniques \cite{ludwig2009new, syha2012three, abdolvand2015deformation}.  Because of the  difficulties in experimentally characterizing three-dimensional microstructures consisting a large numbers of grains, and the fact that isotropic grain boundary energy and mobility are rare, simulations provide the preferred method to investigate isotropic normal grain growth microstructures.

The primary purpose of this paper is to provide the most complete characterization of the geometric and topological features of  two- and three-dimensional isotropic grain growth microstructures to date, as well as the correlations between such features.  To this end, we employ a computer simulation technique \cite{2010lazar, 2011lazar} to generate one of the largest and most accurate steady-state isotropic grain growth microstructure databases. Our geometric and topological characterization of these ideal, normal grain growth  microstructures provides a rigorous, fundamental set of data against which all polycrystalline microstructures and grain growth theories can be compared.

A secondary focus of this paper is to compare two- and three-dimensional steady-state microstructures in order to explore the manner in which the introduction of a third dimension essentially changes the steady-state structure.  This is partly motivated by the substantial investment of the research community in two-dimensional simulations \cite{sahni1983kinetics, grest1988domain, 1997fanA, 1997fanB, 1998weygand, 2006kinderlehrer}, often with the implicit hope that results gleaned from these will shed light on the experimentally more relevant three-dimensional case.  Finally, since three-dimensional experimental samples are often viewed and characterized in cross-section, we also examine the relationship between cross-sections of three-dimensional structures with both the two- and full three-dimensional microstructures.

\section{Simulation}
\label{sec:simulation}

\subsection{Prior approaches}
\label{subset:prior_approaches}

Prior simulations of grain growth fall roughly into three categories.  Monte Carlo Potts models \cite{sahni1983kinetics, grest1988domain, 1989anderson, 1991mulheran} were among the earliest and are still among the most frequently implemented.  Here, space is subdivided into cubical voxels which are assigned a label indicating to which grain they belong.  Voxels are relabeled in a stochastic manner to minimize interfacial energy between neighboring grains.  While the implementation of this method is relatively straightforward, some of the complications that arise include anisotropic boundary conditions \cite{1989anderson, 1991holm}, weakening of vertex angle boundary conditions \cite{1989anderson, 1991holm}, boundary pinning, and altered kinetics \cite{1989anderson, 1991holm, 2005kim}.  Closely related to this model is the cellular automaton model \cite{2001geiger, 2002raabe, 2003janssens, 2006ding}, which allows additional flexibility in defining energetics.

Phase-field models constitute a second category of simulation methods \cite{2002krill, 2006kim}.  In this approach, grains are represented as regions in which the value of a nominally continuous order parameter goes to one.  The evolution of the microstructure is  described using time-dependent Ginzburg--Landau equations \cite{1950ginzburg}, where different order parameters are associated with each grain orientation.  Modifications of the energy functional allows for the incorporation of additional physical effects \cite{1998tikare}.  Limitations of this approach include numerical instability and the high cost of of representing systems with large numbers of grains \cite{2002krill, 2006kim}.  Variations on this approach include  continuum-field \cite{2002krill, 2007vanherpe, 2010dorr} and multi-phase-field \cite{2006kim} models. The simulation of three-dimensional grain growth by diffusion-controlled interface motion \cite{2009elsey, 2011elsey} shares some features with phase-field models, and appears to be unconditionally stable and conceptually and computationally straightforward, allowing for the simulation of a large number of grains.  

A final category of grain growth simulations are front-tracking methods, including vertex models \cite{1990nagai, 1998weygand}, finite element models \cite{2000kuprat, 2008mora, 2010syha} and discretized boundary models as developed by the authors \cite{2010lazar, 2011lazar} and employed here.  In such front tracking approaches, the grain boundary network is explicitly discretized and evolved without devoting memory or processing power to the grain interiors.  While computationally efficient and convenient for the measurement of geometric and topological features, this approach requires that all topological events which occur during normal grain growth be anticipated and handled explicitly.  The complexity associated with mesh management in three dimensions has been a limiting factor in the widespread implementation of these models.

\subsection{Current approach}
\label{subset:current_approach}

Most front-tracking algorithms evolve a system in the direction of steepest descent of an energy proportional to the total grain boundary area \cite{1990nagai, 1995fuchizaki}, which is equivalent to mean curvature flow in a continuous setting.  One notable feature of our simulations \cite{2010lazar, 2011lazar} is that the equations of motion derive directly from the von Neumann--Mullins \cite{1952vonneumann, 1956mullins} relation in two dimensions and the MacPherson--Srolovitz relation \cite{2007macpherson} in three dimensions.  This means that our simulations satisfy the constraints imposed by these exact relations to high accuracy.  The low curvature of most grain boundaries \cite{1990nagai} justifies our use of a restricted discretization to reduce the computational requirements.

In two dimensions, boundaries that separate two neighboring grains are referred to as {\it edges}, and three edges meet at a {\it vertex}.  Each edge is discretized into a piecewise linear curve; the discretization adapts during the microstructure evolution to ensure stability and numerical accuracy.  Equations of motion are obtained from the von Neumann-Mullins relation \cite{1952vonneumann, 1956mullins}, which describes the evolution of individual grains:
\begin{equation}
\frac{dA}{dt} = -2\pi M \gamma \left(1-\frac{n}{6}\right),
\label{evn2d}
\end{equation}
where $A$ is the area of a grain, $n$ is its number of edges, and $M$ and $\gamma$ are scalar constants describing the grain boundary mobility and energy, respectively.  This result is exact for isotropic grain growth in two dimensions, where the total energy of the system is proportional to the sum of all edge lengths and the system evolves in the direction of steepest descent in energy.  This equation may be adapted to give equations of motion for a discretized representation of the system \cite{2010lazar}, which is evolved using an explicit time integration scheme.  Since this algorithm derives from the exact von Neumann--Mullins relation, the behavior of every grain in the system satisfies this condition with minimal error.  Specifically, the error resulting from the use of a discrete time step $\Delta t$ is of order $O(\Delta t^2)$, and the error from the discretization of grain boundaries is small as well \cite{2010lazar}.  Occasionally, when an element of the system becomes sufficiently small, we adjust the network topology by removing an edge or a grain, or by flipping an edge.  Interested readers are referred to \cite{2010lazar} for a more detailed discussion of the method and error analysis.  

The situation is similar in three dimensions, where a grain boundary separating two neighboring grains is referred to as a {\it face}.  Three adjacent faces meet along an {\it edge} and four edges meet at a {\it vertex}.  In simulations, the grain boundary network is discretized by triangulating grain faces, and edges are discretized into piecewise linear curves.  The extension of the von Neumann--Mullins relation to three dimensions is the MacPherson--Srolovitz relation \cite{2007macpherson}.  The relation describes the volume evolution of individual grains as
\begin{equation}
\frac{dV}{dt} = -2\pi M \gamma \left(\mathcal{L}-\frac{1}{6}\mathcal{M}\right),
\label{evn3d}
\end{equation}
where $V$ is the volume of grain, $\mathcal{L}$ is a one-dimensional measure of the grain called the {\it mean width}, and $\mathcal{M}$ is the sum of lengths of all grain edges.  This result is an exact description of isotropic grain growth in three dimensions.  As before, this equation may be adapted to give equations of motion for a discretized representation of the system \cite{lazar2011evolution}.  An explicit time integration scheme is used to evolve the system.  Since the equations of motion are derived from the exact MacPherson--Srolovitz relation, the evolution of each grain volume satisfies this condition with minimal numerical error \cite{2011lazar}.  Occasionally the size of a grain, face, or edge will become sufficiently small, requiring that the topology of the system be adjusted appropriately.  Interested readers are referred to \cite{2011lazar} and \cite{lazar2011evolution} for a more detailed analysis of the method.  

One difficulty in using simulations to characterize the steady-state is that the system must be evolved from the initial system to the point that steady-state is achieved.  This requires that the initial model contain several times more grains than eventually contribute to the measured steady-state microstructure statistics.  Moreover, since the steady-state condition is not precisely defined, identifying the point in time when it is reached is still a matter of some contention.  Data for our 2D system came from a simulation which began from a Voronoi tessellation of 10,000,000 points uniformly distributed in the unit square with periodic boundaries.  Data for our 3D system came from 25 simulations, each of which began from a Voronoi tessellation of 100,000 points uniformly distributed in the unit cube with periodic boundaries.  This number of grains is well beyond the range of contemporary front-tracking models \cite{2000wakai}, and rivals the largest grain growth simulations performed to date by any method \cite{2006kim, 2006thomas, 2011elsey}.

\subsection{The Canonical Steady-State}
\label{subsec:canonical_steady_state}

One reason that steady-state grain-growth microstructures are so important is that despite the fact that the evolution equation (mean curvature flow) is very simple, the resultant microstructures are particularly robust; i.e., a wide range of initial microstructures all eventually evolve to a statistically identical microstructure.  To illustrate this point, we constructed  three very different initial microstructures and allowed them to evolve for a time sufficient for their statistical properties to stabilize (i.e., reach the steady-state behavior).  

\begin{figure*}
\begin{center}$
\begin{tabular}{ccc}
\includegraphics[height=0.229\linewidth]{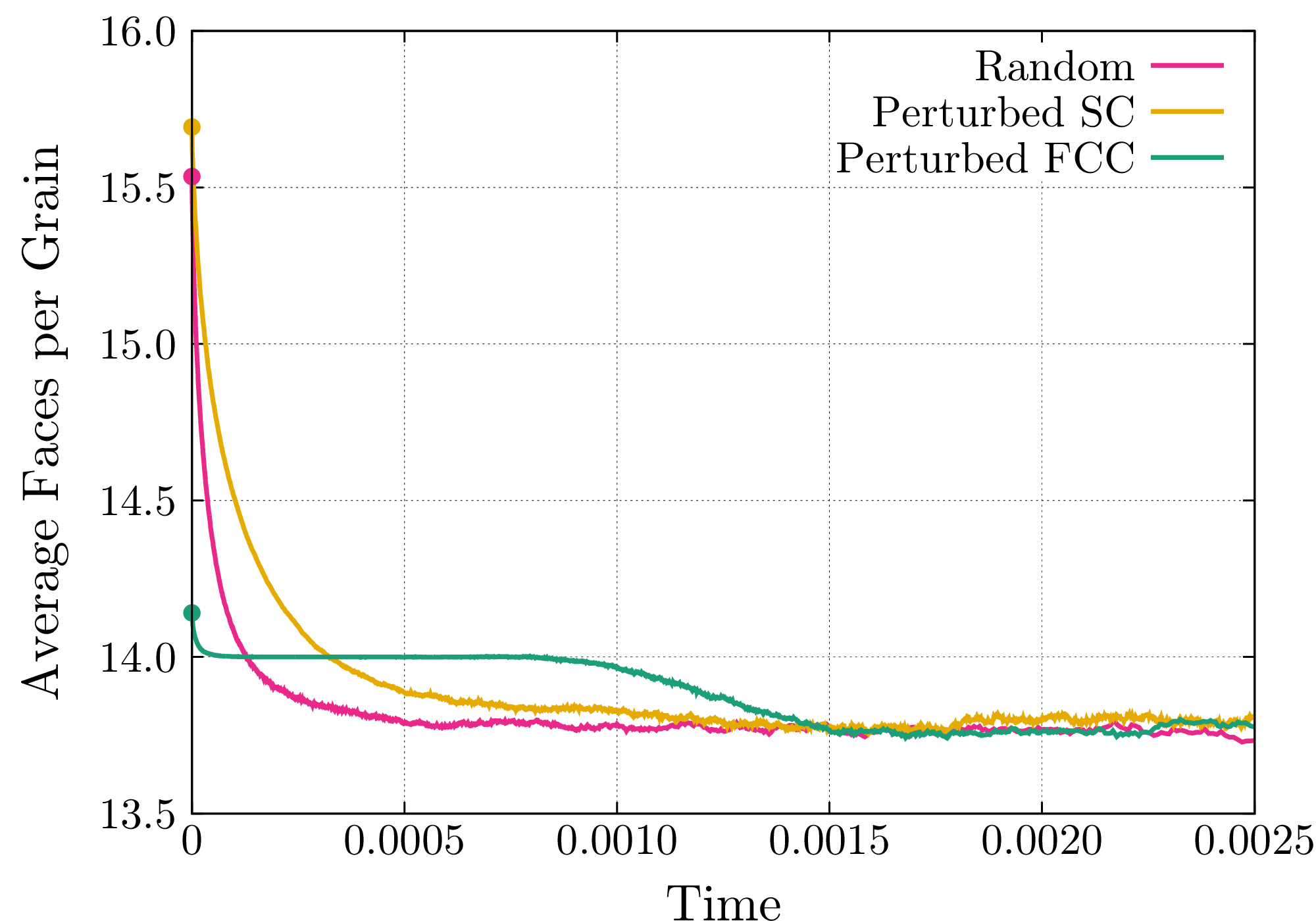} &
\includegraphics[height=0.229\linewidth]{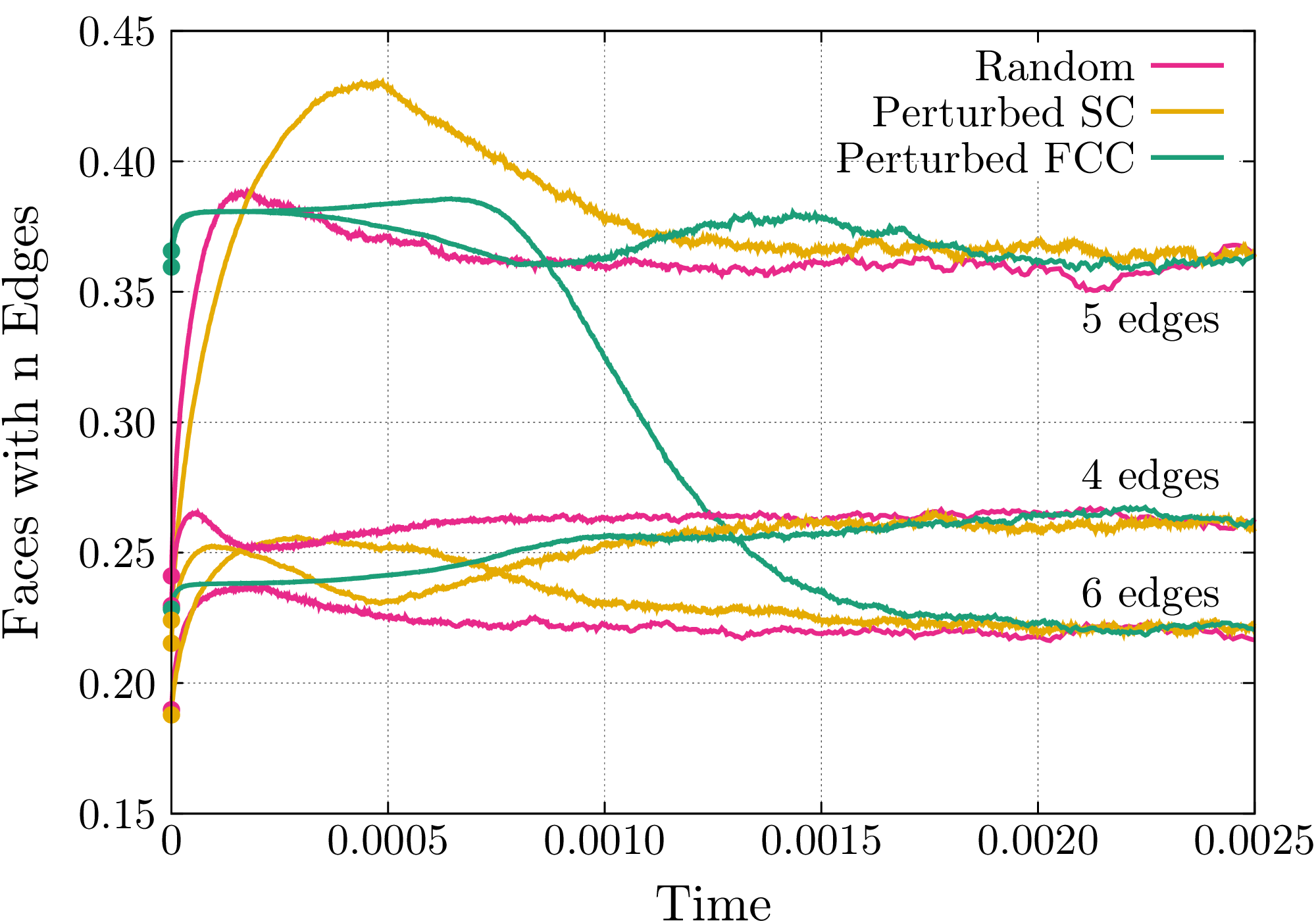}  &
\includegraphics[height=0.229\linewidth]{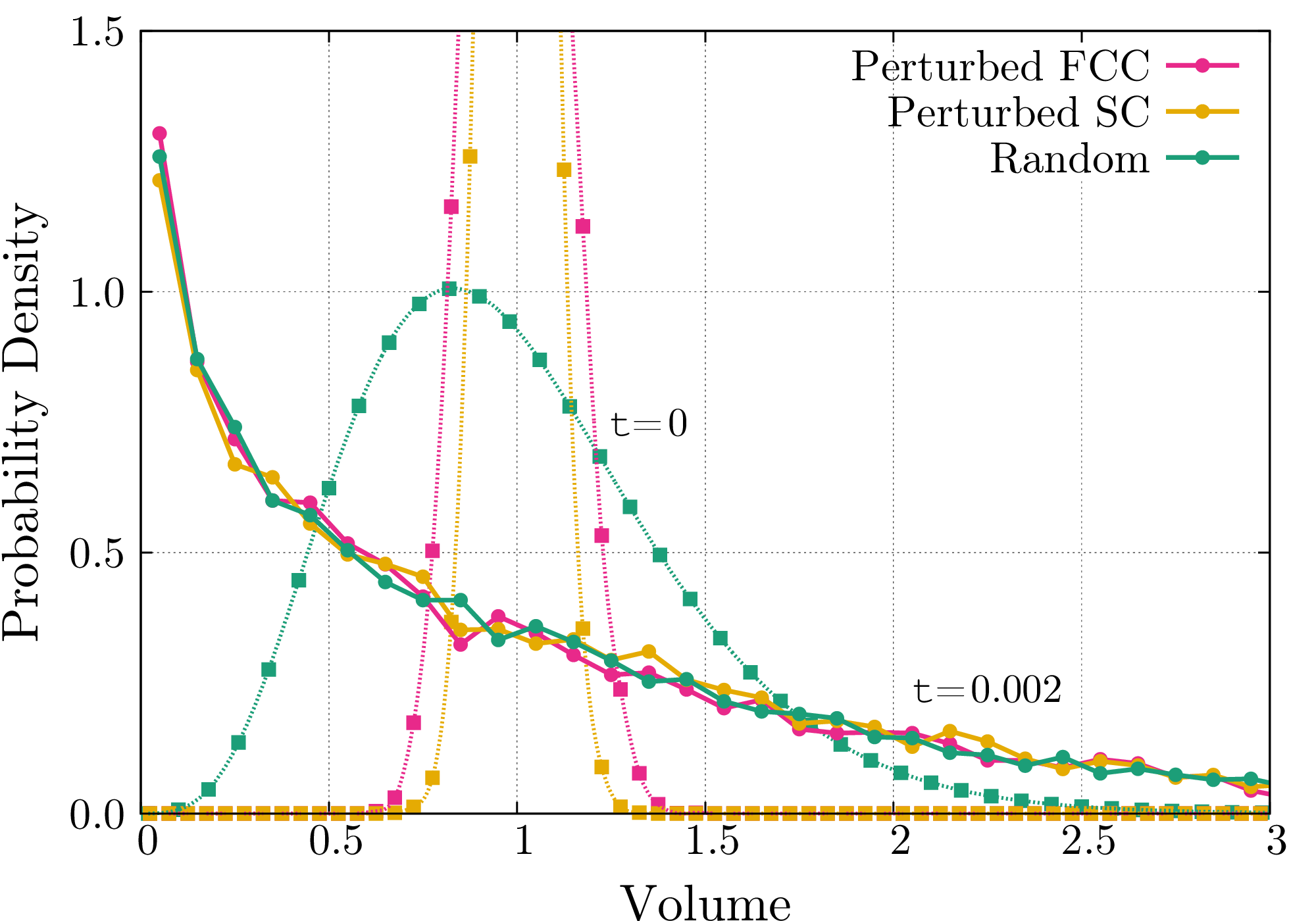}\\
(a)&(b)&(c)
\end{tabular}$
\end{center}
\vspace{-10pt}
\caption{(a) The evolution of the average faces per grain $\langle F \rangle$ in the three systems (circles indicate $\langle F\rangle$ at $t=0$), converging to the steady-state value 13.766.  (b) The evolution of the fraction of faces with 4, 5 and 6 edges in the three systems, converging to steady-state values $p(4) = 26.2\%$, $p(5)=36.4\%$, and $p(6)=22.0\%$. (c) Initial (squares) and final (circles) distributions of grain volumes (normalized by the mean volume) in the three systems.}
\label{leveling}
\end{figure*}

Each of the three initial conditions considered here were constructed as Voronoi tessellations of a set of points in space.  The first system of points we considered resulted from a random Poisson process with 100,000 points in the unit cube; i.e., the $x$, $y$, and $z$ coordinates of each point were chosen with uniform probability in the domain $[0,1]$.  To construct the second system, we generated a set of points on a $30 \times 30 \times 30$ simple cubic (SC) lattice in the unit cube, for a total of 27,000 points.  Then, each point was given an independent, random displacement from its lattice position according to a three-dimensional Gaussian distribution with a standard deviation equal to one tenth the SC lattice parameter.  The third system was constructed in a similar manner as for in the perturbed SC case, except that the original coordinates were chosen to lie on a $20 \times 20 \times 20$ face-centered cubic (FCC) lattice with 4 points per unit cell, for a total of 32,000 points; in this case the points were perturbed away from their FCC positions according to a three-dimensional Gaussian distribution with a standard deviation equal to one fifth of the FCC lattice parameter.  In all three cases, we evolved the systems according to the above algorithm for a sufficiently long time (measured in dimensionless units of $1/M\gamma$) that the statistical properties all achieved steady-state to within statistical error.

\setlength{\fboxsep}{0pt}
\begin{figure}
\begin{center}$
\begin{tabular}{ccc}
\fbox{\includegraphics[trim={3cm 3cm 3cm 3cm},clip, height=0.31\linewidth]{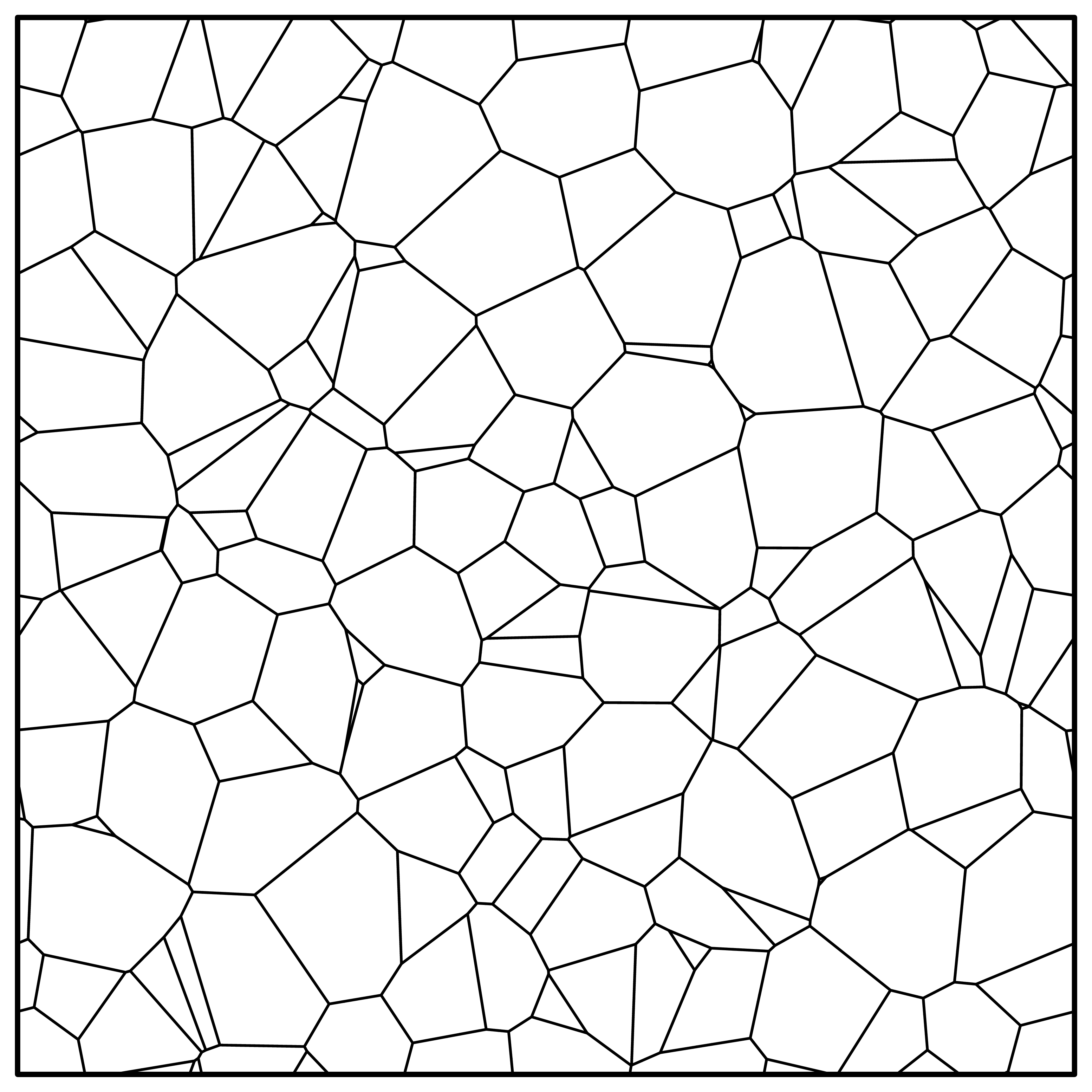}} &
\fbox{\includegraphics[trim={8cm 8cm 8cm 8cm},clip, height=0.31\linewidth]{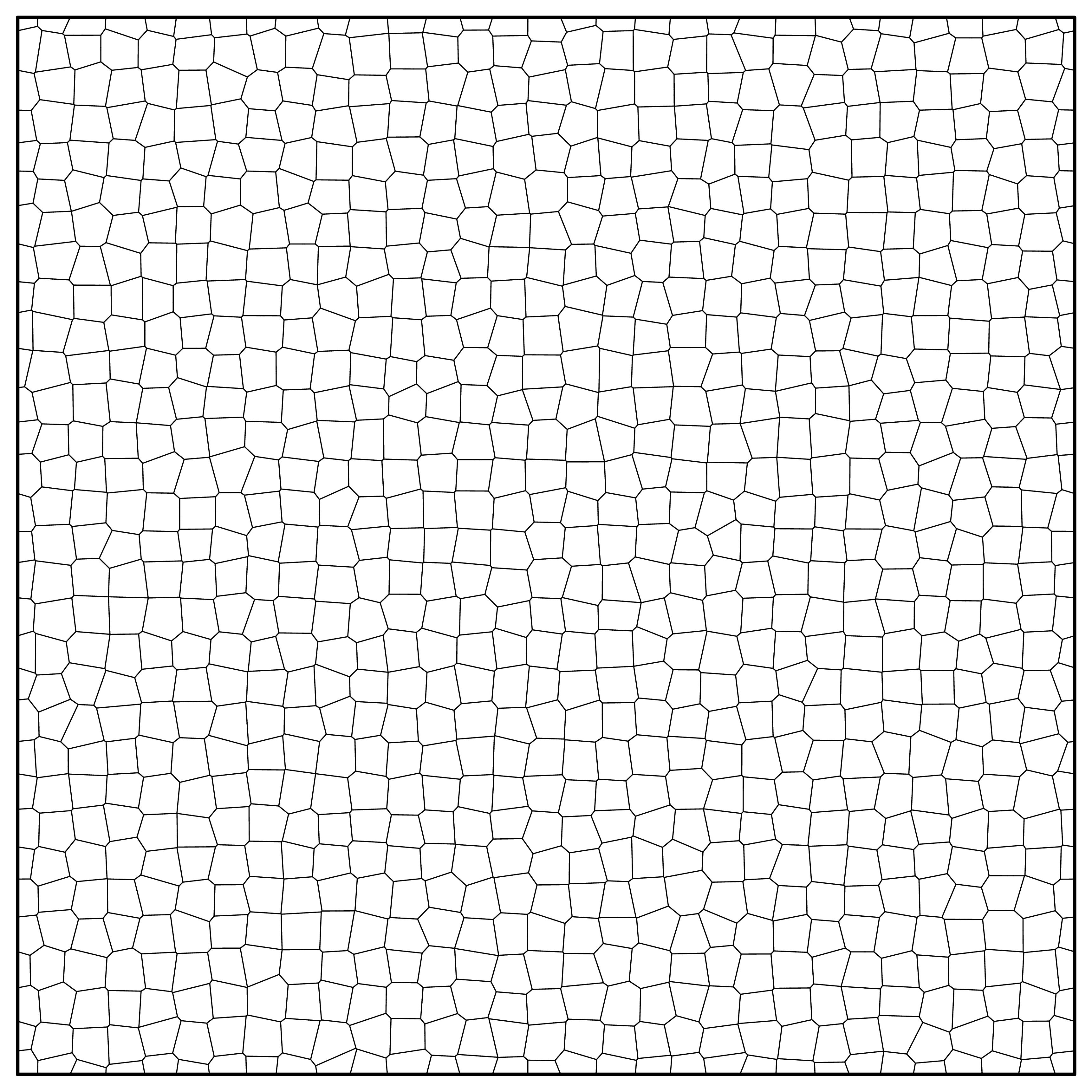}} &
\fbox{\includegraphics[trim={8cm 8cm 8cm 8cm},clip, height=0.31\linewidth]{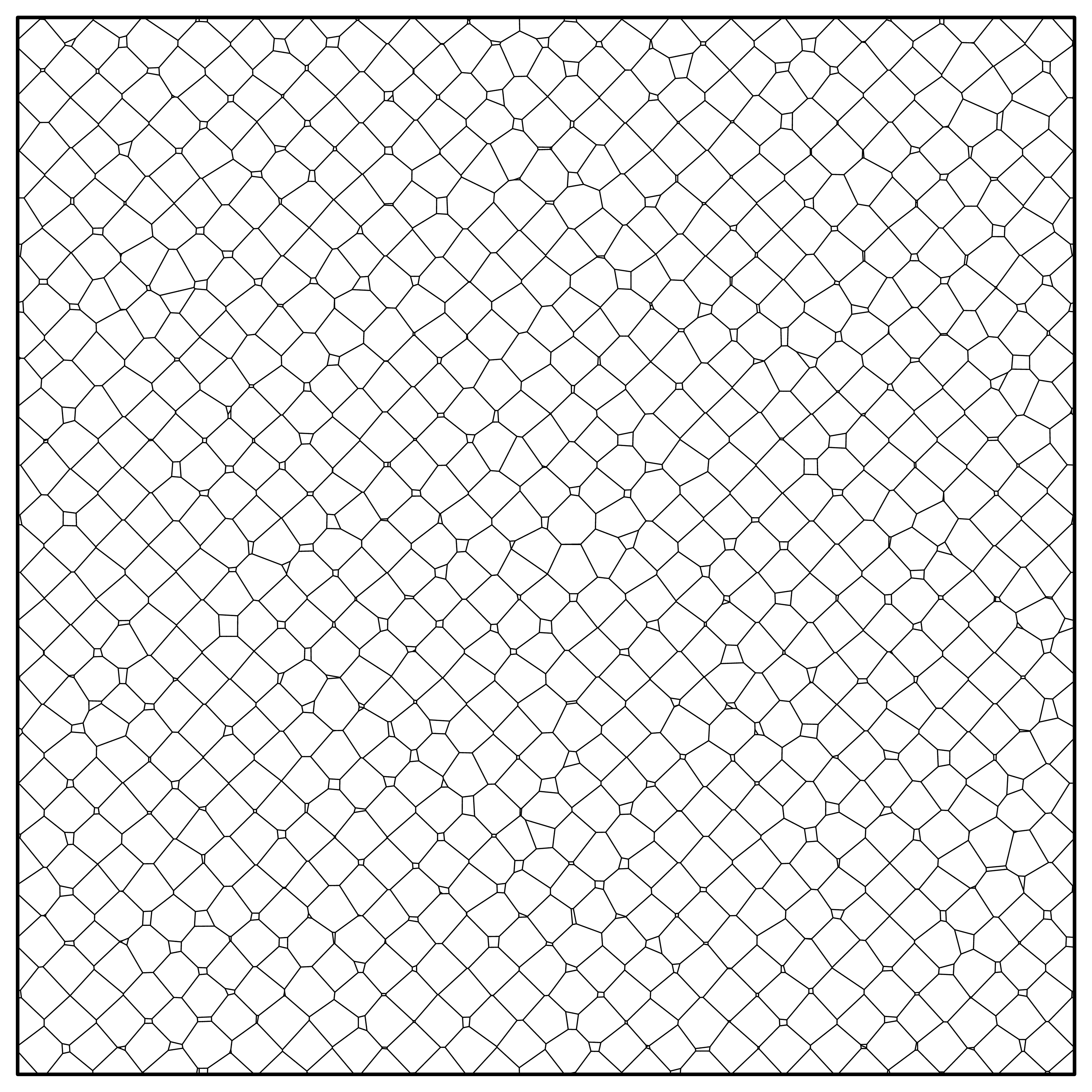}}\vspace{0.5mm}\\
\fbox{\includegraphics[trim={1cm 1cm 1cm 1cm},clip, height=0.31\linewidth]{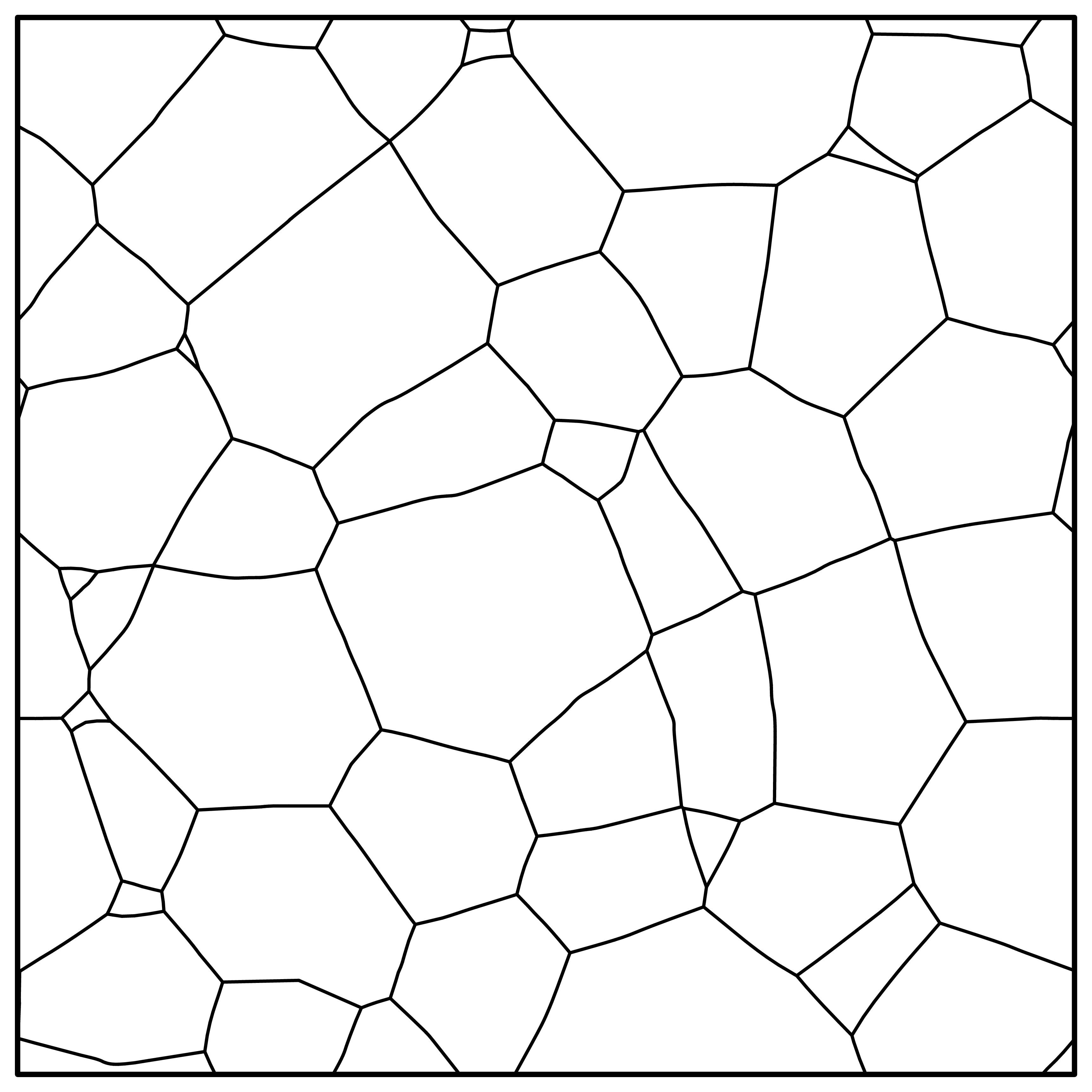}} &
\fbox{\includegraphics[trim={1cm 1cm 1cm 1cm},clip,height=0.31\linewidth]{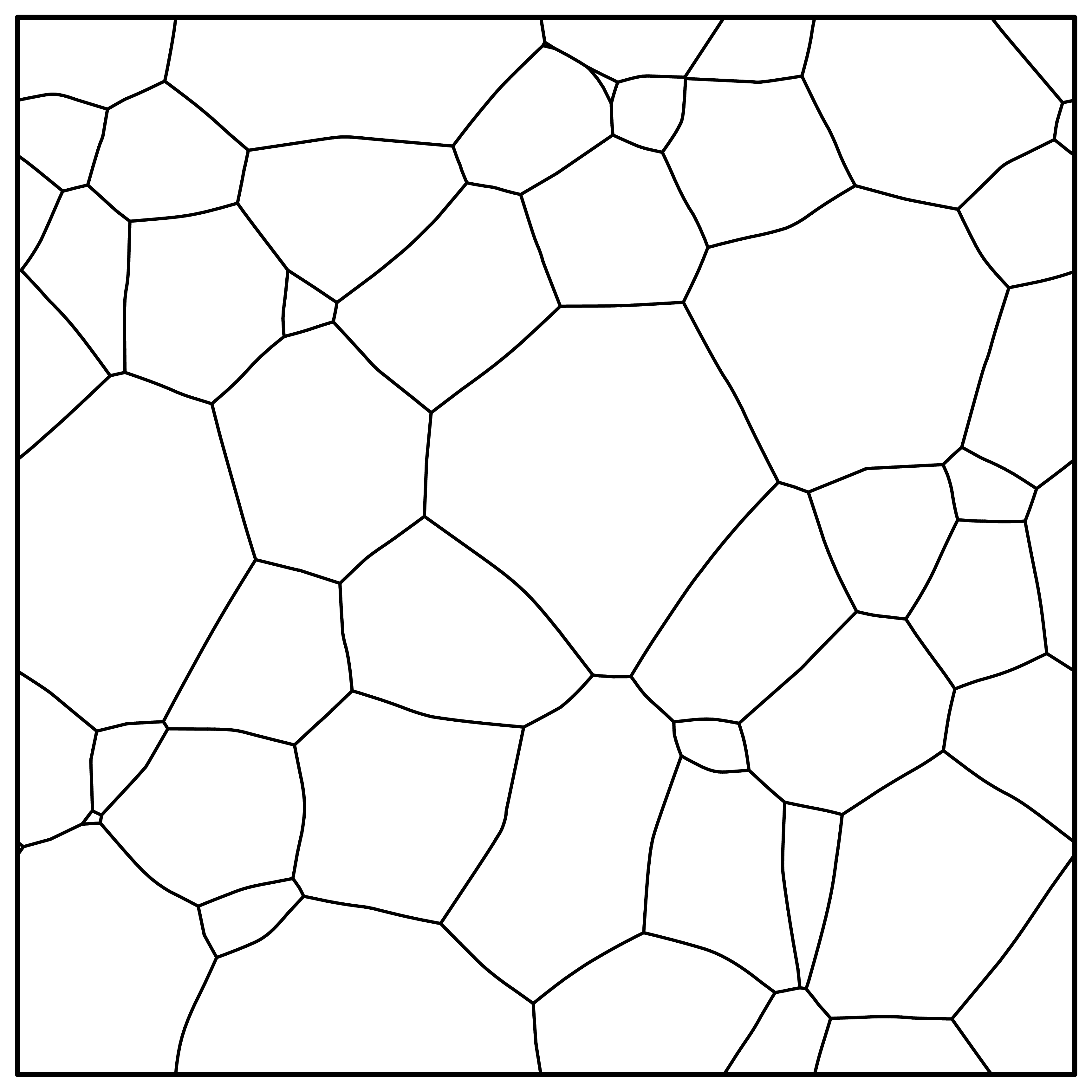}} &
\fbox{\includegraphics[trim={1cm 1cm 1cm 1cm},clip, height=0.31\linewidth]{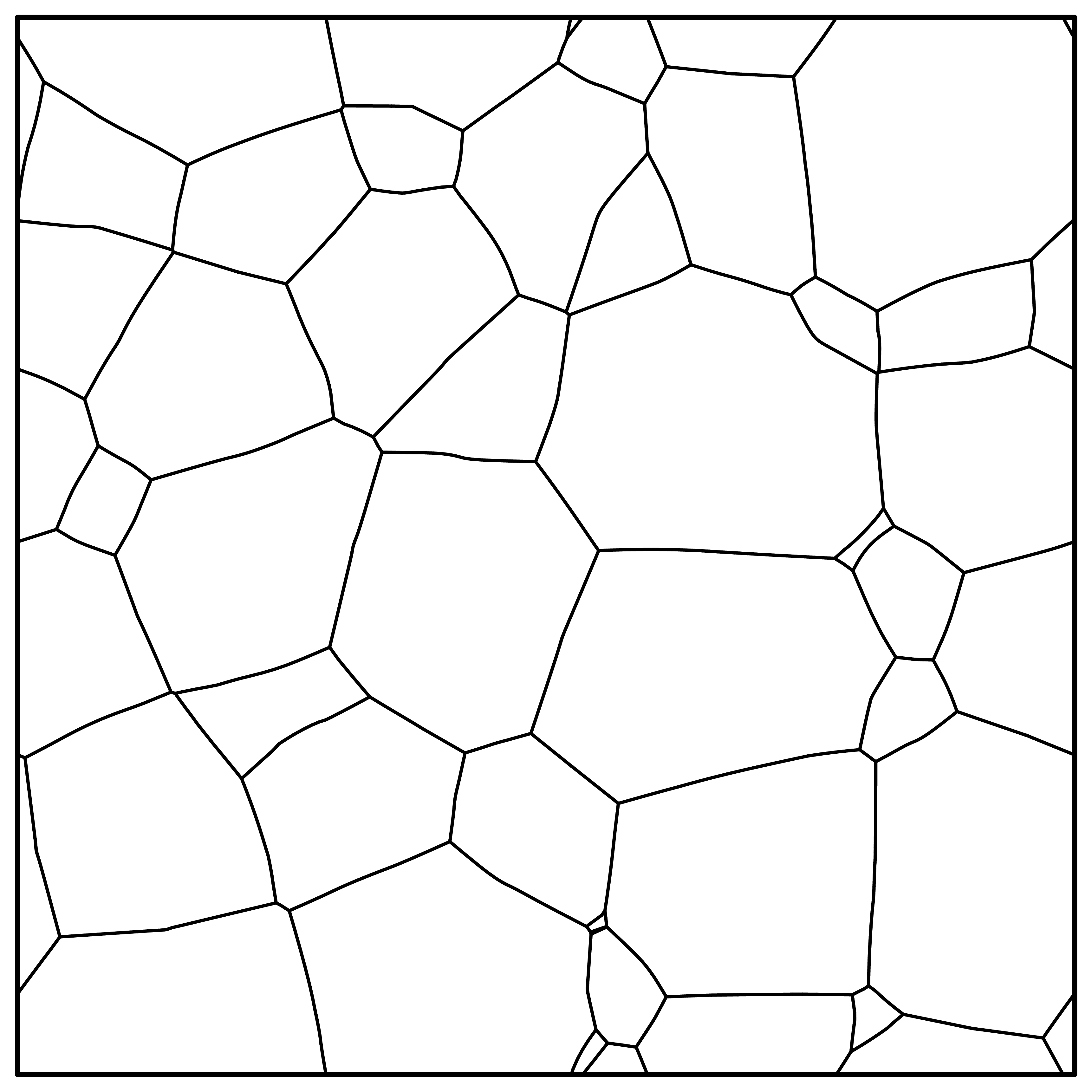}}\\
(a)&(b)&(c)
\end{tabular}$
\end{center}
\vspace{-10pt}
\caption{Cross-sections of three distinct initial microstructures (top) and the steady-state grain growth microstructure resulting from it (bottom).  Each system began as a Voronoi tessellation from a set of points: (a) random distributed points, (b) a perturbed simple cubic lattice, and (c) a perturbed face centered cubic lattice.}  
\label{new}
\end{figure}

Figure~\ref{leveling}(a) shows how the average number of faces per grain $\langle F \rangle$ changes as the three systems coarsen under curvature flow.  The initial value of $\langle F \rangle$ is different in the three systems; in particular $\langle F \rangle_{PV}= 15.535$, $\langle F \rangle_{SC}= 15.693$, and $\langle F \rangle_{FCC}= 14.654$ (as indicated by the symbols at $t=0$).  Over time, these values gradually decrease and all converge to 13.766 faces per grain.  In a similar manner, Fig.~\ref{leveling}(b) shows the evolution of  the fraction of faces with $n=4$, $5$ and $6$ edges $p(n)$ in the three systems.  Here again we see that despite significant differences in their initial values, $p(n)$ for all three systems  converge to the same values, $p(4) = 26.2\%$, $p(5)=36.4\%$, and $p(6)=22.0\%$.  Finally, Fig.~\ref{leveling}(c) shows the initial and steady-state grain volume distributions, normalized by the mean grain volume, for all three systems.  Despite having started with drastically different grain size distributions, all three systems evolve to the same steady-state.  The statistical properties of the three systems are effectively identical after $t = 0.002$, supporting the notion that initial conditions are eventually ``forgotten'' after a system has coarsened by grain growth for a sufficient time.

The convergence of different random microstructures to the same statistical steady state can also be observed directly; typical initial and final cross sections from the three types of initial conditions are shown in Figure.~\ref{new}. Notice that despite the different initial appearances and degrees of order, the final three microstructures could easily be different areas of a single structure.

\section{Characterization}
\label{sec:characterization}

This section reports on data from microstructures initialized as Poisson-Voronoi (PV) cells (i.e., Voronoi tessellations of randomly distributed points) and evolved by the grain growth (curvature flow) algorithm until a statistical steady-state was achieved. A single two-dimensional simulation began with 10,000,000 grains and achieved steady-state when approximately 4\% of the grains remained.  Samples taken from several points in the subsequent evolution gave two-dimensional statistics from a total of 1,000,000 grains. In three dimensions a series of 25 simulations was performed, each beginning with 100,000 grains. Statistical steady-state was achieved when just over 10\% of the grains remained, and samples taken from the resulting configurations gave three-dimensional statistics from a total of 269,555 grains.  For the cross-sections of three-dimensional microstructures, we took a series of cross-sections parallel to the faces of the cubic simulation cell. These were spaced roughly five grain diameters apart to reduce correlations between neighboring cross-sections, giving statistics from a total of roughly 100,000 grains.  Though there appears to be no consensus in the literature as to the precise fraction of the initial grains remaining when the steady-state is reached \cite{2006thomas}, we believe our conditions are conservative.

We report three types of descriptions of steady-state microstructures.  The first are geometric and topological features of individual grains, e.g., distributions of grain sizes and number of neighbors.  We refer to these as point quantities to emphasize that they measure features associated with individual grains, as opposed to collective arrangements of multiple grains.  The second type of description are correlations of point quantities between neighboring grains, where distance between neighboring grains is measured between centers of mass using the standard Euclidean distance.  Third, we investigate correlations of point quantities between neighboring grains, where distance between grains is measured using a topological quantity that we call bond distance; we provide evidence that this measurement of proximity is preferable to the more commonly-used notions of nearest and next nearest neighbors. Finally, given substantial evidence that the steady-state grain-growth microstructure is well-defined, we argue that the steady-state is a more reasonable proxy for real polycrystalline materials than the frequently-used Poisson Voronoi construction.

Where possible, we compare analogous data from two-dimensional simulations, three-dimensional simulations and cross-sections of three-dimensional simulations, denoted 2D, 3D, and 3DX, respectively.  Error bars in all plots indicate one standard error from the mean, though any systematic error is neglected.

\subsection{Point quantities}
\label{sec:point_quantities}

The most frequently reported statistics describing steady-state grain growth microstructures are of features of individual grains.  In this section, we consider distributions of point quantities including the number of faces, surface area, volume, edge lengths, perimeters, mean width, a measure of compactness, and others.  A discussion of Lewis's law \cite{1928lewis} and Feltham's law \cite{1957feltham}, correlations between the number of edges and the size of a grain in two dimensions, and the generalization of these to the three-dimensional structures, is included as well.

\subsubsection{Edges and faces}
\label{subsec:edges_faces}

\begin{figure}
\begin{center}
\subfigure[]{\label{faces_a}\resizebox{0.85\linewidth}{!}{\includegraphics{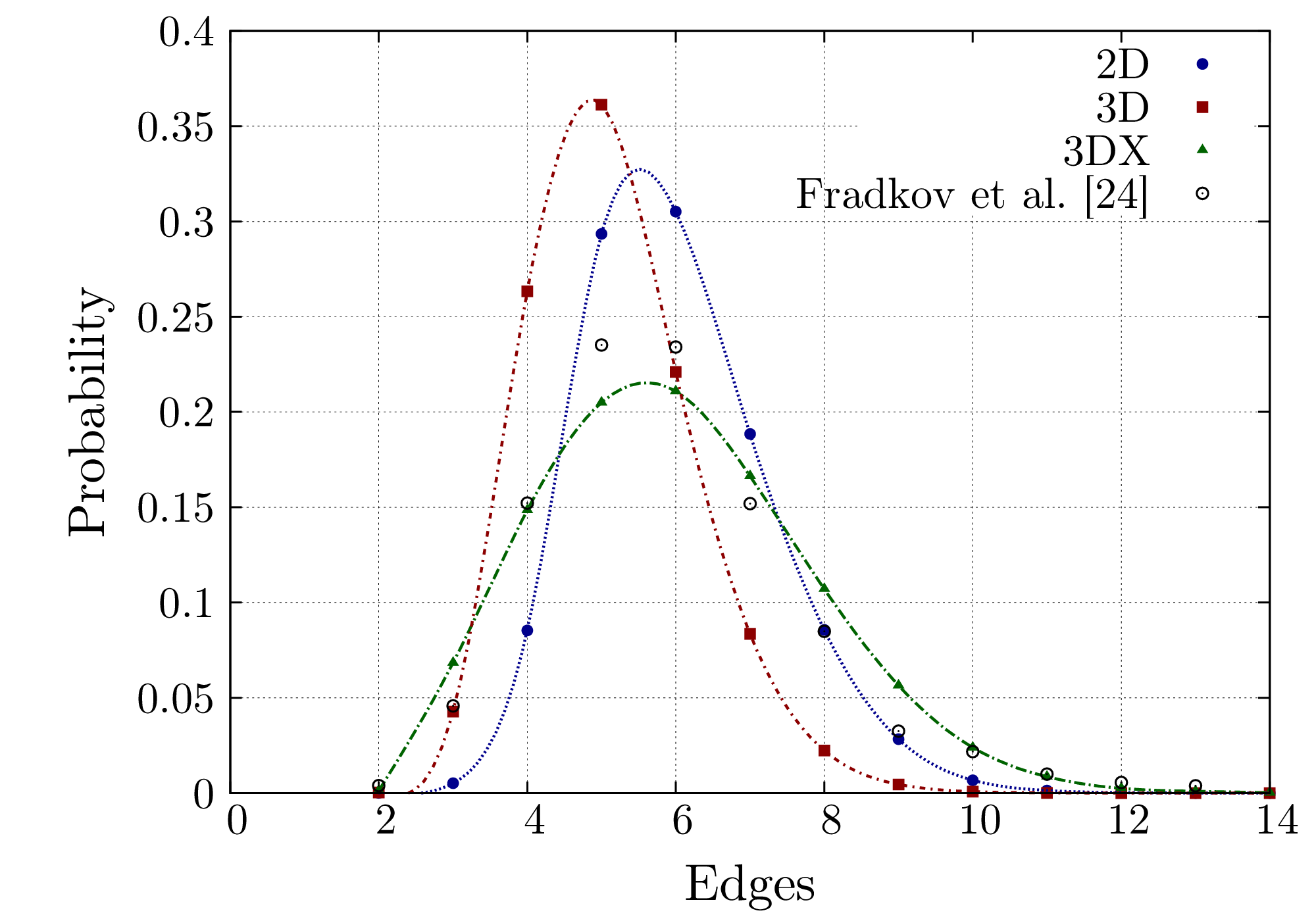}}} \\
\subfigure[]{\label{faces_b}\resizebox{0.85\linewidth}{!}{\includegraphics{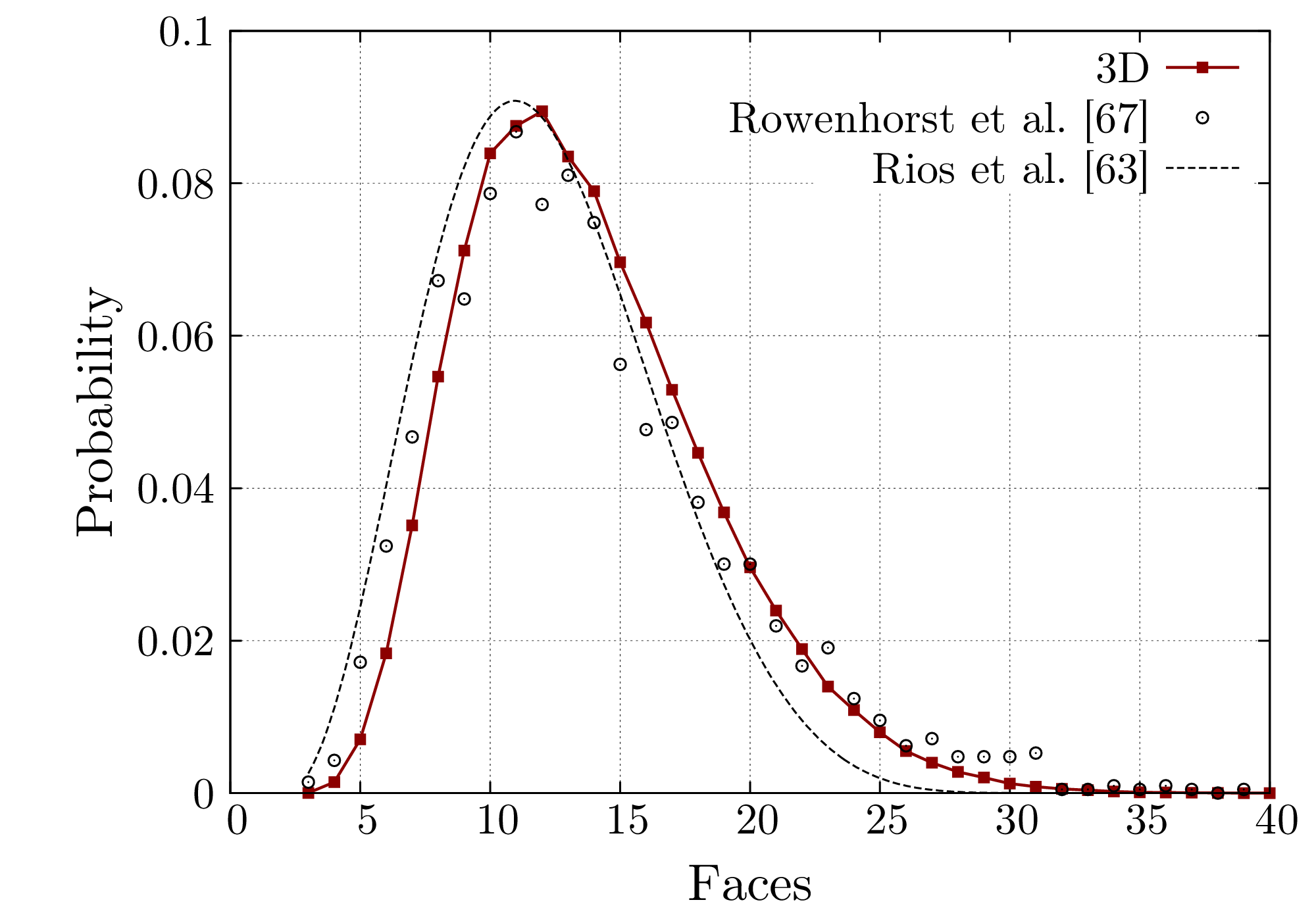}}}
\caption{(a) Edges per grain in 2D and 3DX, and of faces in 3D.  Splines provide a guide for the eye; error bars are smaller than the markers.  Experimental results from a  3DX aluminum sample \cite{1985fradkov-b} are provided for comparison.  (b) Faces per grain in 3D; error bars are smaller than the markers.  Experimental results from a titanium alloy \cite{2010rowenhorst} and a numerical prediction \cite{2008rios} are provided for comparison.}
\label{edges_and_faces}
\end{center}
\end{figure}

The von Neumann--Mullins relation indicates that the number of edges of a grain in the 2D system completely determines the rate of change of a grain's area, suggesting that the distribution of the number of edges is important to the evolution of the microstructure as a whole.  This distribution is reported in Fig.~\ref{faces_a} for the 2D and 3DX systems.  Several analogous quantities may be identified in three dimensions, including the number of edges per face or faces per grain, reported in Figs.~\ref{faces_a} and \ref{faces_b}.  One consequence of Euler's Theorem \cite{1986rivier} is that the average number of edges per grain in the 2D and 3DX systems must be precisely six.  Simulation data shows $6.000 \pm 0.001$ edges per face in 2D, with a standard deviation $1.273 \pm 0.001$; analogous data shows $6.000 \pm 0.006$ edges per face in 3DX, and a standard deviation of $1.813 \pm 0.006$.  In 3D, the corresponding constraints on the average number of edges per face and the average number of faces per grain is weaker \cite{1984weaire, 1989fortes}.  Simulation data indicates $5.1283 \pm 0.0006$ faces per grain in 3D, and a standard deviation of $1.1292 \pm 0.0006$.  In 3D, the average number of faces per grain $\langle F \rangle$ determines the average number of edges per face $\langle E \rangle$ through the relation $\langle F \rangle = 12 / (6 - \langle E \rangle)$.  Since $\langle E \rangle = 5.1283 \pm 0.0006$, we have that $\langle F \rangle \approx 13.766 \pm 0.009$.  A precise justification of these average values in 3D is not known, though several studies predict the smaller value of $13.397$ \cite{1982rivier, 1992kusner, 1994dehoff, 2001hilgenfeldt, 2005glicksman} for the average number of faces  and one predicts $13.564$ \cite{1958coxeter}.  Likewise, there is no known analytic function that gives the distributions in Figs.~\ref{faces_a} and (b), despite some recent attempts \cite{2007rios, 2008rios} for the 3D case.  Nevertheless, the striking similarity in the shapes of the distributions for edges per grain in 2D and for faces per grain in 3D suggests the existence of a shared fundamental cause.  Meanwhile, the difference in the distributions of edges per grain in the 2D and 3DX systems, along with considerable other evidence in the literature, makes clear the difficulty of directly comparing two-dimensional simulations with cross-sections of experimental samples.  This is the case despite suggestive results reported by Anderson et~al.~\cite{1989anderson} where these distributions nearly overlap.

Our results for the 2D system compare well with the distribution of edges per grain as given by other recent simulations \cite{1999fayad, 2006kinderlehrer, 2009elsey}, and provide further evidence that the distribution peaks at six sides rather than five.  This is in contradiction to early simulation \cite{1989anderson} and experimental \cite{1985fradkov-b} results.  Simulation results for the 3DX case occur less frequently in the literature due to the increased computational requirements, though our distribution appears to be consistent with more recent studies \cite{1999weygand, 2009elsey}.  Curiously, the corresponding experimental results \cite{1954beck, 1957feltham, 1969aboav} for cross-sections of three-dimensional samples consistently give distributions with more five-sided faces than six-sided faces.  Since the apparent number of edges in cross-section depends sensitively on the grain geometry, this discrepancy may depend on how a sample was prepared, and may not be generically representative of steady-state grain growth microstructure.

For the 3D system, the distribution of edges per face in our simulations is consistent with that reported for grains in an aluminum alloy \cite{1952williams_a} as well as with the results of several other simulations \cite{1999weygand, 2000wakai, 2002krill}.  Finally, the distribution of faces per grain has been a subject of frequent study, as it was believed that a grain's number of faces guides its growth in 3D in the way in which a grain's number of edges guides it in 2D.  Our findings agree with those of simulations implemented using a variety of approaches \cite{2011elsey, 2006thomas, 2006kim}, as well as with the statistics derived from a population of thousands of grains \cite{2010rowenhorst} obtained from a $\beta$-titanium alloy by serial sectioning techniques.  There is, therefore, ample support for our findings.  One of the more recent theoretical predictions \cite{2008rios} generally performs well, but shows statistically significant deviation from our data -- it predicts more grains with few faces and fewer grains with many faces than we observe.

Regardless of the similarities between Figs.~\ref{faces_a} and (b), we note that there is an important difference in what they measure.  In two dimensions, the combinatorial type of a grain is completely described by its number of faces.  That is, any grain with $n$ edges has exactly the same combinatorial structure as any other grain with $n$ edges.  However, three-dimensional grains are much more complicated, and many distinct polyhedra share the same number of faces \cite{2012lazar}.  For example, grains with 8 faces can have one of 14 distinct types, while those with 9 faces can have one of 50.  As the number of faces increases, the number of distinct combinatorial types increases faster than exponentially.  Therefore, the information indicated in Fig.~\ref{faces_b} is much less descriptive of a three-dimensional system than Fig.~\ref{faces_a} is of a two-dimensional one.

\subsubsection{Areas and volumes}
\label{subsec:areas_volumes}

\begin{figure}
\begin{center}
\subfigure[]{\label{radii_a}\resizebox{0.85\linewidth}{!}{\includegraphics{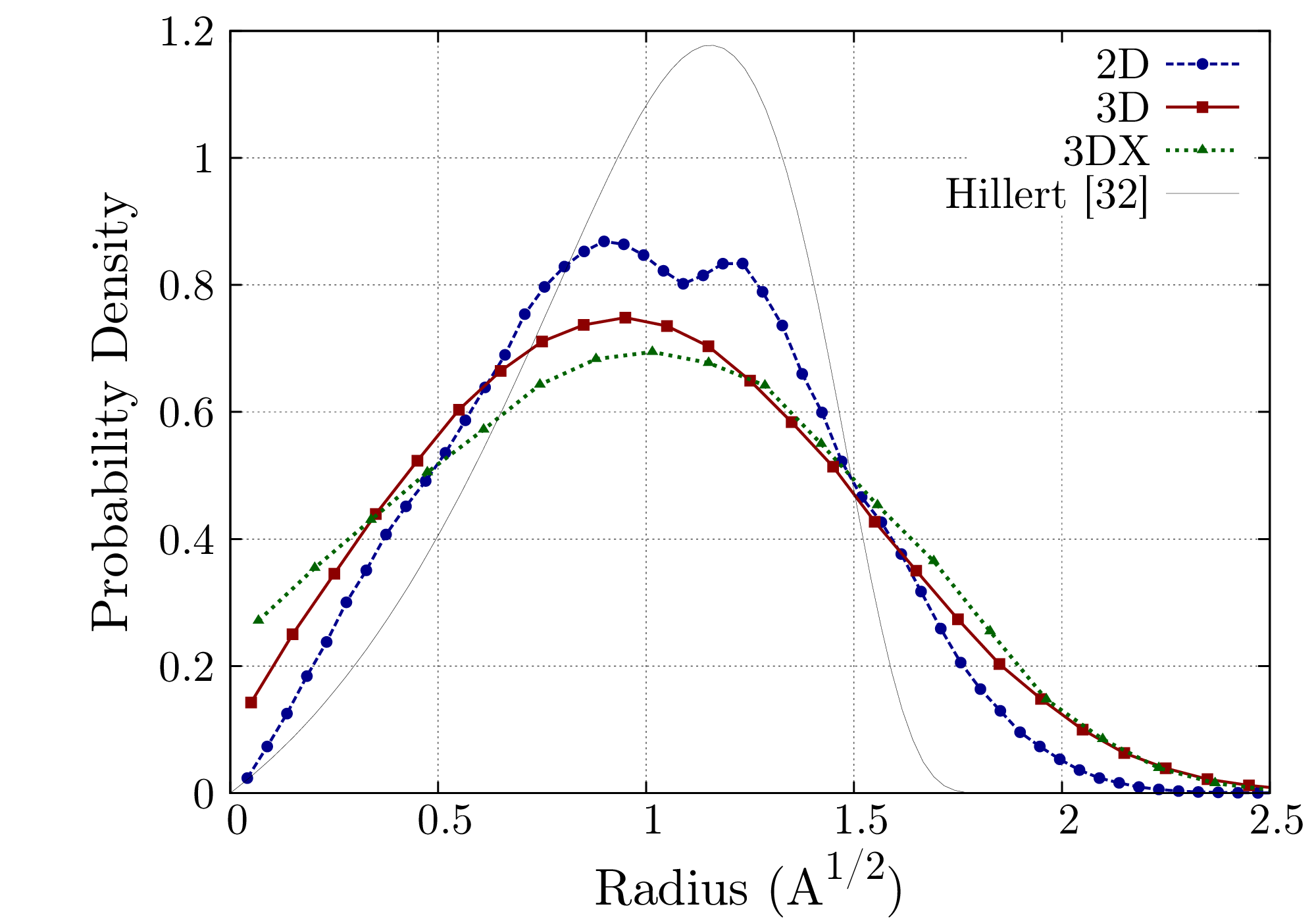}}} \\
\subfigure[]{\label{radii_b}\resizebox{0.85\linewidth}{!}{\includegraphics{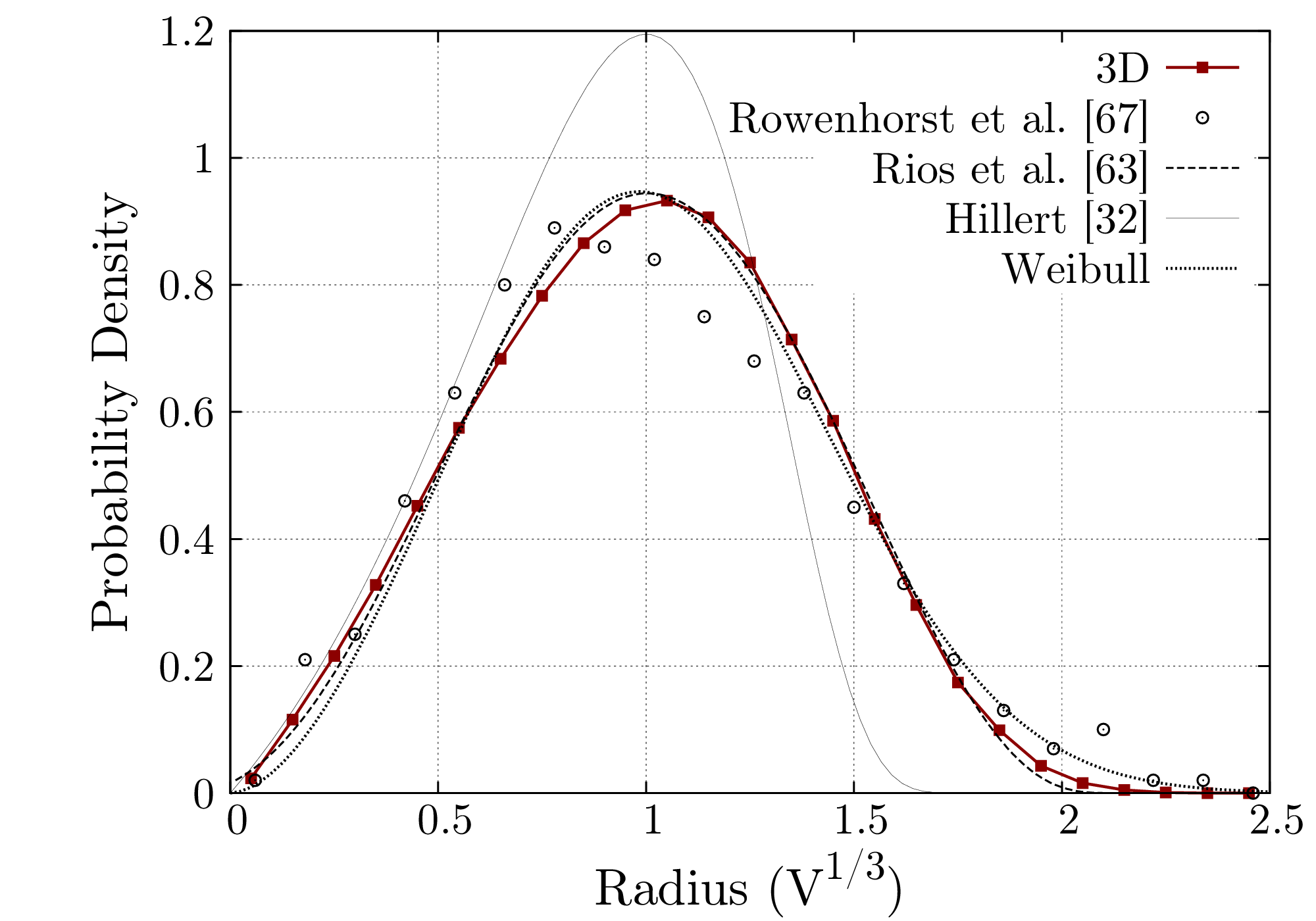}}}
\caption{(a) Effective grain radii for 2D and 3DX, and effective face radii for 3D.  (b) Effective grain radii for 3D.  Predictions \cite{2008rios, 1965hillert} and experimental results \cite{2010rowenhorst} are provided for comparison.  Effective radii given in units of average effective radius; error bars are smaller than the marker size.}
\label{radii}
\end{center}
\end{figure}

One of the more accessible and physically relevant descriptions of a microstructure is the distribution of grain sizes, though that can be measured in different ways.  The most common measurement of size is the effective grain radius, i.e., the square root of the area of a two-dimensional grain or the cube root of the volume of a three-dimensional grain.  Figure \ref{radii_a} gives the distributions of the effective grain radii for the 2D and 3DX systems, as well as of the effective face radii for the 3D system; the prediction of Hillert \cite{1965hillert} for the distribution of grain radii in the 2D system is plotted as well for comparison.  While Hillert's prediction is still often regarded in the literature as a reliable reference, it does not agree with our results.  Figure \ref{radii_b} gives the corresponding distribution for the effective grain radii in the 3D simulation and from thousands of grains in an experimental $\beta$-titanium alloy \cite{2010rowenhorst}.  The Hillert \cite{1965hillert}, Weibull \cite{1999fayad}, and Rios \cite{2008rios} distributions are provided for comparison, where the Weibull distribution is given by
\begin{equation}
f(r) = \frac{k r^{k - 1}}{\lambda^k} \exp \bigg( \frac{-r^k}{\lambda^k} \bigg),
\label{weibull}
\end{equation}
with fitting parameters $k = 2.762 \pm 0.001$ and $\lambda = 1.1570 \pm 0.0003$ as determined by least squares.  The coefficients of determination $R^2$ -- 0.767, 0.991 and 0.998, respectively -- indicate that the Rios distribution follows our simulations most closely.  However, the Weibull distribution performs only slightly less well and has the advantage of a simple algebraic expression.

Several features of the distributions in Fig.~\ref{radii_a} deserve further mention.  One is that the distributions for the 3DX and 3D systems do not pass through the origin, meaning that there is a finite probability of finding a grain or a face with an arbitrarily small effective radius.  The other is that the distribution for the effective radii of the 2D system shows more than one peak; the most pronounced being to either side of an effective radius of one.  A shoulder occurs just below an effective radius of $0.5$ as well, though this is more subtle.  Given the relatively small size of our errors, these features cannot be attributed to noise, and another explanation is required.

\begin{figure}
\begin{center}
\subfigure[]{\label{dist_areas}\resizebox{0.85\linewidth}{!}{\includegraphics{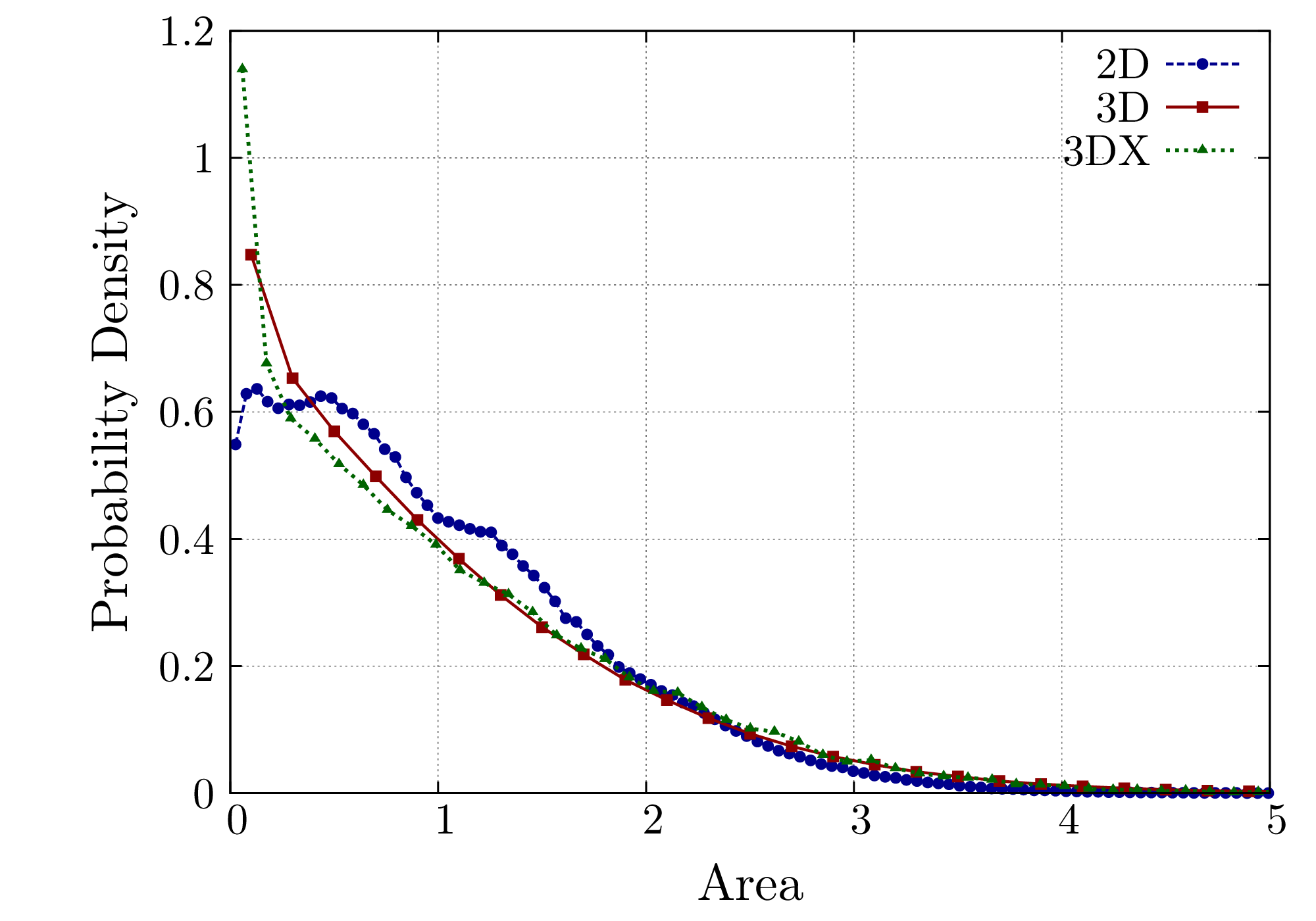}}} \\
\subfigure[]{\label{dist_areas_broken}\resizebox{0.85\linewidth}{!}{\includegraphics{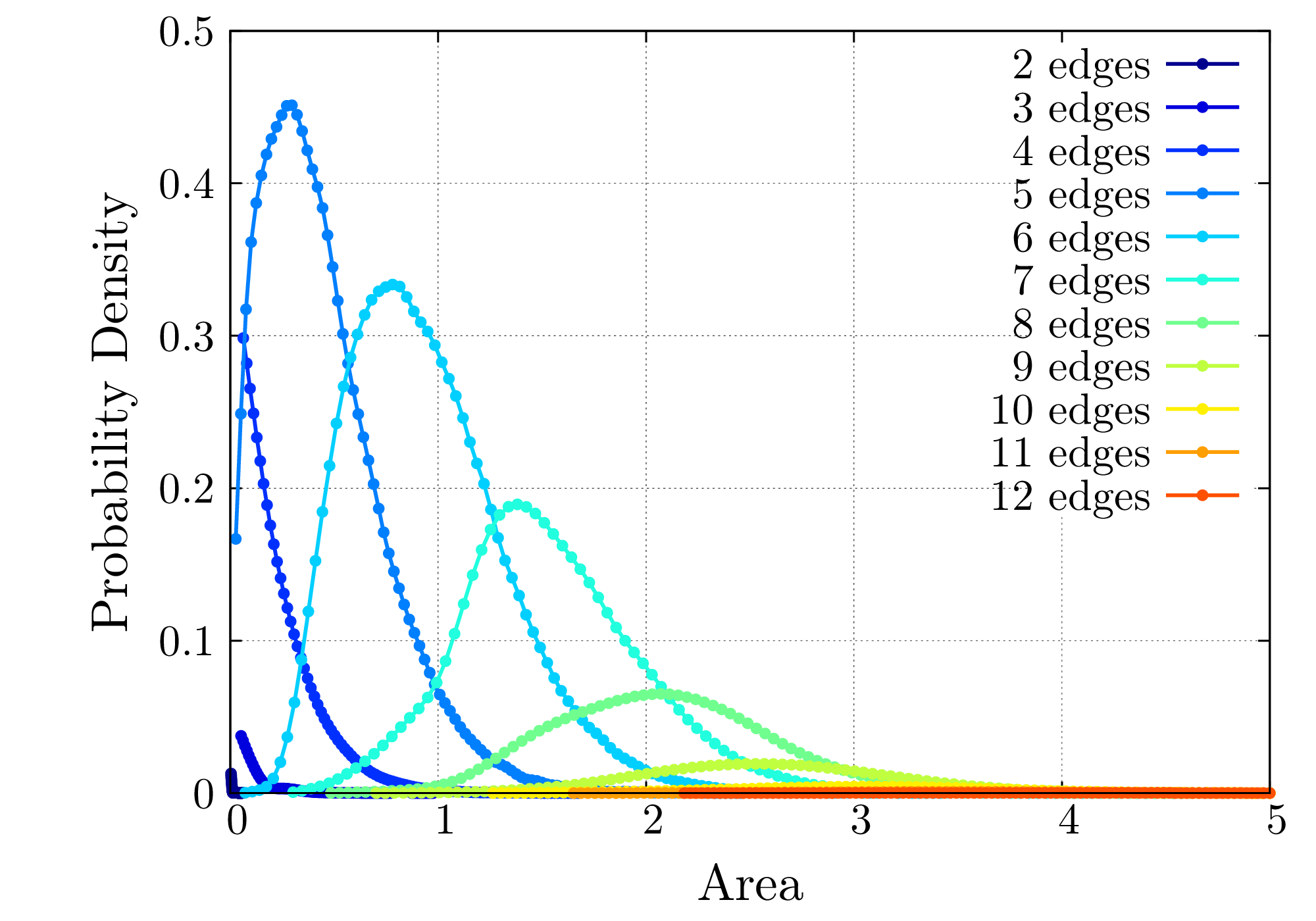}}}
\end{center}
\vspace{-15pt}
\caption{(a) Grain areas in 2D and 3DX, and face areas in 3D.  (b) Probability density functions for grain areas in 2D with a fixed number of edges.  Areas are given in normalized units; error bars are smaller than the marker size. }
\label{areas}
\end{figure}

We first discuss the observation of finite probability at zero size. This discussion is simplified by changing the independent variables of the probability density functions in Fig.~\ref{radii}.  Specifically, a probability density function $f(r)$ of the radius $r=a^{1/2}$ in Fig.~\ref{radii_a} is converted into the corresponding probability density of the area $a$ 
\begin{equation}
g(a) = \frac{a^{-1/2}}{2} f(a^{1/2})
\end{equation}
and is plotted in Fig.~\ref{dist_areas}.  From this vantage, the positive probability of finding a grain or a face with an arbitrarily small effective radius in the 3DX or 3D systems is simply the result of the probability density function of areas in these systems decaying as $a^{-1/2}$ near the origin.  

The explanation for the peaks in the distribution of the effective radii for the 2D system, and the corresponding ones in the probability density function of the areas, is more involved.  For this purpose, we separate the grains of the 2D system into classes with a fixed number of edges.  The probability density functions for the populations of grains in the separate classes appear in Fig.~\ref{dist_areas_broken}, where they may be observed to be smooth functions.  Adding the separate curves in Fig.~\ref{dist_areas_broken} to obtain the distribution in Fig.~\ref{dist_areas} results in peaks precisely where the distribution for grains with four sides intersects that for grains with five, the distribution for grains with five sides intersects that for grains with six, and the distribution for grains with six sides intersects that for grains with seven.  Presumably other peaks occur at the remaining intersections of these distributions, but the number of grains involved may be small enough that the peak is not easily distinguished from the background.  This phenomenon has been recognized in the literature before \cite{1990nagai, 2006kim}, though the apparent importance of the combinatorial structure of the grains to the system as a whole does not seem to be widely appreciated.  This may be due to the relative scarcity of sufficiently precise data to adequately resolve the peaks, leading to several suggestions that the areas follow an exponential distribution \cite{1989anderson, 2000wakai}.

\begin{figure}
\begin{center}
\resizebox{0.85\linewidth}{!}{\includegraphics{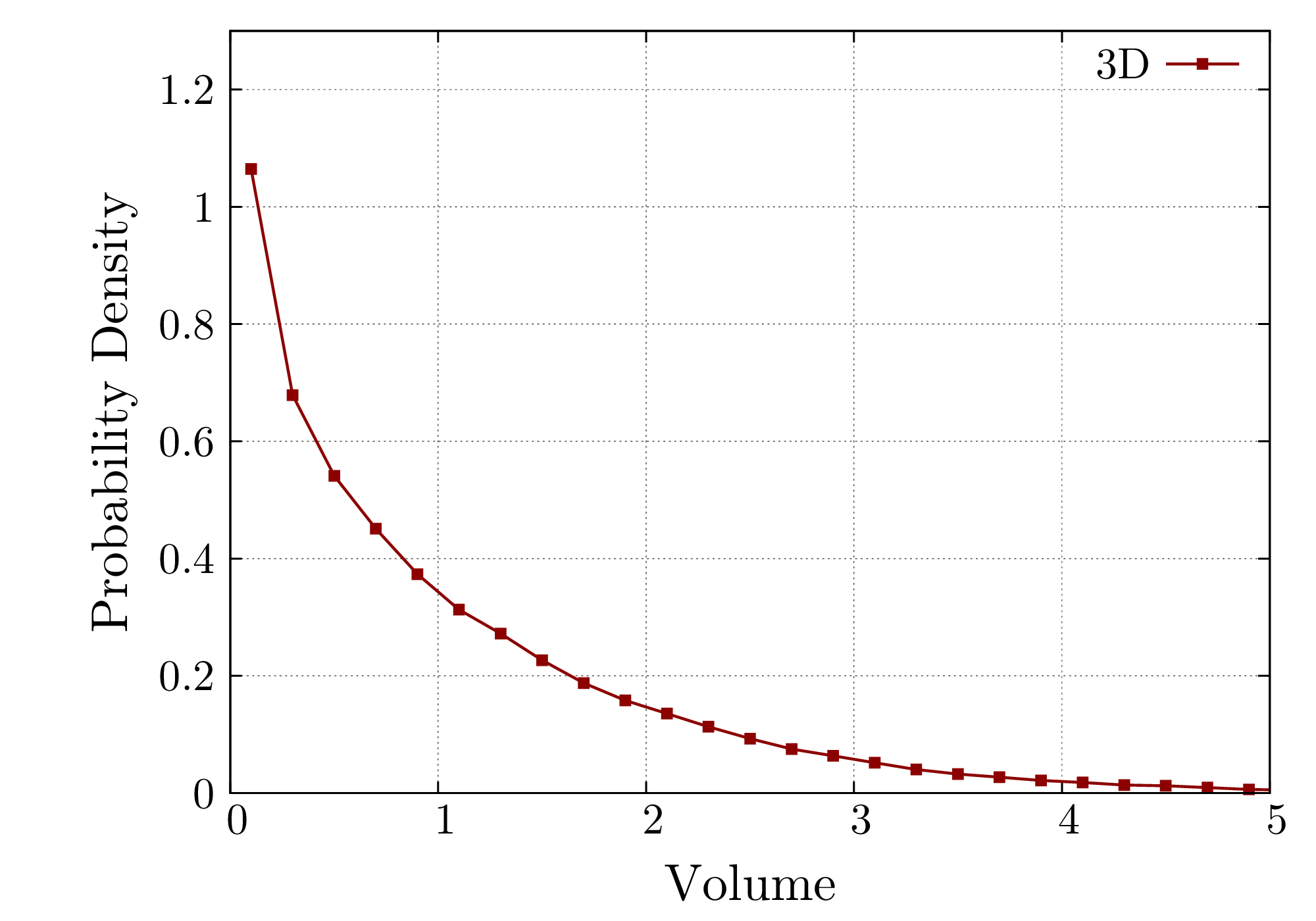}}
\caption{Grain volumes, given in units of average grain volume; error bars are smaller than the marker size.}
\label{volumes}
\end{center}
\end{figure}

Following the transformation from Fig.~\ref{radii_a} to Fig.~\ref{dist_areas}, the probability density function $f(r)$ for the 3D system in Fig.~\ref{radii_b} is converted into a function of volume  $v$
\begin{equation}
g(v) = \frac{v^{-2/3}}{3} f(v^{1/3})
\label{change_to_volume}
\end{equation}
and is plotted in Fig.~\ref{volumes}.  The most notable feature of the resulting probability density function is the resemblance to an exponential distribution.  This striking result has been reported previously \cite{1989anderson, 2000wakai, 2009wang}.  An exponential distribution of volumes implies a Weibull distribution for effective grain radii with $k = 3$ and $\lambda = 1$ in Equation \ref{weibull}, found by inverting Equation \ref{change_to_volume}.   The reason why a (near) exponential distribution is observed remains unknown.

A second notable feature of the grain volume distribution is the absence of multiple peaks analogous to those for the area distribution of grains in the 2D system.  One possible explanation is that the separate distributions corresponding to those in Fig.~\ref{dist_areas_broken} more nearly overlap \cite{2009wang}, making the transitions less abrupt.  Alternatively, the MacPherson--Srolovitz relation indicates that the combinatorial type of a grain in a 3D system does not directly govern its volume evolution, suggesting that a division of the grains into distinct classes based on combinatorial type may not be especially meaningful.

We note that many grain size distributions have been reported (e.g., see recent simulation data for 2D \cite{2006kinderlehrer, 2006kim, 2009elsey}, 3DX \cite{2000wakai, 2009elsey, 2011elsey} and 3D \cite{2002krill, 2009wang, 2011elsey}).  While these results appear consistent between the different studies, the very large samples represented in the data presented here bring new features to the fore, including the statistically significant deviation of the grain volume distribution from a pure exponential.

\subsubsection{Lewis's law, Feltham's law, and generalizations}
\label{subsec:lewis_feltham}

\begin{figure}[t]
\begin{center}
\subfigure[]{\label{edges_areas}\resizebox{0.85\linewidth}{!}{\includegraphics{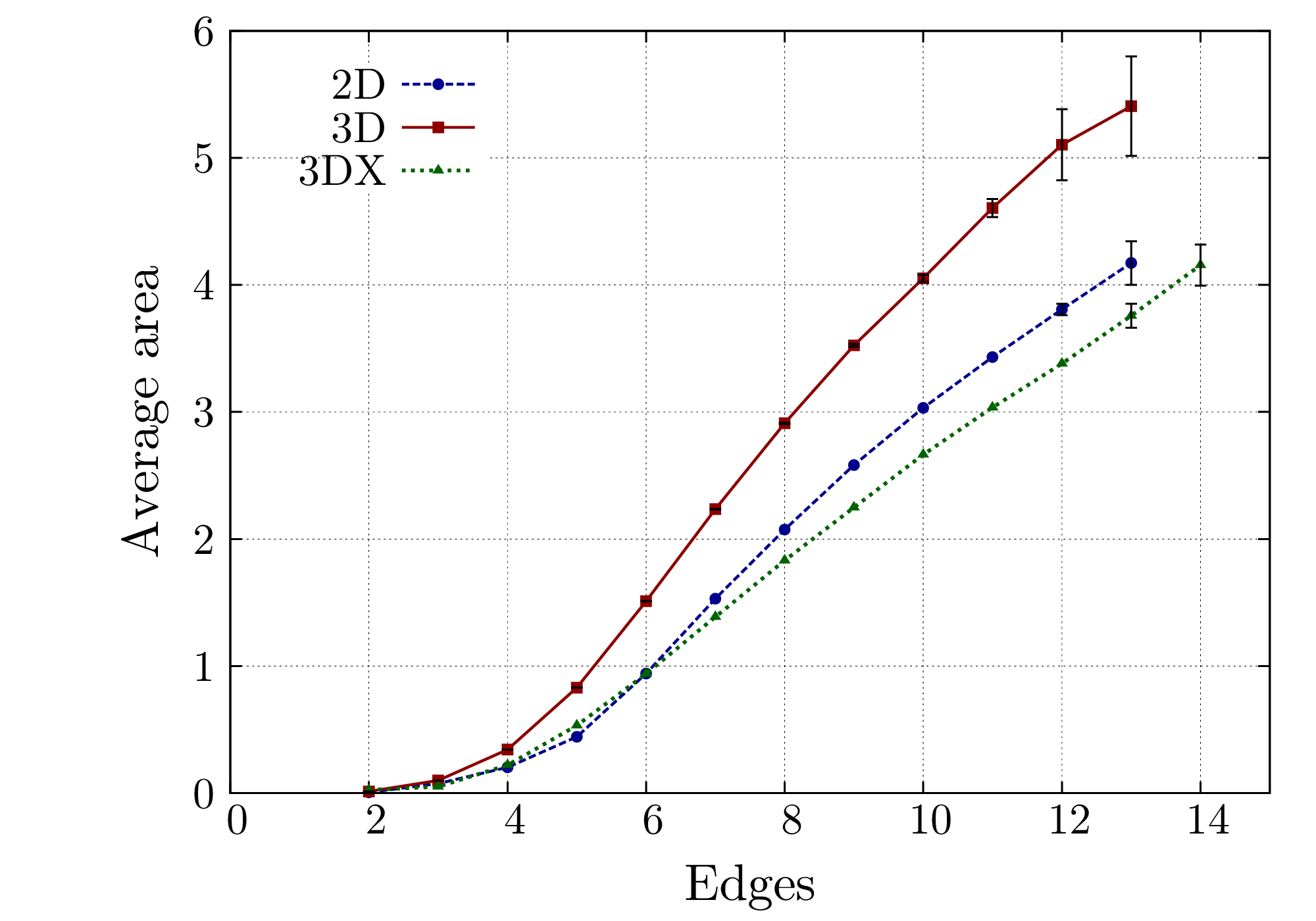}}} \\
\subfigure[]{\label{edges_perims}\resizebox{0.85\linewidth}{!}{\includegraphics{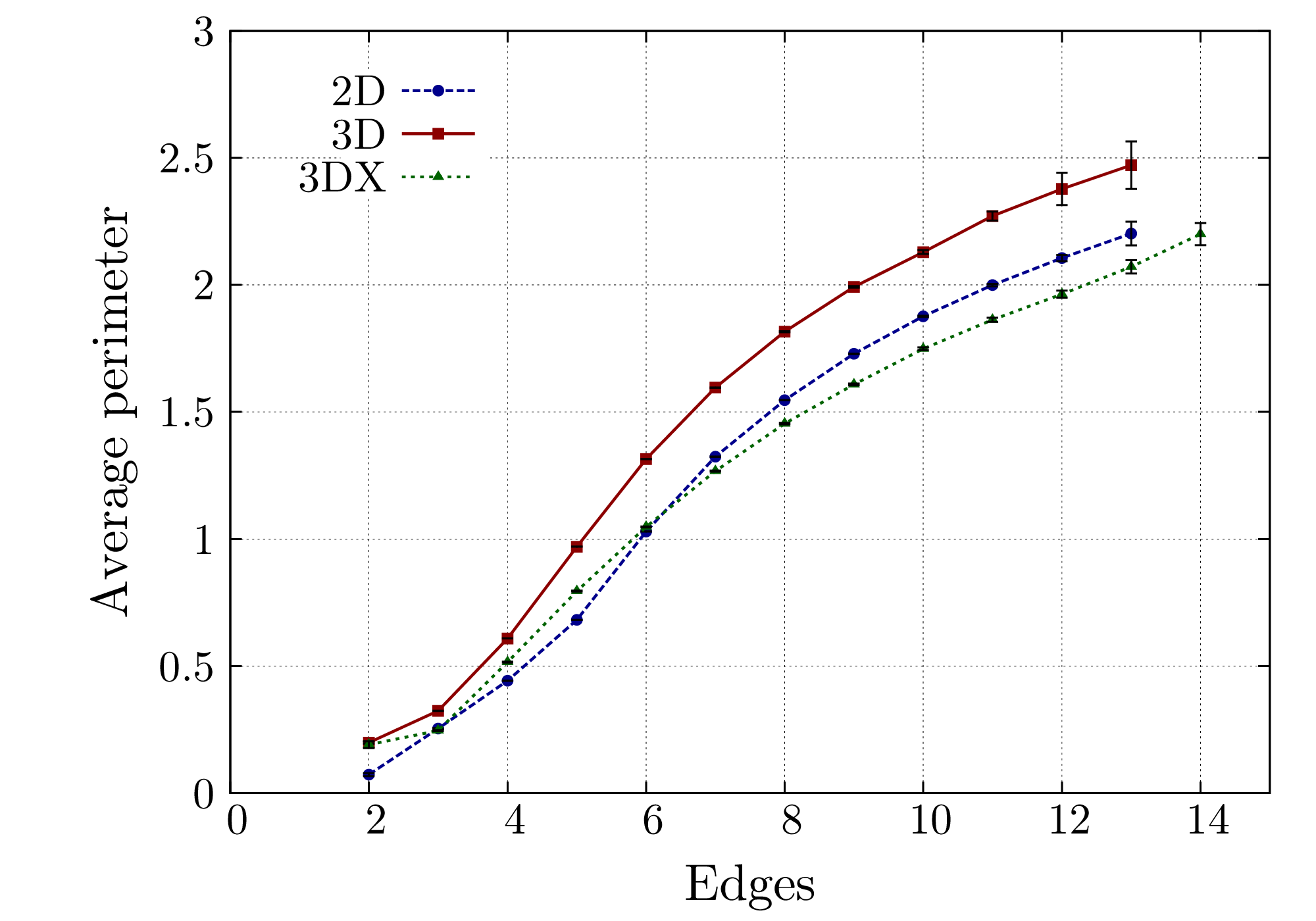}}}
\end{center}
\vspace{-15pt}
\caption{(a) Average area of grains (2D and 3DX) or faces (3D) with a given number of edges.  (b) Average perimeter of grains (2D and 3DX) or faces (3D) with a given number of edges.  Areas and perimeters given in normalized units; some error bars are smaller than the marker size.}
\label{edges_perim_area}
\end{figure}

While the importance of combinatorially distinct populations of grains to the peaks in the grain area distribution function in 2D is not widely recognized, the importance of combinatorially distinct classes has been considered in other contexts.  When investigating a section of cucumber epidermis, Lewis \cite{1928lewis} noticed that cell areas appeared to be proportional to their number of sides $n$, for values of $n$ ranging from about four to seven.  This relationship is often referred to as Lewis' law, and plays an important part in the application of the maximum entropy formalism to the structural properties of cellular networks \cite{1982rivier, 1985rivier, 1986rivier}.  While it is widely believed that Lewis' law is a general result, holding over a wide range of $n$ (neglecting Lewis' original qualification), the areas of the grains in the 2D system do not appear to follow Lewis' law over an appreciable range; see Fig.~\ref{edges_areas}.  Similarly, the same figure indicates that Lewis' law neither applies widely to grain areas in 3DX nor to face areas in the 3D system.  This conclusion is supported by other literature \cite{1993stavans, 1998weygand}, as well.

\begin{figure*}
\begin{center}
\begin{tabular}{ccc}
\subfigure[]{\label{faces_vs_a}\resizebox{0.33\linewidth}{!}{\includegraphics{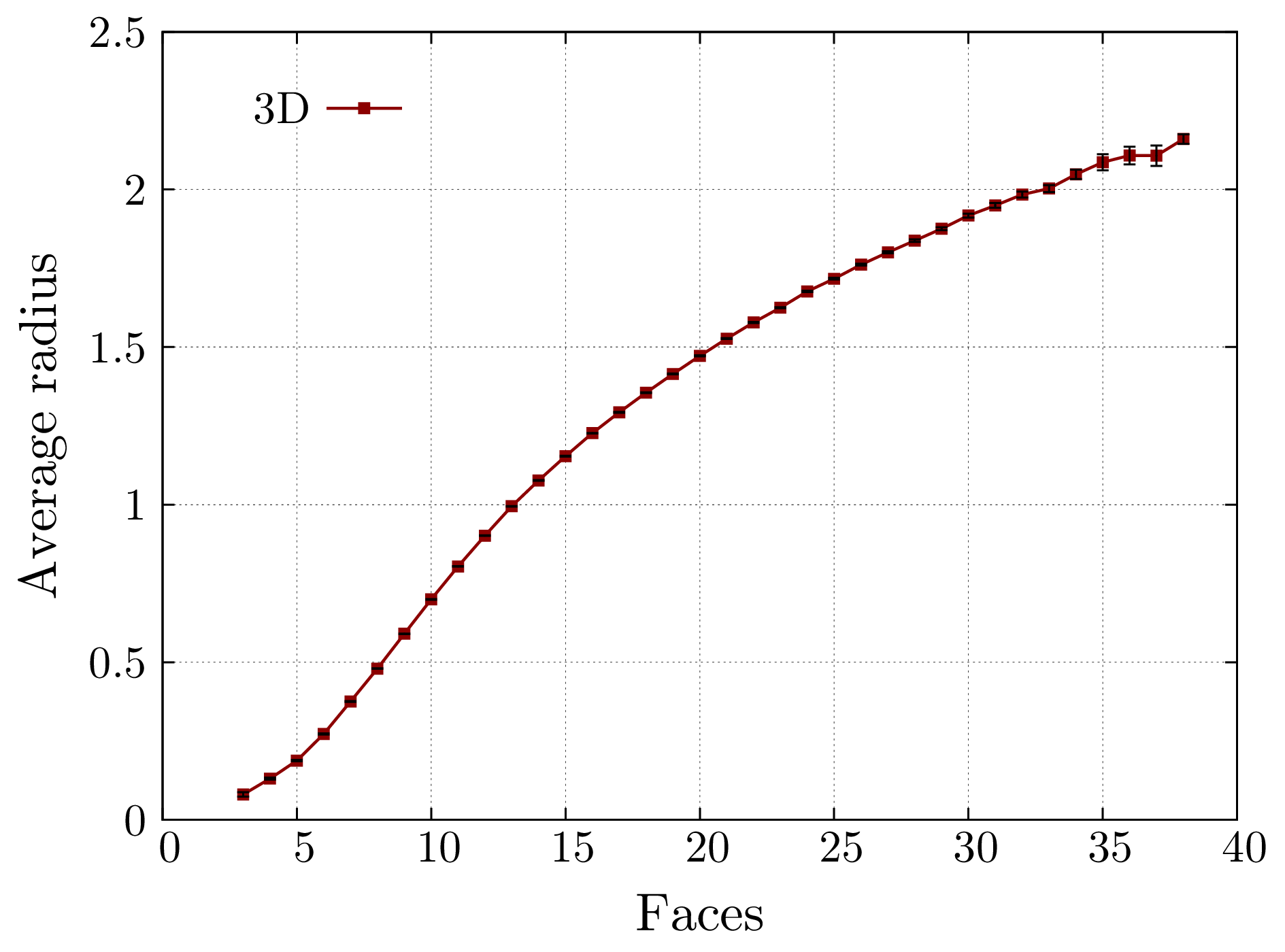}}} &
\subfigure[]{\label{faces_vs_b}\resizebox{0.33\linewidth}{!}{\includegraphics{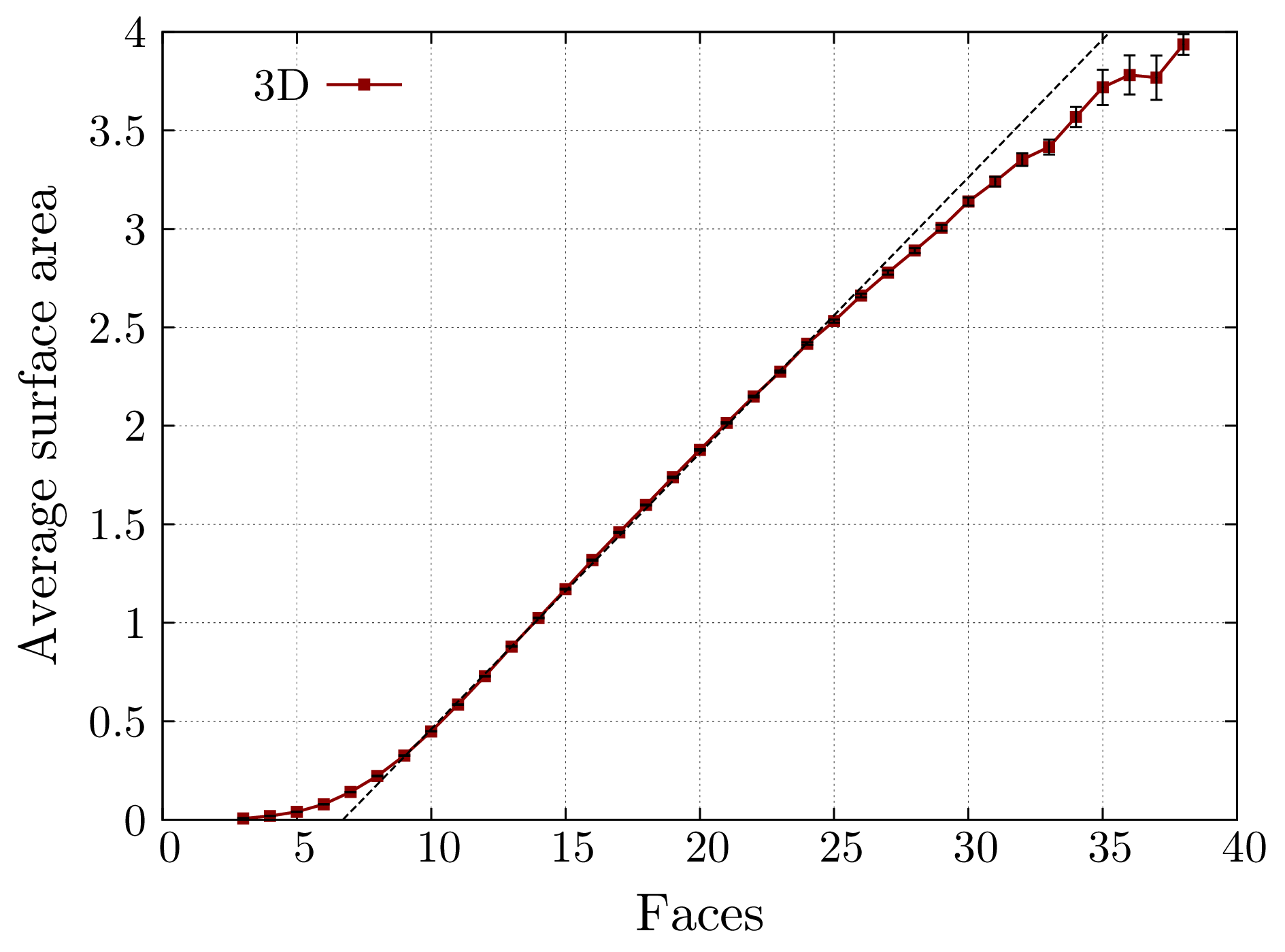}}} &
\subfigure[]{\label{faces_vs_c}\resizebox{0.33\linewidth}{!}{\includegraphics{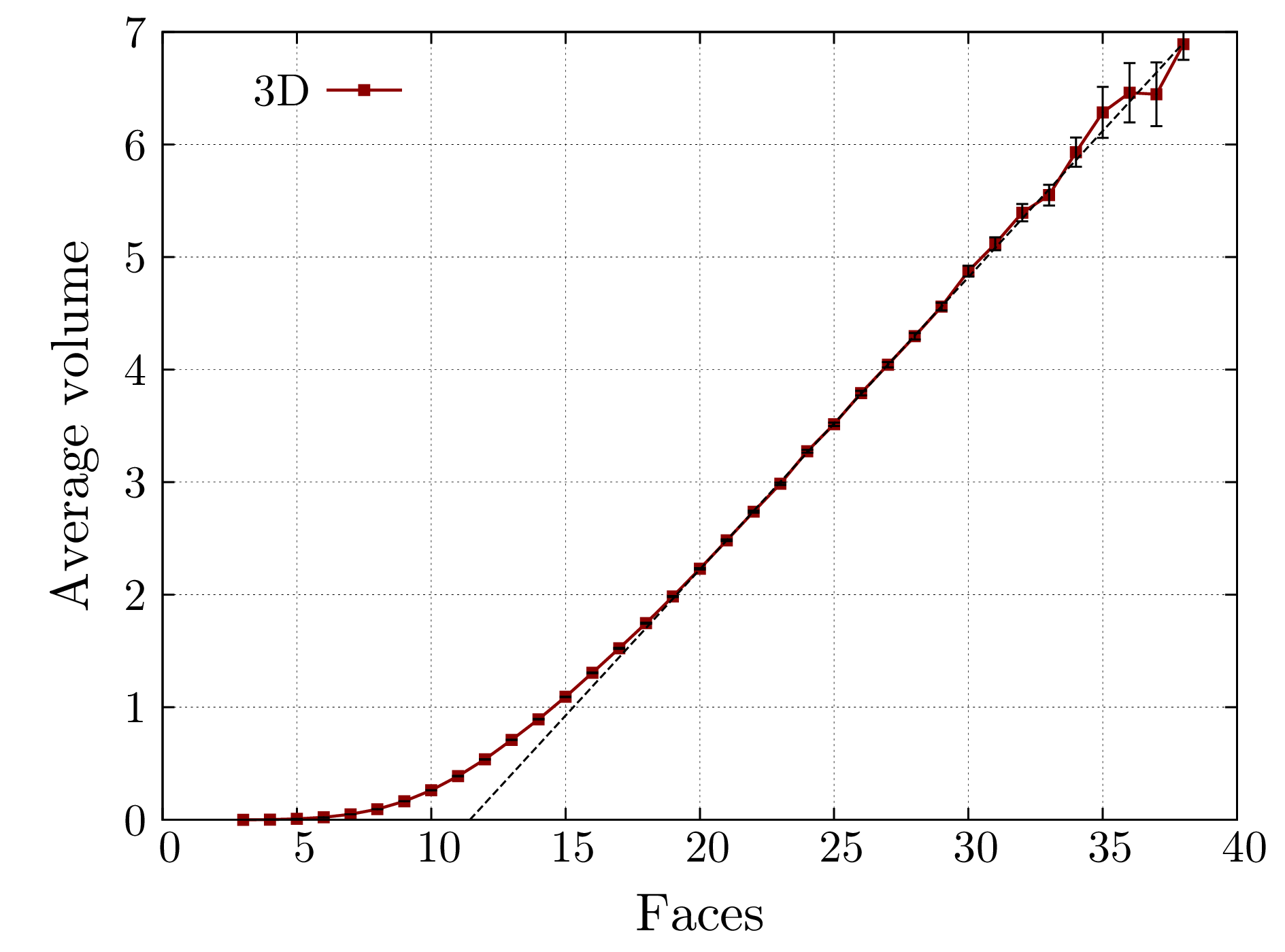}}}
\end{tabular}
\caption{(a) Average effective radius, (b) surface area and (c) volume of grains with a given number of faces for 3D.  Using least-squares fitting, the average volume of grains with $n$ faces for $n \ge 20$ follows the linear relation $a \cdot n + b$ with $a = 0.260 \pm 0.001$ and $b = 2.97 \pm 0.03$.  Effective radius, surface area and volume given in normalized units; some error bars are smaller than the marker size.}
\label{volumes_and_surface_areas}
\end{center}
\end{figure*}

The situation is slightly better for the extension of this relation to 3D in Fig.~\ref{faces_vs_c}, where the average grain volume appears to increase linearly with the number of faces $n$ for $n \ge 20$.  This is consistent with the literature \cite{1995fuchizaki}, though Fig.~\ref{faces_b} indicates that this extension of Lewis' law would hold only for a relatively small fraction of the grains in the system.

Feltham \cite{1957feltham} provided experimental data on annealed tin in support of the claim that the average linear measure of a two-dimensional grain increases linearly with its number of edges; this is referred to as Feltham's law \cite{1992ling, 1995szeto, 1997dealmeida, 1998thiele}.  Figure \ref{edges_perims} indicates how perimeters of grains in the 2D and 3DX systems, and of faces in the 3D system, increase with number of edges.  Of course, a single system cannot simultaneously follow Lewis' law and Feltham's law, and the respective accuracy of the two is a question that has received some attention \cite{1995szeto}.  With regard to our simulations though, Feltham's law does not appear to hold for any of the cases in Fig.~\ref{edges_perims}.  Nor does Feltham's law hold for the analogous plot of the effective radii of grains in the 3D system in Fig.~\ref{faces_vs_a}.  Our statistically significant conclusion for an accurate normal grain growth simulation is supported by several other simulations in 2D \cite{1998saito, 1998weygand} and 3D \cite{2000wakai, 2011elsey}, but not by all \cite{1989anderson,1998saito, 2005kim}.

Finally, Fig.~\ref{faces_vs_b} considers a relationship that lies between Lewis' law and Feltham's law, namely, the dependence of the average surface area of a grain in 3D on its number of faces.  There is a portion of the plot that could be considered linear, though the noticeable deviations for low and high values of $n$ discourages any proposal that this is a fundamental feature of the system.

\subsubsection{Edge lengths, perimeters, and surface areas}
\label{subsec:edge_perim_area}

\begin{figure}[t]
\begin{center}
\subfigure[]{\label{dist_edge}\resizebox{0.85\linewidth}{!}{\includegraphics{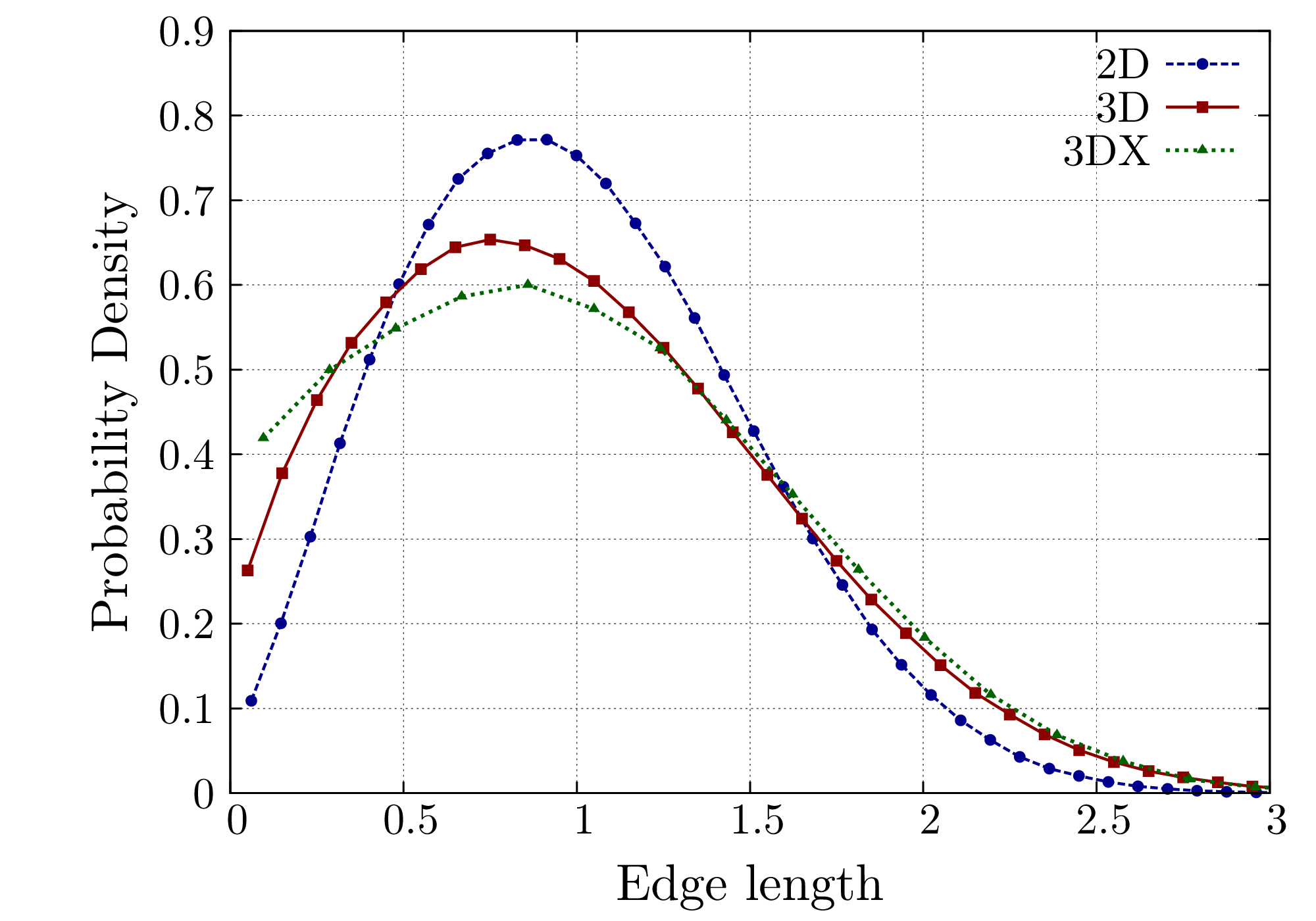}}} \\
\subfigure[]{\label{dist_perim}\resizebox{0.85\linewidth}{!}{\includegraphics{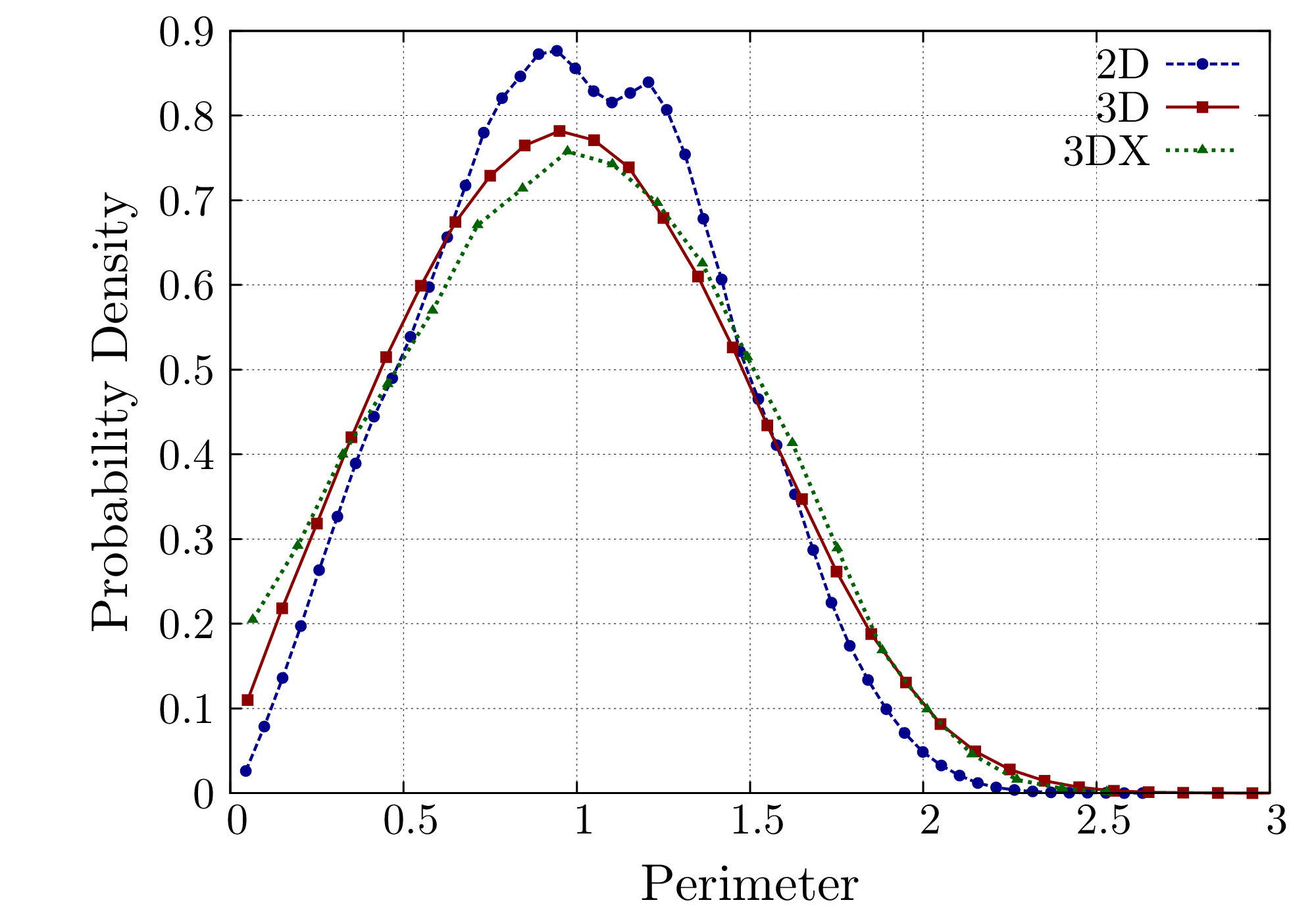}}}
\end{center}
\vspace{-15pt}
\caption{(a) Edge lengths in 2D, 3DX and 3D.  (b) Grain perimeters in 2D and 3DX and face perimeters in 3D.  Edge lengths and perimeters given in normalized units; error bars are smaller than the marker size.}
\label{edges_and_perimeters}
\end{figure}

Having studied the most common measures of grain size in Section \ref{subsec:areas_volumes}, we now consider the size of features on the boundaries of grains.  Figure \ref{dist_edge} shows the distribution of edge lengths for the 2D, 3DX and 3D systems.  As noted previously with regard to Fig.~\ref{radii_a}, the structure of a genuinely two-dimensional structure is noticeably distinct from a cross-section of a three-dimensional one.  A second striking feature is that there is a finite probability of finding an arbitrarily short edge in all three systems.  However, since the  2D and 3D evolution equations do not reference the features of individual edges, there is no reason to question the existence of arbitrarily short edges in normal grain growth situations.  Although this distribution appears infrequently in the literature, Thomas \cite{2006thomas} performed a Monte Carlo simulation of a 3D system and reported an edge length distribution substantially different from ours.  We suspect that this is attributable to the difficulty of measuring edge lengths for voxelized grains, particularly for short edges.

The absence of the grain perimeter in the evolution equations may be used to explain the finite probability of an arbitrarily small perimeter for a grain in the 3DX system and for a face in the 3D system as well, as indicated in Fig.~\ref{dist_perim}.  This does not apply for the perimeters of grains in the 2D system though, since Fig.~\ref{edges_perims} indicates that a grain with a small perimeter will generally have few edges, and therefore becomes smaller (to the point of vanishing).  The most conspicuous feature of Fig.~\ref{dist_perim} though is the resemblance to Fig.~\ref{radii_a}.  Indeed, they should be related by a rescaling of the independent variable, since both give a probability density function of a linear measure of the size of a grain.  This implies that the peaks in the distribution for the 2D system result from the separation of the grains into combinatorially distinct populations.

\begin{figure}[t]
\center
\includegraphics[width=0.85\linewidth]{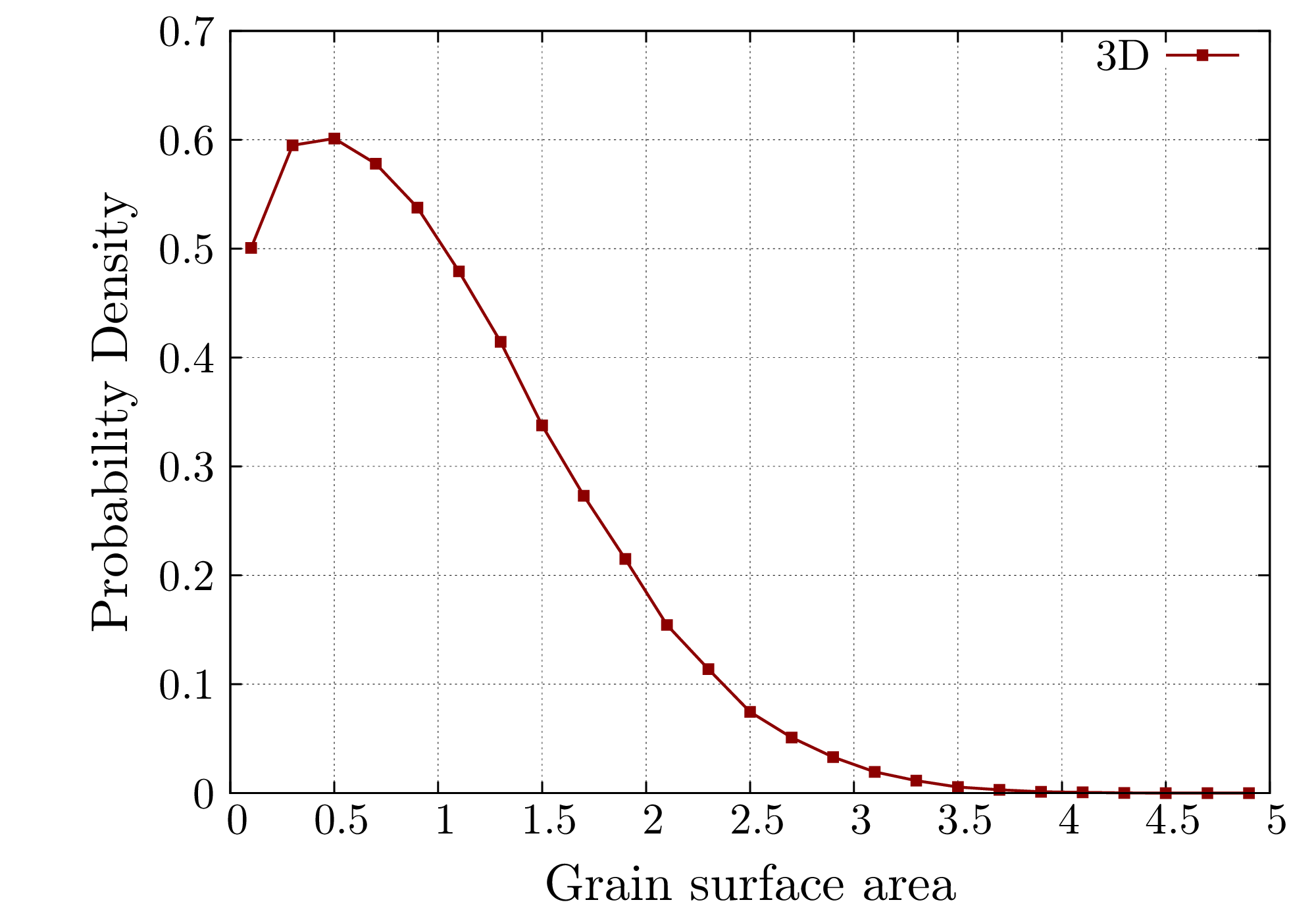}
\caption{Grain surface areas, in normalized units; error bars are smaller than the marker size.}
\label{3D_metric_normalized_surface_area}
\end{figure}

The remaining boundary element is unique to three-dimensional grains.  Figure \ref{3D_metric_normalized_surface_area} shows the distribution of surface areas of grains for the 3D system; similar distributions of grain surface areas have been reported previously \cite{2000wakai, 2006thomas}.  Although the bin for smallest grain surface area shows a substantial probability density, the trend in this distribution does not necessarily imply that the probability density at zero grain surface area is finite.

\subsubsection{Isoperimetric ratios}
\label{subsec:isoperimetric}

\begin{figure}
\begin{center}
\subfigure[]{\label{roundness_a}\resizebox{0.85\linewidth}{!}{\includegraphics{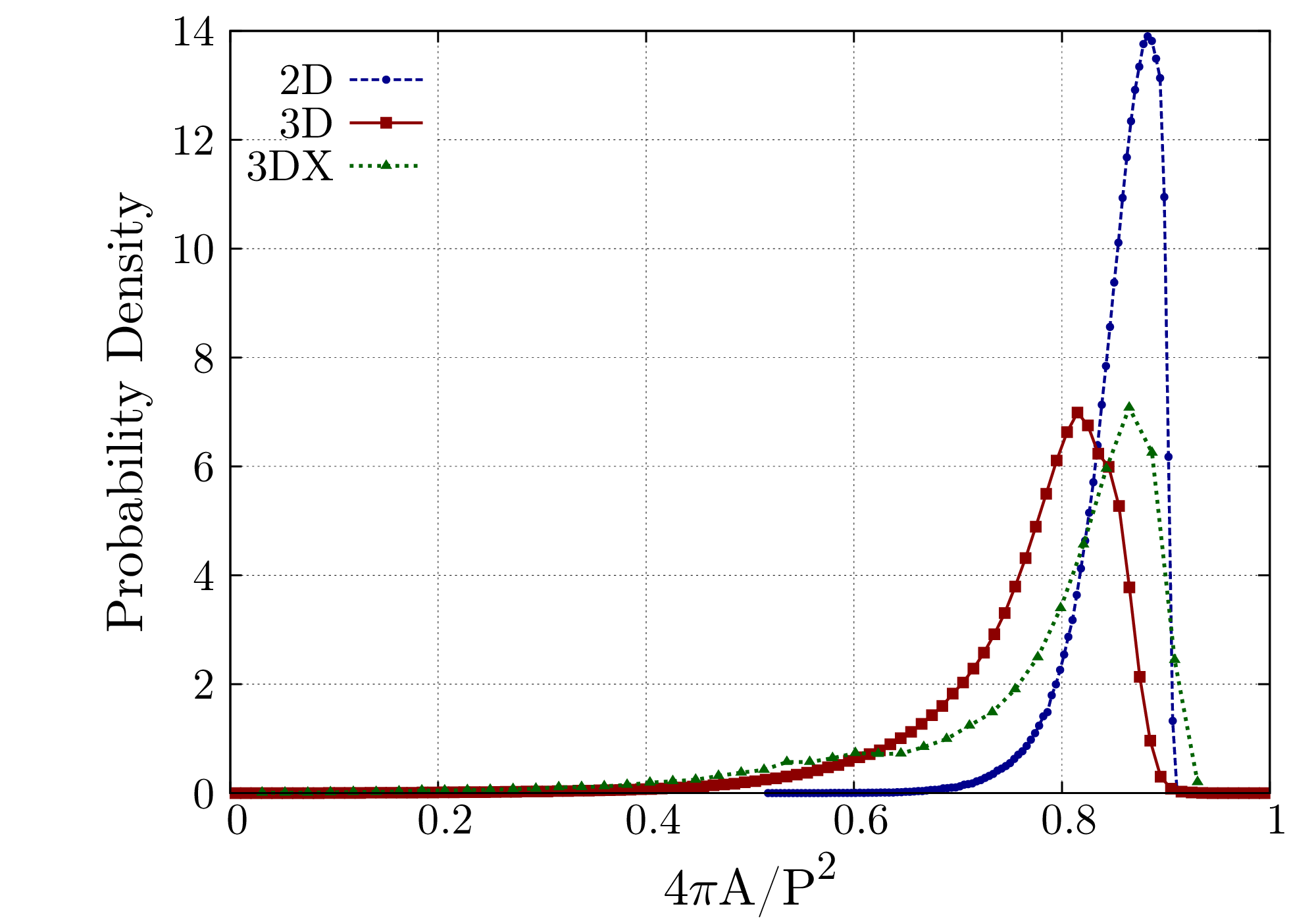}}} \\
\subfigure[]{\label{roundness_b}\resizebox{0.85\linewidth}{!}{\includegraphics{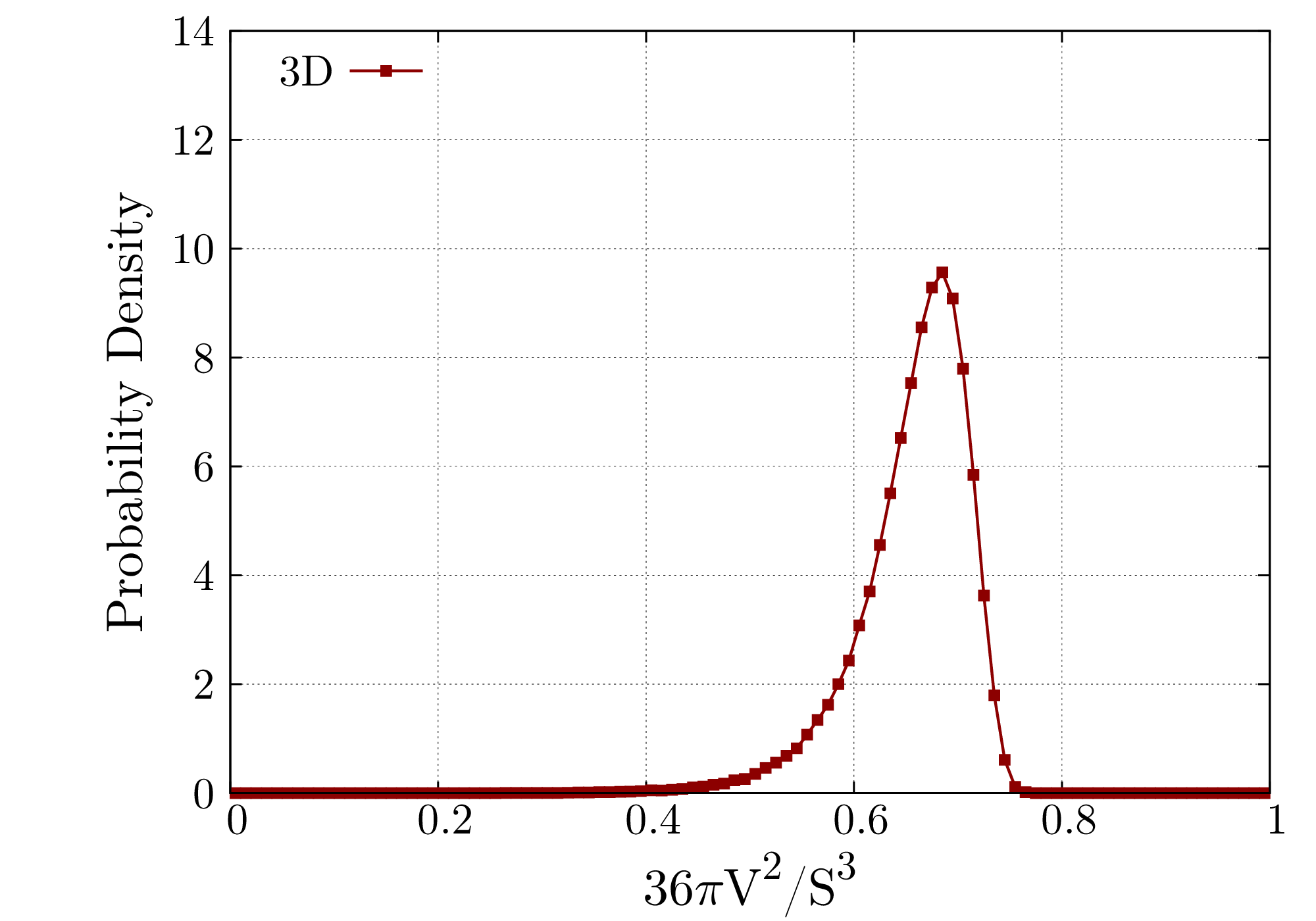}}}
\end{center}
\vspace{-15pt}
\caption{(a) Isoperimetric ratios for grains in 2D and 3DX, and for faces in 3D.  (b) Isoperimetric ratios for grains in 3D.  Error bars are smaller than the marker size.}
\label{roundness_measures}
\end{figure}

Another way to characterize individual grains is by the deviation of their shape from a circle or sphere in two or three dimensions, respectively.  For a grain in two dimensions or a face in three dimensions, the isoperimetric ratio is defined as the ratio of the area $A$ of a grain to the area of a circle with the same perimeter $P$, or $4 \pi A / P^2$.  Figure \ref{roundness_a} shows this quantity for grains in 2D and 3DX, and for grain faces in 3D.  Similarly, for a grain in three dimensions the isoperimetric ratio is defined as the ratio of the square of the volume $V$ of a grain to the square of the volume of a sphere having the same surface area $S$, or $36 \pi V^2 / S^3$.  Figure \ref{roundness_b} shows the distribution of this measure for grains from the 3D system.

The isoperimetric inequality states that the isoperimetric ratio is bounded above by one, with equality only for a sphere.  Clearly, all of the data peak close to unity, indicating that grains in normal grain growth microstructures are quite close to spheres (this is consistent with the microstructures seen in Fig.~\ref{cell_systems}).  The average isoperimetric ratios of grains in the 2D and 3DX systems, and of faces in the 3D system, are $0.85628 \pm 0.00004$, $0.7817 \pm 0.0004$ and $0.76803 \pm 0.00005$, respectively.  That grains of the 2D system are considerably more circular than grains of the 3D and 3DX systems should be expected from energetic considerations, since the energy of the 2D system is proportional to the total boundary length.  Evolution along the steepest descent in energy therefore drives the grains of the 2D system to be more circular, subject to the constraint that they remain space filling.  For comparison, we note that the isoperimetric ratio for (space-filling) regular hexagons is $\approx 0.907$.

Meanwhile, the average isoperimetric ratio of grains in 3D is $0.6571 \pm 0.0002$.  This should not be directly compared with the corresponding quantity in two dimensions though, since the isoperimetric ratio in two and three dimensions is the quotient of quantities with dimensions of the second and sixth powers of length, respectively.  Accounting for this difference by taking the cube root of $0.6571$ gives $0.8694$, implying that  that grains in three dimensions are slightly more spherical than those in two dimensions are circular, as may be expected from the increase in geometrical degrees of freedom with dimension.  For comparison, we note that the isoperimetric ratio for (space-filling) truncated octahedra is $\approx 0.753$.

While the energetic argument, above, applies to normal grain growth structures, the grains of other cellular microstructures are often substantially less spherical.  For instance, the cells in Voronoi tessellations of Poisson distributed points have average isoperimetric ratios of $0.7281 \pm 0.0001$ and $0.53320 \pm 0.00005$ in two and three dimensions, respectively.

\subsubsection{Mean width and related quantities}
\label{subsec:mean_width}

\begin{figure}
\begin{center}
\subfigure[]{\label{mean_width_a}\resizebox{0.85\linewidth}{!}{\includegraphics{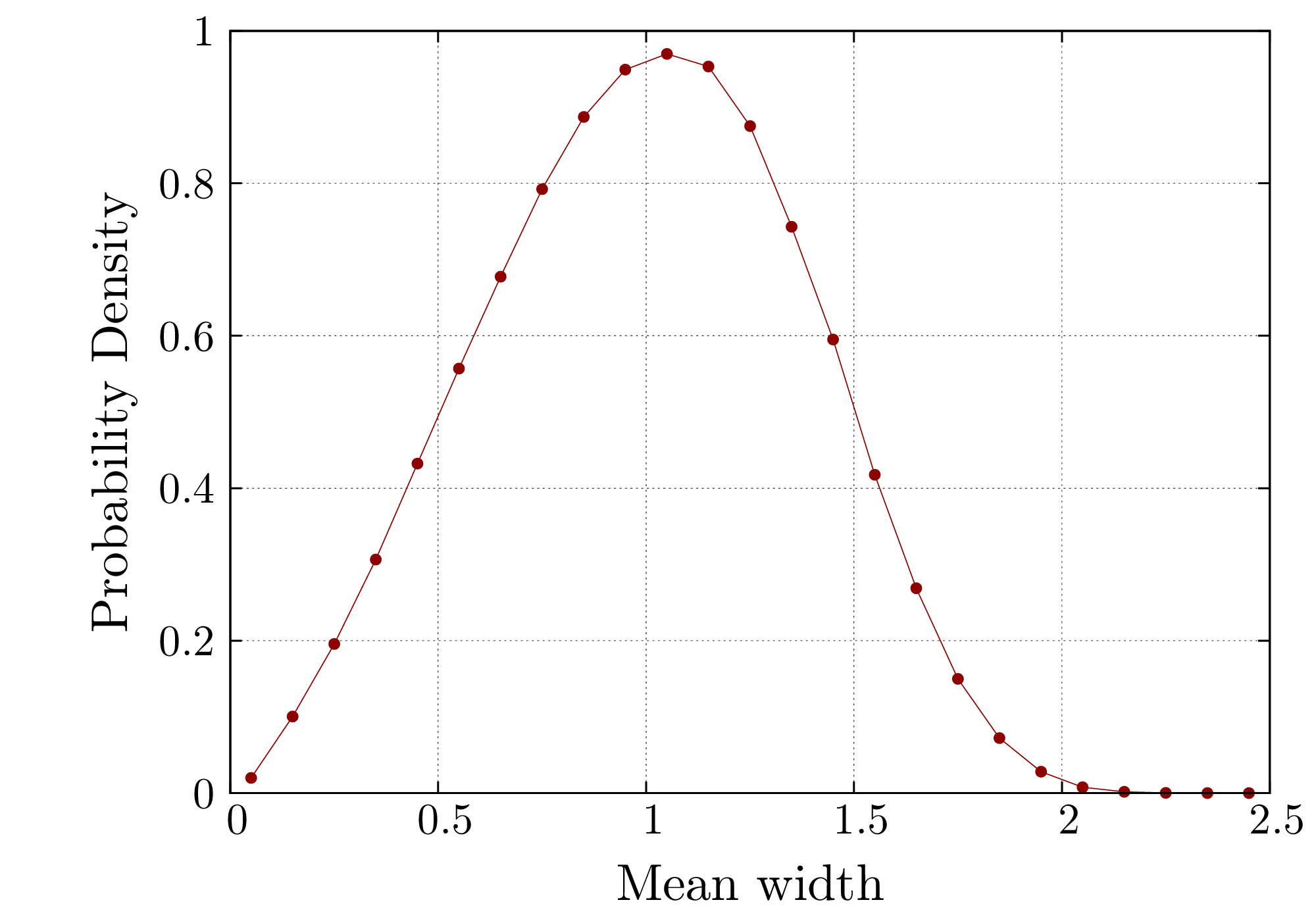}}} \\
\subfigure[]{\label{mean_width_b}\resizebox{0.85\linewidth}{!}{\includegraphics{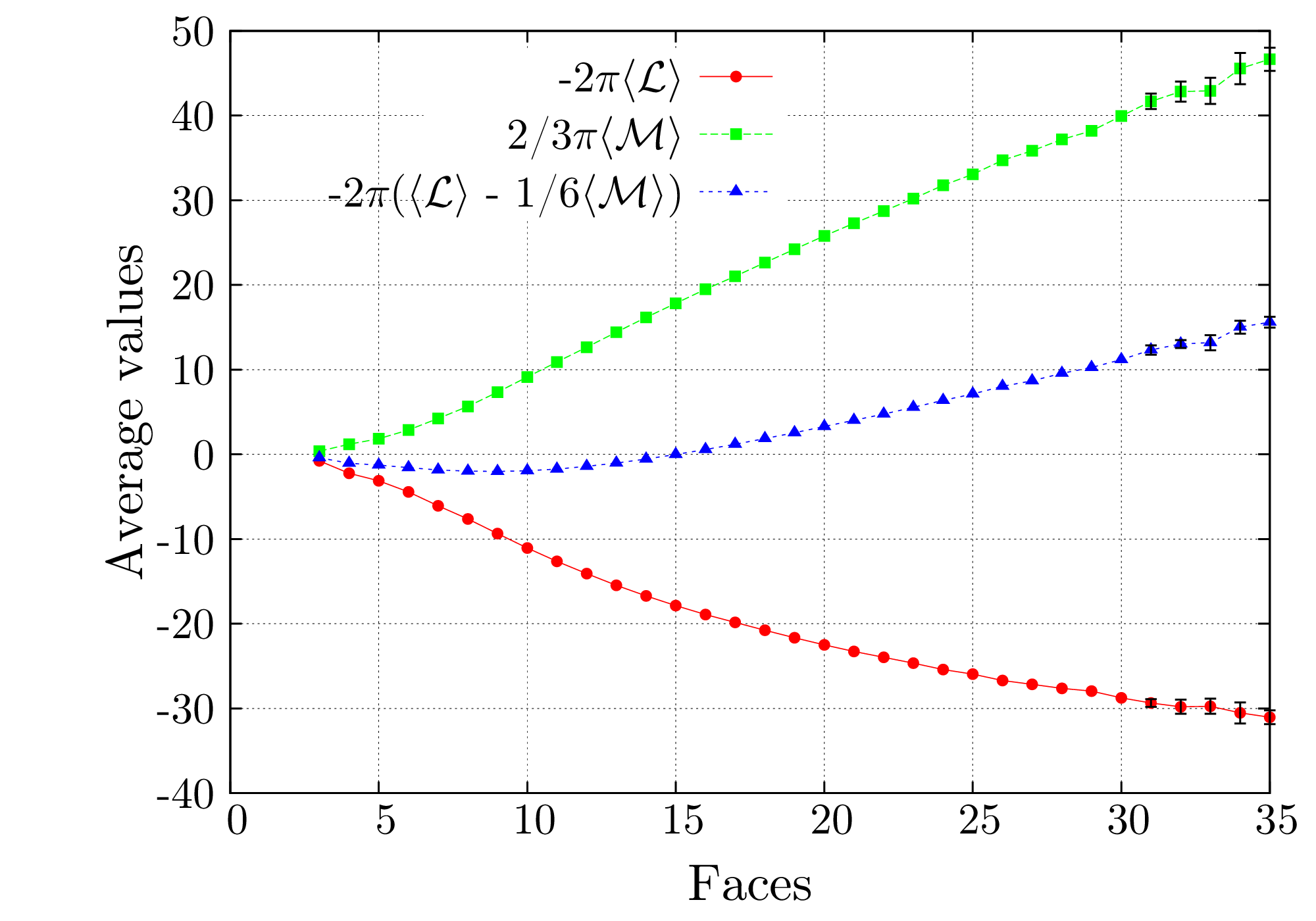}}}
\caption{(a) Mean widths of grains, in normalized units; error bars are smaller than the marker size.  (b) Average mean width, average sum of triple edge lengths bounding a grain, and average sum of those quantities over all grains with a given number of faces.}
\label{mean_width_stuff}
\end{center}
\end{figure}

One measure of particular importance in view of the MacPherson--Srolovitz relation is the mean width of grains in three dimensions. The probability distribution for the mean width is plotted in Fig.~\ref{mean_width_a}.  Since the importance of the mean width for normal grain growth has only recently been appreciated \cite{2007macpherson}, no experimental studies and relatively few simulations (e.g., see \cite{2008mora}) report this quantity.  Inspection of Fig.~\ref{mean_width_a} reveals that this distribution nearly coincides with the plot in Fig.~\ref{radii_b}.  While the coincidence of these probability density functions does not necessarily imply that the mean width of a particular grain scales with the cubed root of its volume, this observation may help to formulate a closed set of equations involving the MacPherson--Srolovitz relation (Eq.~\ref{evn3d}), much in the same way that the Hillert theory of grain growth \cite{1965hillert} is based upon the von Neumann--Mullins relation.

The sum of the bounding triple edge lengths appears together with the mean width in the MacPherson-Srolovitz relation for grain growth in 3D.  To  explore the interaction of the mean width $\mathcal{L}$ of a grain and the sum of the bounding triple edge lengths $\mathcal{M}$ in determining the growth rate, we consider the averages of these quantities over all grains with a fixed number of faces in Fig.~\ref{mean_width_b}.  When scaled appropriately, these quantities may be summed to determine the average growth rate of grains with a given number of faces (plotted in blue).  An analysis of regular polyhedra \cite{2005glicksman} gives a prediction for the mean width of grains as a function of their number of faces, though the difference of the prediction from our results reveals that grains with small number of faces often differ substantially from regular polyhedra.  Following the same line of reasoning, a prediction \cite{2009glicksman} for the rate of change of volume of a grain roughly follows the blue curve in Fig.~\ref{mean_width_b} for grains with many faces.  The difference between this prediction from our results for grains with few faces is quite significant though, since grains with few faces often evolve the most rapidly.

Notice that the expected mean width and sum of the triple edges of a grain increase with its number of faces for grains with many faces.  For grains with few faces, the mean width dominates the MacPherson--Srolovitz equation, and the grain shrinks.  For grains with many faces, the sum of the triple edges dominates the equation and the grain grows.  The crossover occurs for grains with approximately $15$ faces.  Curiously, there does not appear to be agreement between any of the simulations results \cite{2006thomas, 2008mora, 2010syha} corresponding to the blue curve in Fig.~\ref{mean_width_b}.  The origin of this inconsistency is not known, though the frequent disagreement for grains with small numbers of faces suggests that the cause may be an inconsistent treatment of small grains.

\subsection{Quantities correlated over metric distance}
\label{sec:metric_dist}

Many features of a microstructure depend not only on properties of its individual grains, but also on their arrangement relative to one another.  A variety of functions may be used to characterize these geometric features, including the two-point correlation function for two-phase materials \cite{2009jiao, 2008fullwood} and the radial pair correlation function for liquids or amorphous solids \cite{2002weeks, 2010lee}.  The quantity measured in this section is similar to the usual radial pair correlation function, but is applied to the grains of a microstructure rather than to atoms of a liquid.

\begin{figure*}
\begin{center}
\begin{tabular}{ccc}
\subfigure[]{\label{corr-a}\resizebox{0.31\linewidth}{!}{\includegraphics{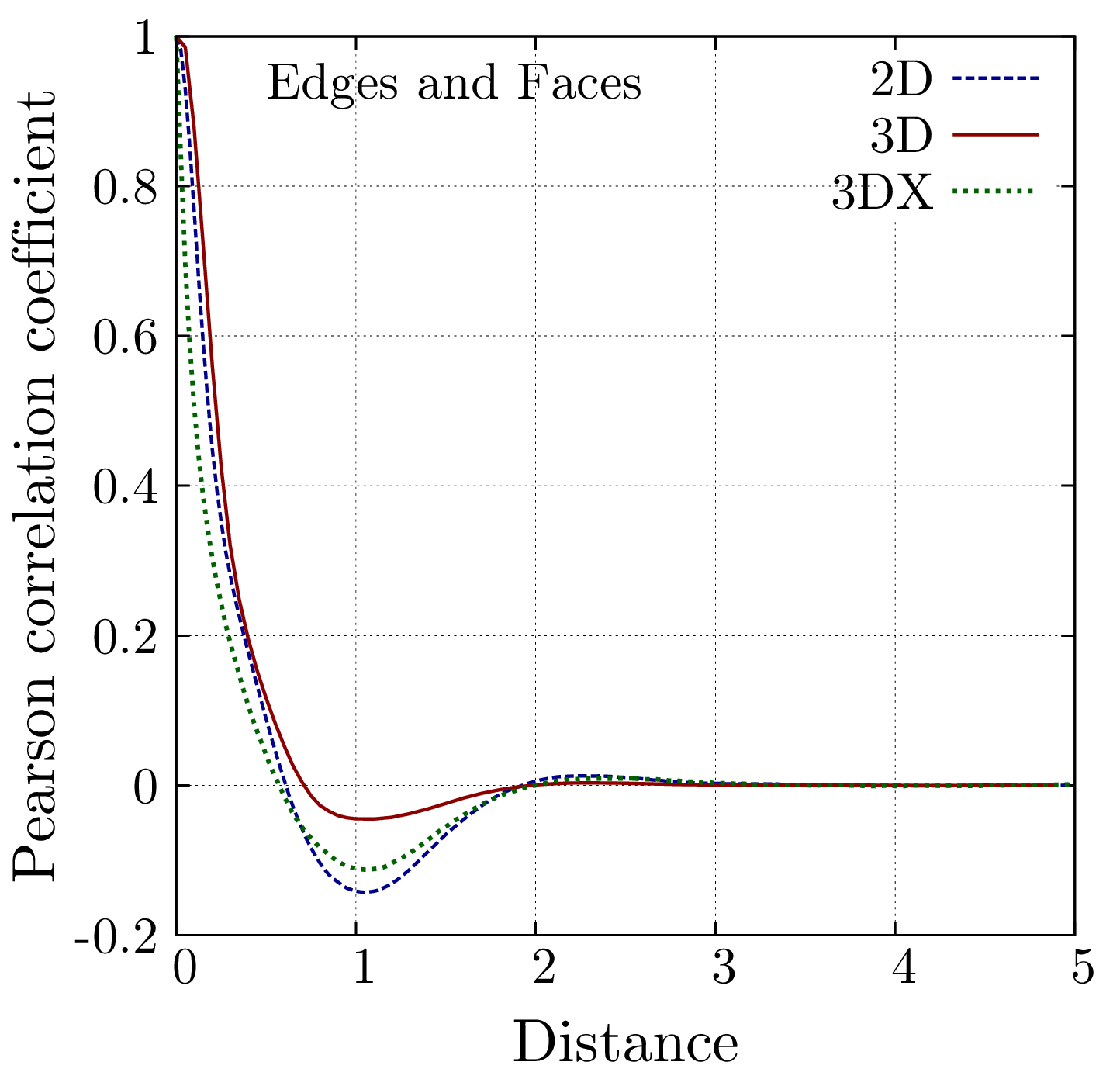}}} &
\subfigure[]{\label{corr-b}\resizebox{0.31\linewidth}{!}{\includegraphics{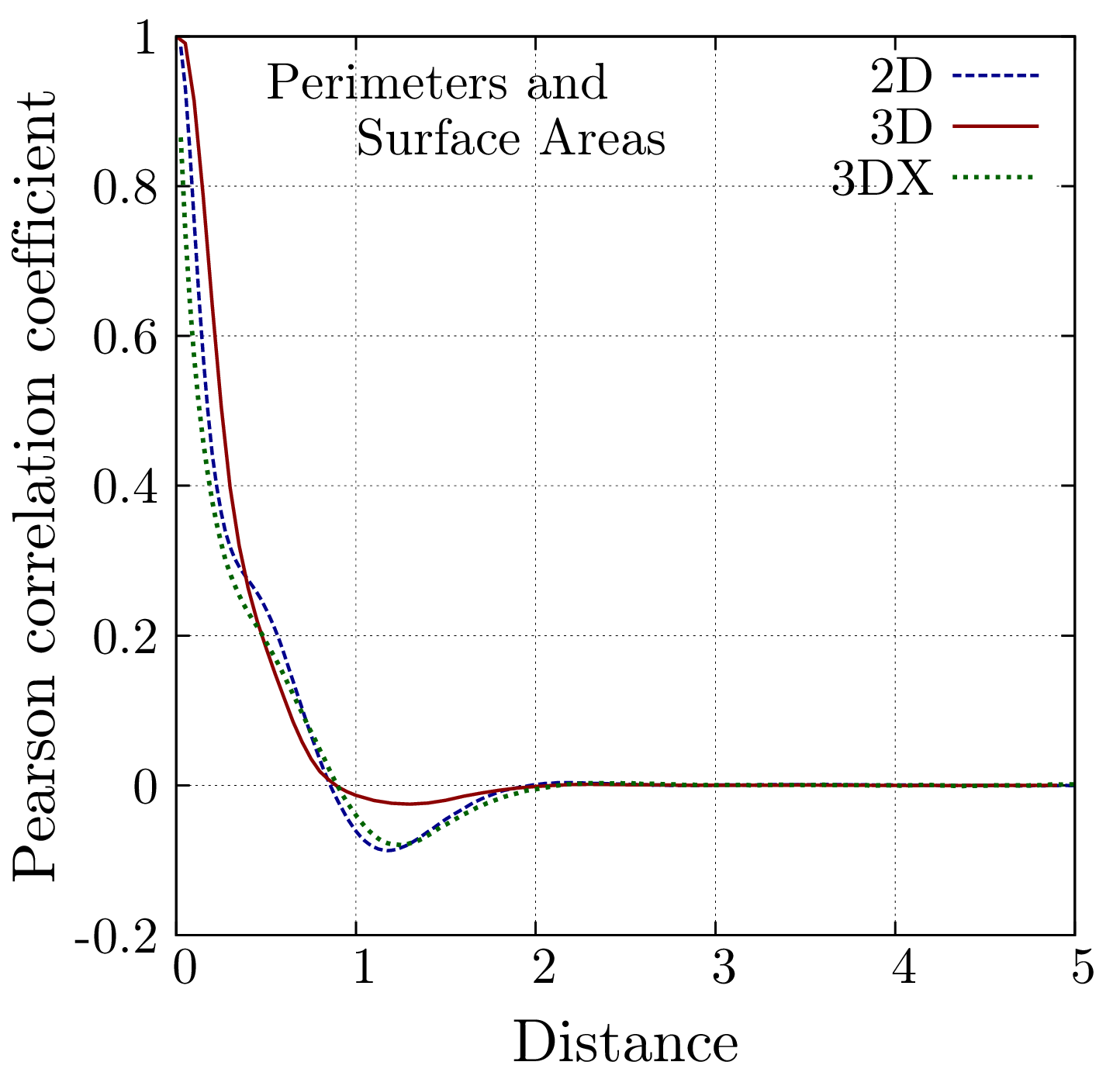}}} &
\subfigure[]{\label{corr-c}\resizebox{0.31\linewidth}{!}{\includegraphics{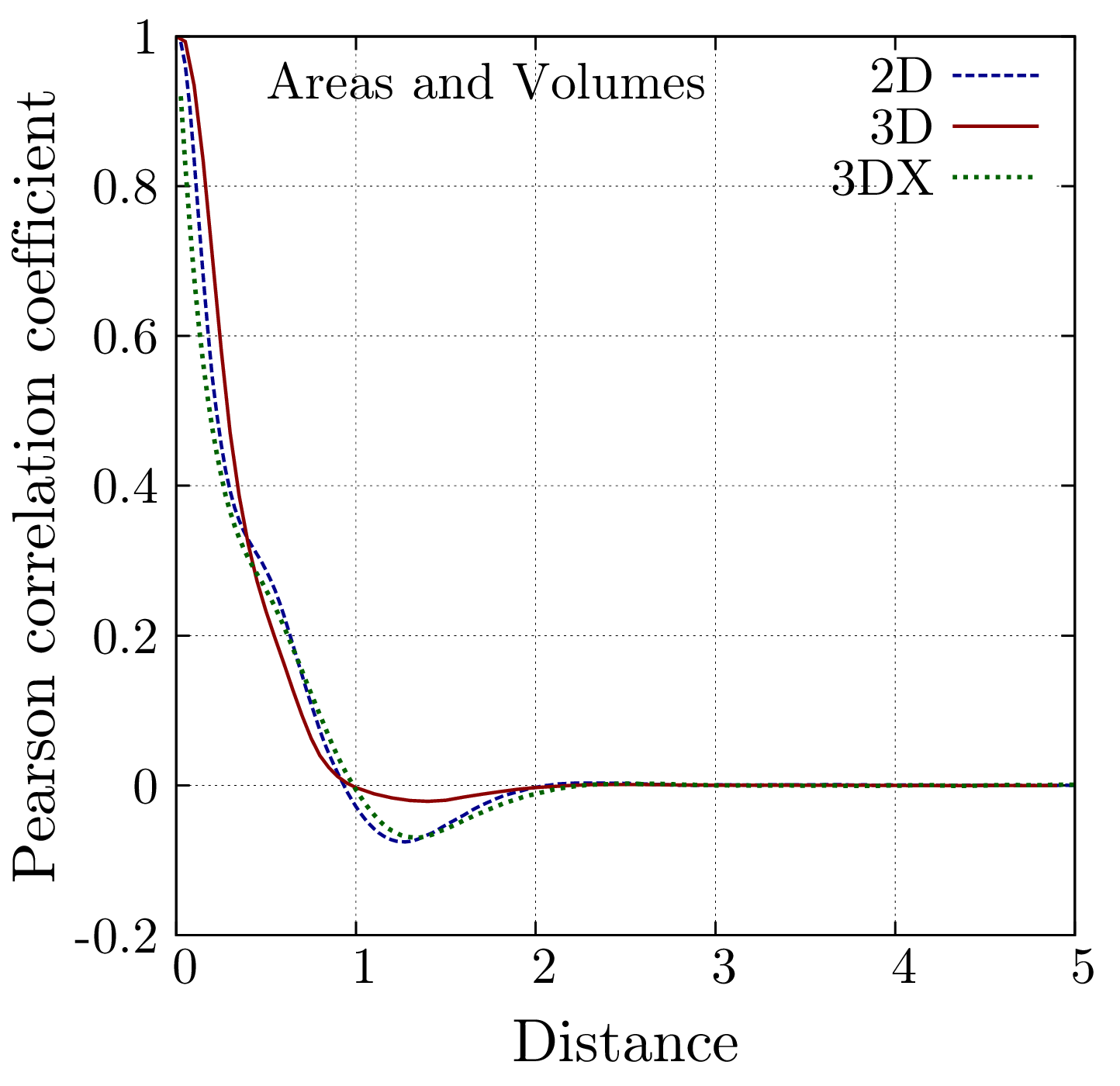}}}
\end{tabular}
\vspace{-5pt}
\caption{(a) Correlation between the number of edges (2D, 3DX) or faces (3D) of a central grain with those of surrounding grains. (b) Correlation between the perimeter (2D, 3DX) or surface area (3D) of a central grain with those of surrounding grains. (c) Correlation between the area (2D, 3DX) or volume (3D) of a central grain with those of surrounding grains.}
\label{correlations}
\end{center}
\end{figure*}

Specifically, let $x_i$ and $y_j$ be values of a property of interest (e.g., volume, surface area) for the $i$th and $j$th grains of the microstructure.  The indicator function $\chi_{ij}(\rho)$ is equal to one when the $j$th grain intersects a sphere of radius $\rho$ constructed around the center of mass of the $i$th grain, and vanishes otherwise.  Notice that $\chi_{ij}(\rho)$ is not necessarily symmetric in $i$ and $j$.  Finally, let $\langle x \rangle$ and $\langle y \rangle$ indicate the weighted average values of $x_i$ and $y_i$, calculated using the expressions $\sum_{ij}\chi_{ij}(\rho)x_{i}/\sum_{ij}\chi_{ij}(\rho)$ and $\sum_{ij}\chi_{ij}(\rho)y_{j}/\sum_{ij}\chi_{ij}(\rho)$, respectively.  The Pearson correlation coefficient between the values of this property for a central grain and the grains residing at some distance $\rho$ is given by 
\begin{equation}
r_{xy}(\rho) = \frac{\sum\limits_{ij}\chi_{ij}(\rho)(x_{i} - \langle x \rangle)(y_{j} - \langle y \rangle)}{\sqrt{ \big[ \sum\limits_{ij}\chi_{ij}(\rho)(x_{i} - \langle x \rangle)^2 \big] \big[ \sum\limits_{ij}\chi_{ij}(\rho)(y_{j} - \langle y \rangle)^2 \big] } },
\end{equation}
where the sums include all allowed values of $i$ and $j$.  The correlation coefficient $r_{xy}$ is plotted in Fig.~\ref{correlations} for several properties as a function of $\rho$.

The correlation between the numbers of edges of a pair of grains for 2D and 3DX, and between the numbers of faces of a pair of grains for 3D, is given in Fig.~\ref{corr-a}.  We normalize distance by the square root of the average grain area or cube root of the average grain volume, as appropriate.  The most conspicuous feature of this plot is the strong peak at small distances, caused by the perfect correlation of the number of sides of a central grain with itself.  As the distance from a central grain increases this correlation oscillates, becoming negative at a distance of about one average grain diameter and positive at about two average grain diameters.  This correlation rapidly weakens, to the point that there is no noticeable correlation between the number of sides of a central grain with grains three (average) grain diameters or more away.

Figure \ref{corr-b} shows the correlation between perimeters of neighboring grains in 2D and 3DX, and between surface areas of neighboring grains in 3D.  Analogously, Fig.~\ref{corr-c} gives the correlation between areas of neighboring grains in 2D and 3DX, and between volumes of neighboring grains in 3D.  Despite the general consistency of the behavior described above throughout the plots in Fig.~\ref{correlations}, a comparison of the correlations in the 2D, 3DX and 3D systems reveals several significant features.  For example, the dimensionality of the system appears to strongly influence the correlations, to the point that the curves for the 2D and 3DX systems nearly overlap.  This is particularly striking given the differences in the statistical features of the 2D and 3DX systems reported in Section \ref{sec:point_quantities}, and the differences in the origin of the structures.  Meanwhile, the correlations for the 3D system are consistently weaker than for the two-dimensional systems.  This is consistent with the general principle that more degrees of freedom are accessible to systems in higher dimensions.  Finally, the shoulder appearing in the plots of the 2D and 3DX systems at a distance of about half a grain diameter does not appear to have been noticed previously in the literature.

Some features of these figures may be explained by separating the contributions into a part from the central grain, a part from the nearest neighbors, and so forth.  With reference to Fig.~\ref{corr-a}, at a distance of roughly one average grain diameter, the number of sides of a central grain is negatively correlated with that of its surrounding grains.  Assuming that many of the grains intersected by a sphere of radius one average grain diameter share a side with the central grain, this negative correlation indicates a property of the nearest neighbors.  Specifically, this provides some support for the Aboav--Weaire law \cite{1970aboav, 1974weaire} or the empirical observation that grains with many sides are generally surrounded by grains with few sides, and {\it vice versa}.

A small positive correlation in the number of sides is observed at a distance of two average grain diameters.  Following the above reasoning, one expects that a sphere of this radius should generally intersect the second nearest neighbors, indicating that the average number of sides of the second nearest neighbors is weakly positively correlated with the central grain.  This may be explained as an iterated effect of the Aboav--Weaire law.  That is, the nearest neighbors of a grain with many sides will generally have few sides, and the nearest neighbors of any one of these nearest neighbors with few sides will generally have many sides.  Since the set of nearest neighbors of the nearest neighbors shares many grains with the set of second nearest neighbors of the central grain, the Aboav--Weaire law implies that there should be a weak positive correlation in the number of sides of a central grain with the grains at about two average grain diameters.

As stated though, the Aboav--Weaire law relates the number of sides of grains sharing a particular topological relationship, not necessarily a particular spatial separation.  The assumption that the majority of grains at a distance of one average grain diameter are nearest neighbors of the central grain, and that many of the grains at a distance of two average grain diameters are second nearest neighbors of the central grain, appears to be reasonable.  Nevertheless, the use of a spatial separation introduces a complication that could be avoided by considering the topological relationship of the grain directly.

\subsection{Quantities correlated over topological distance}
\label{sec:top_distance}

\begin{figure}
\center
\includegraphics[height=5.cm]{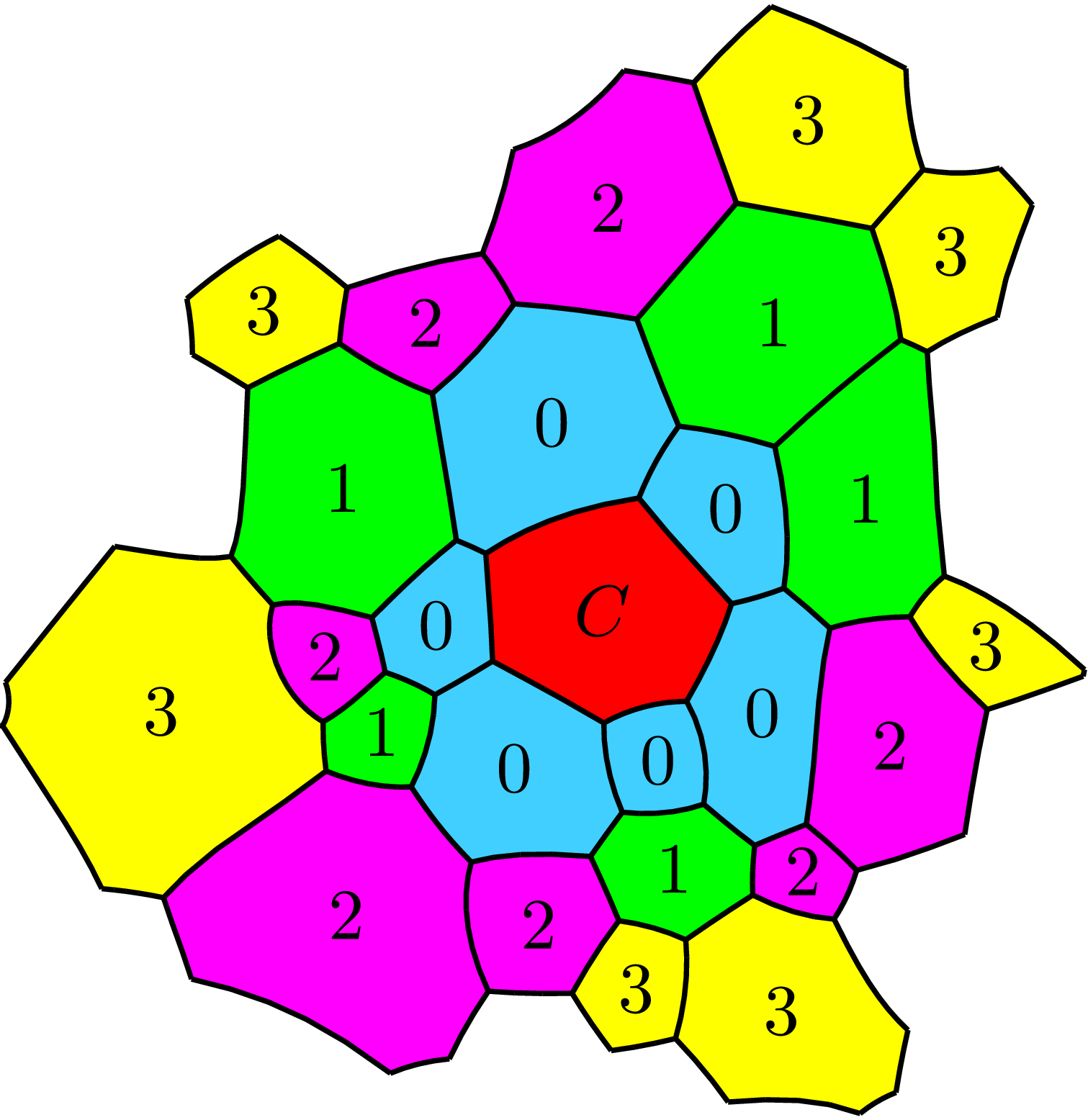}
\caption{Bond distance calculated for neighboring grains in a two-dimensional microstructure.  The central grain is labeled $C$, and numbers indicate bond distances to other grains.}
\label{bond_distance}
\end{figure}

An interpretation of the correlation functions reported in Section \ref{sec:metric_dist} relies on the separation of contributions from the central grain, from the grains with which it shares a side, and from grains further away.  This separation suggests that the features of the structure be studied directly for pairs of grains with a given topological relationship, rather than indirectly for pairs of grains with a given spatial separation.  Several relevant notions of topological distance appear in the literature, including shell distance \cite{1993fortes, 1996aste, 1996asteB} and bond distance \cite{1998ohlenbusch, 1999dubertret}.  The bond distance between two grains is the minimum number of edges in a path between their boundaries.  As a reference, Fig.~\ref{bond_distance} indicates the bond distances of neighboring grains around a central grain in a simulated two-dimensional microstructure.  The shell distance between two grains is the minimum number of grain boundaries that must be crossed to go from the interior of one to the interior of the other.

As in Section \ref{sec:metric_dist}, let $x_i$ and $y_j$ be values of a property of interest (e.g., the number of sides of a grain) of the $i$th and $j$th grains.  The indicator function $\chi_{ij}(d)$ is equal to one when the $j$th grain is at bond distance $d$ from the $i$th grain, and vanishes otherwise.  The joint probability $f(x, y | d)$ is calculated from all pairs of $x_i$ and $y_j$ for which $\chi_{ij}(d)$ does not vanish.  The probability $f(x | d)$ is calculated by summing $f(x, y | d)$ over all possible values of $y$.  The conditional probability for $y$ given $x$ is $f(y | x, d) = \frac{f(y, x | d)}{f(x | d)}$.  In the following figures, we give the mean and standard error of the mean of this conditional probability distribution as functions of the value $x$ of the central grain and for $d=0$, $1$ and $2$.

We now consider the number of edges or faces in two or three dimensions, respectively.  Earlier studies investigated the relationship between the number of sides of a central grain to the average number of sides of neighboring grains (i.e., bond distance 0). This stems from the experimental observation by Aboav \cite{1970aboav} on a cross-section of polycrystalline MgO that a grain with many sides is typically surrounded by grains with few sides, and {\it vice versa}. This equation, frequently known as the Aboav--Weaire law \cite{1970aboav, 1974weaire}, attempts to capture this relationship by expressing the average number of sides of a neighbor $m_n$ as a function of the number of sides of a central grain with $n$ sides:
\begin{equation}
m_n = \langle n \rangle - a + \frac{\langle n \rangle a + \mu_2}{n},
\label{aboav_weaire_fortes}
\end{equation}
where $\langle n \rangle$ is the average number of sides per grain in the system, $\mu_2$ is the variance of the distribution of the number of sides, and $a$ is a fitting parameter depending on the structure \cite{1980aboav, 1989fortes}.

\begin{figure*}
\begin{center}
\begin{tabular}{ccc}
\subfigure[]{\label{edge-area-a}\resizebox{0.31\linewidth}{!}{\includegraphics{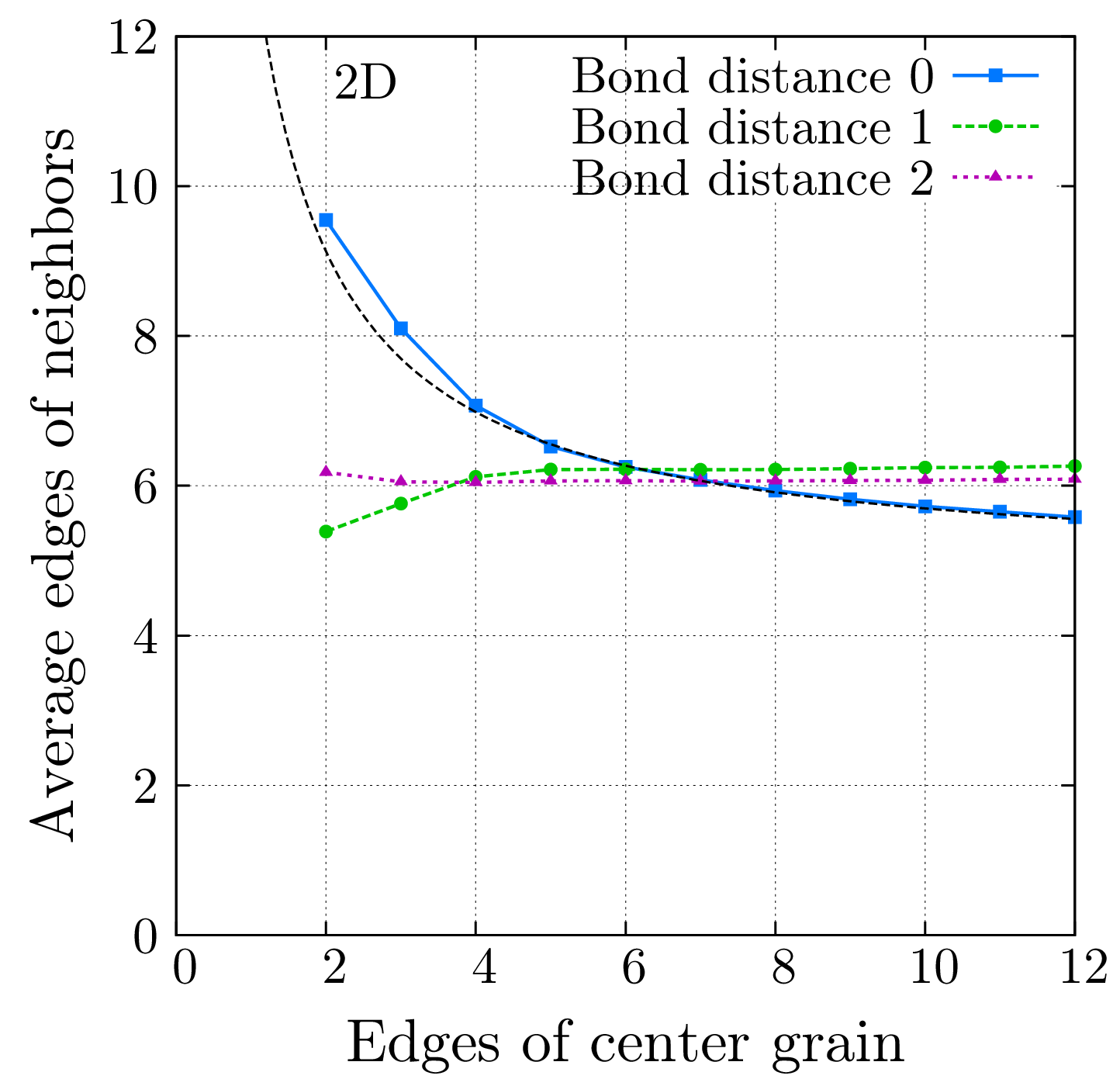}}} &
\subfigure[]{\label{edge-area-b}\resizebox{0.31\linewidth}{!}{\includegraphics{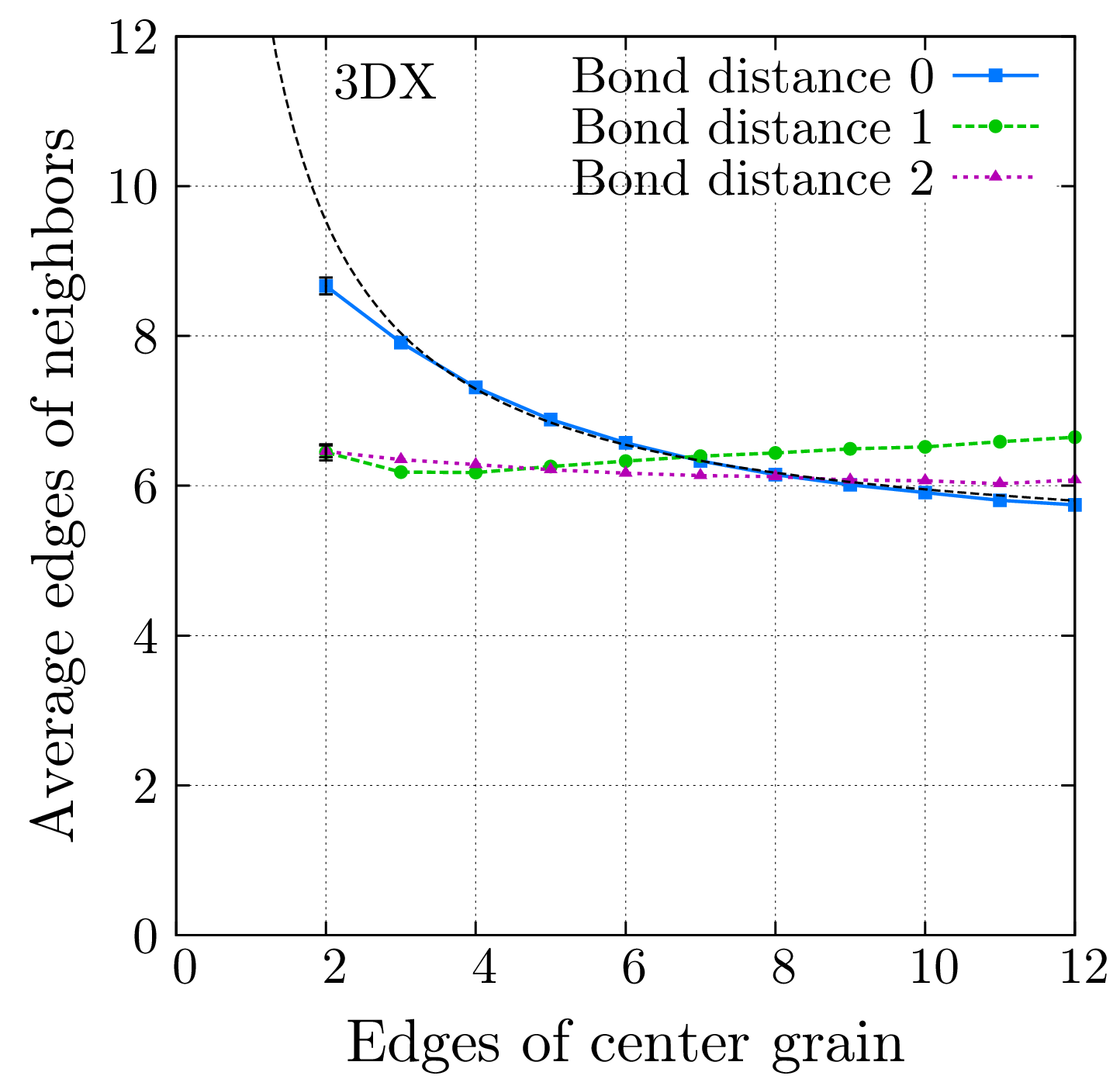}}} &
\subfigure[]{\label{edge-area-c}\resizebox{0.31\linewidth}{!}{\includegraphics{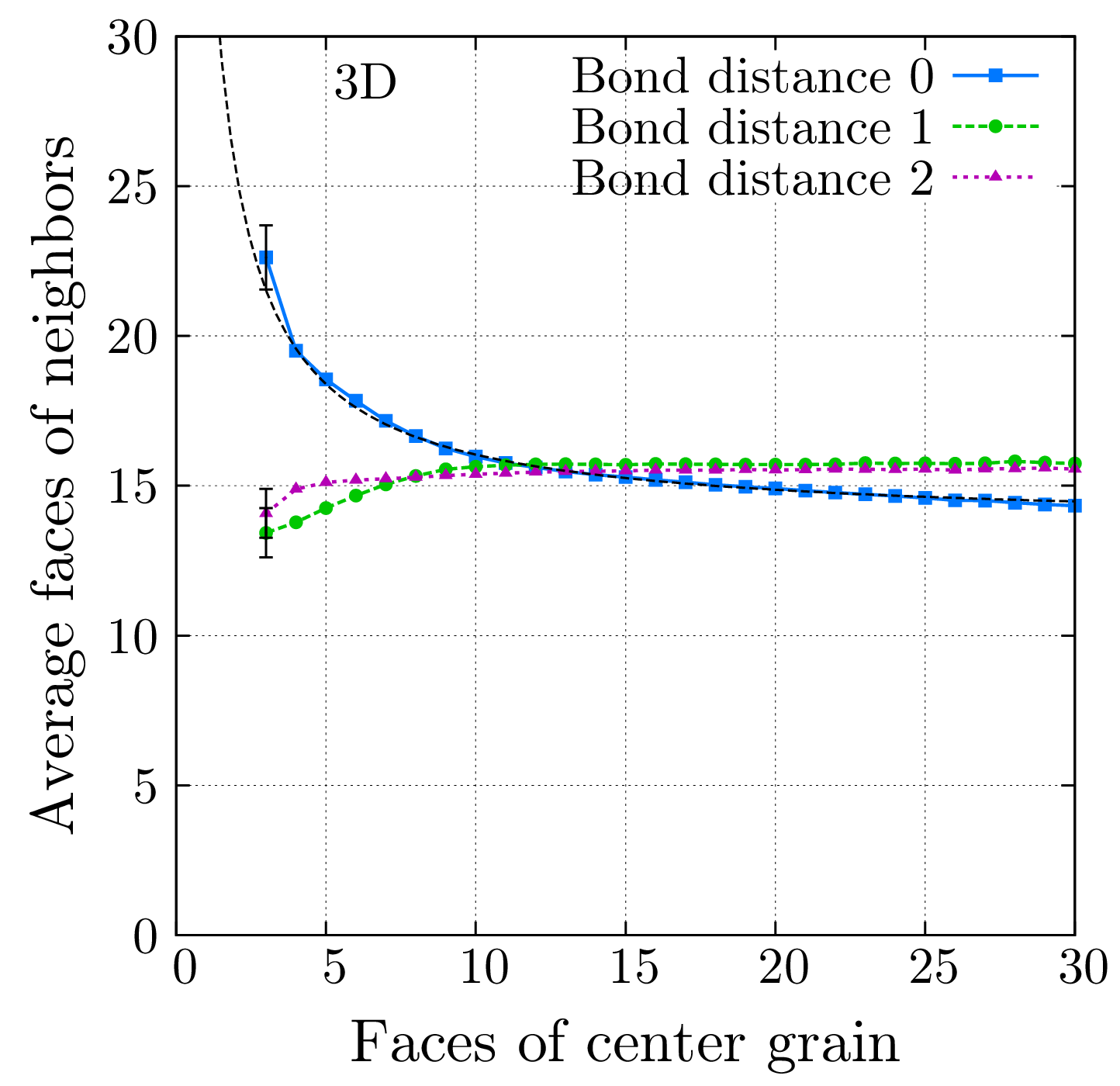}}}
\end{tabular}
\vspace{-5pt}
\caption{Average number of edges or faces of neighboring grains as a function of the number of edges or faces of a central grain for bond distances 0, 1, and 2; where not indicated, error bars are smaller than the marker size.  Predictions of Aboave-Weaire for $d=0$ are shown as black dashed curves, and are obtained using $\mu_2$ from Fig.~\ref{edges_and_faces} and $a$ fitted to the data.}
\label{topological_sides}
\end{center}
\end{figure*}

The correlation between the number of sides of grains and the Aboav--Weaire law appear in Fig.~\ref{topological_sides}. For bond distance 0, the fitting parameters $a$ are determined by least squares to be $a = 1.159 \pm 0.004$, $0.943 \pm 0.007$ and $0.074 \pm 0.007$ for the 2D, 3DX and 3D systems, respectively. While the resulting predictions agree with bond distance $0$ averages reasonably well, there is substantial deviation for central grains with small number of sides, especially in 2D and 3DX. Indeed, these deviations should be expected when applying Eq.~\ref{aboav_weaire_fortes}, since the usual Aboav--Weaire law neglects several terms that are present in a more complete analysis \cite{2012mason_a}.

Neighboring grains at bond distance $1$ generally display a behavior that is opposite to that for bond distance $0$, i.e., they tend to have few sides when the central grain has few and more sides when the central grain has many.  By contrast with the grains at two average grain diameters in Section \ref{sec:metric_dist} though, this may not be explained as the iterated effect of the Aboav--Weaire law.  Consider that both grains at bond distances $1$ and $2$ are nearest neighbors of grains at bond distance $0$ in Fig.~\ref{bond_distance}, and that the average number of sides of grains at bond distance $2$ is almost independent of the number of sides of the central grain in Fig.~\ref{topological_sides}.  If the behavior of the grains at bond distance $1$ could be explained as an iterated effect of the Aboav--Weaire law, then the same behavior should be observed for grains at a bond distance $2$.  This is contrary to the simulation results and some other explanation is required.

\begin{figure}
\center
\includegraphics[width=\linewidth]{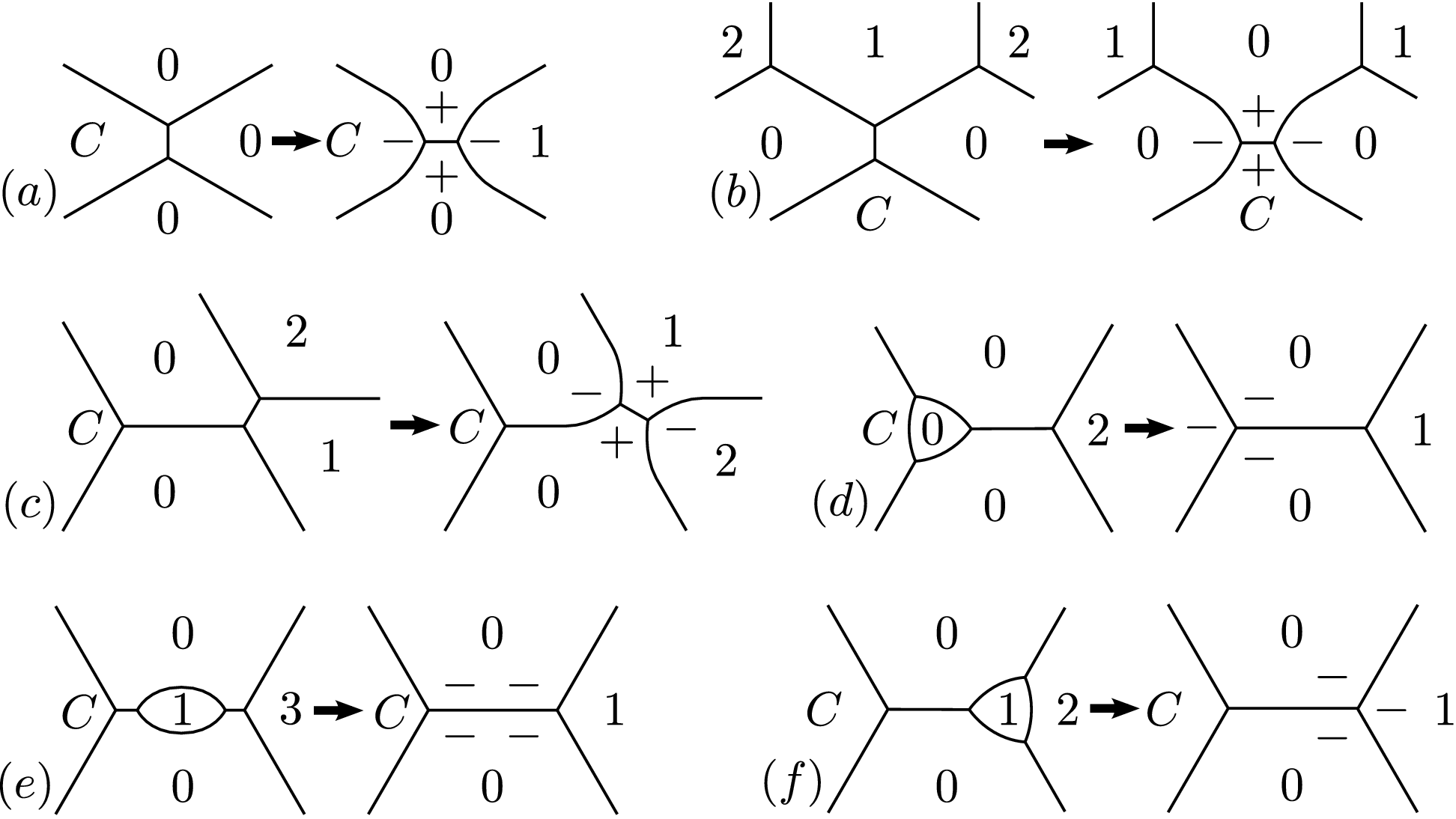}
\caption{All events that change the set of grains at bond distance $1$ in two dimensions.  Numbers indicate the bond distance of the grain from a central grain $C$; $+/-$ signs show whether the grain containing the symbol gains or loses an edge during the event occurring in the direction of the arrow.}
\label{transitions}
\end{figure}

The analysis of the average number of sides of grains at bond distance $1$ is made more manageable by identifying all events that can change the set of grains at this bond distance.  The elementary topological transitions implemented in our simulations \cite{2010lazar, 2011lazar} give six distinct events in two dimensions (see Fig.~\ref{transitions}) and seven distinct events in three dimensions.  Although our discussion focuses on two-dimensional events, we assert that the three-dimensional case is analogous.  The numbers in Fig.~\ref{transitions} indicate the current bond distance of a grain from a central grain $C$, while the $+/-$ sign within each grain indicates whether the grain has gained or lost an edge as a result of the transition.  Events (a), (b) and (c) in Fig.~\ref{transitions} apply an edge-flip to one of the edges in the vicinity of the central grain; events (d), (e) and (f) involve the collapse of a two or three-sided face.  Since the relative fraction of such faces in the structure is small (see Fig.~\ref{faces_a}), these events occur infrequently relative to (a), (b) and (c).

Statistics shown in Fig.~\ref{topological_sides} derive from normal grain-growth microstructures, in which grains with few sides tend to become smaller.  In 2D, this is a direct consequence of the von Neumann-Mullins relation (Eq.~\ref{evn2d}); in 3D this is a result of the trends in Fig.~\ref{mean_width_b} along with the MacPherson--Srolovitz relation (Eq.~\ref{evn3d}).  A central grain with few sides will, therefore, tend to become smaller and shed neighbors that replace the grains at bond distance $1$, as with event (a) in Fig.~\ref{transitions}.  Assuming that the neighbors of the central grain most likely to be lost are those themselves becoming smaller and losing sides, grains at a bond distance $1$ around a central grain with few sides should have few sides as well.

\begin{figure*}
\begin{center}
\begin{tabular}{ccc}
\subfigure[]{\label{area-top-corr-a}\resizebox{0.31\linewidth}{!}{\includegraphics{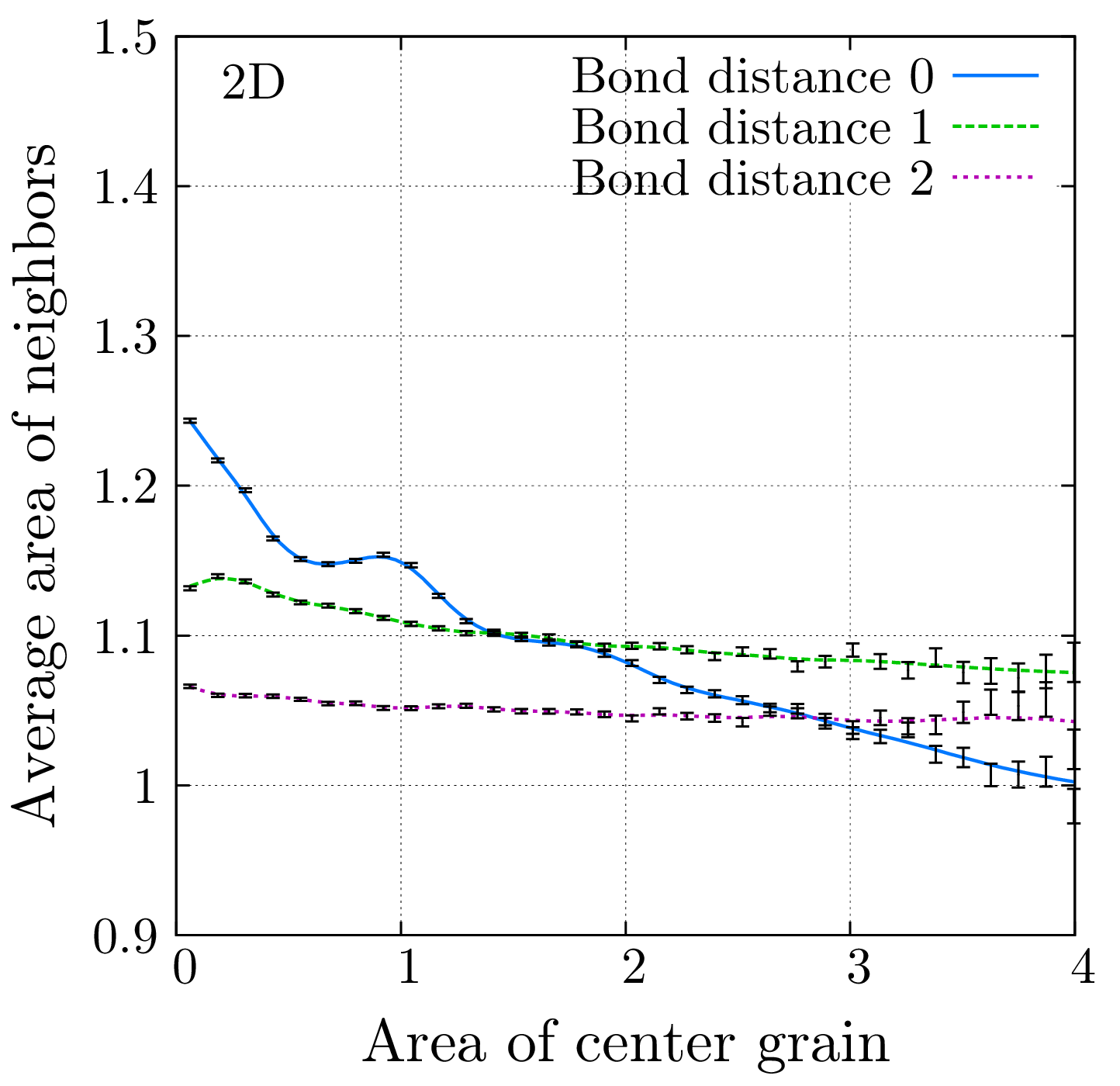}}} &
\subfigure[]{\label{area-top-corr-b}\resizebox{0.31\linewidth}{!}{\includegraphics{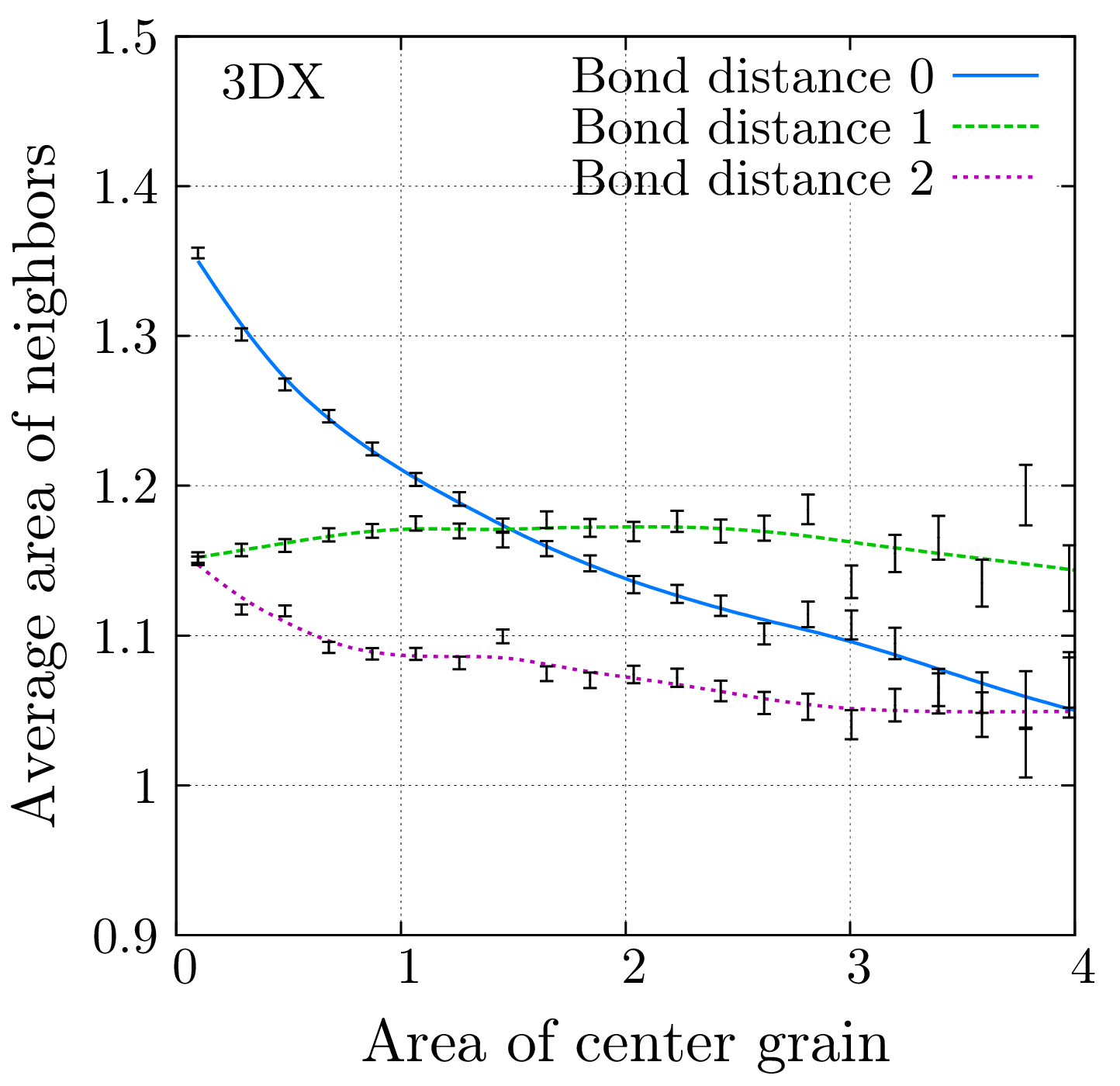}}} &
\subfigure[]{\label{area-top-corr-c}\resizebox{0.31\linewidth}{!}{\includegraphics{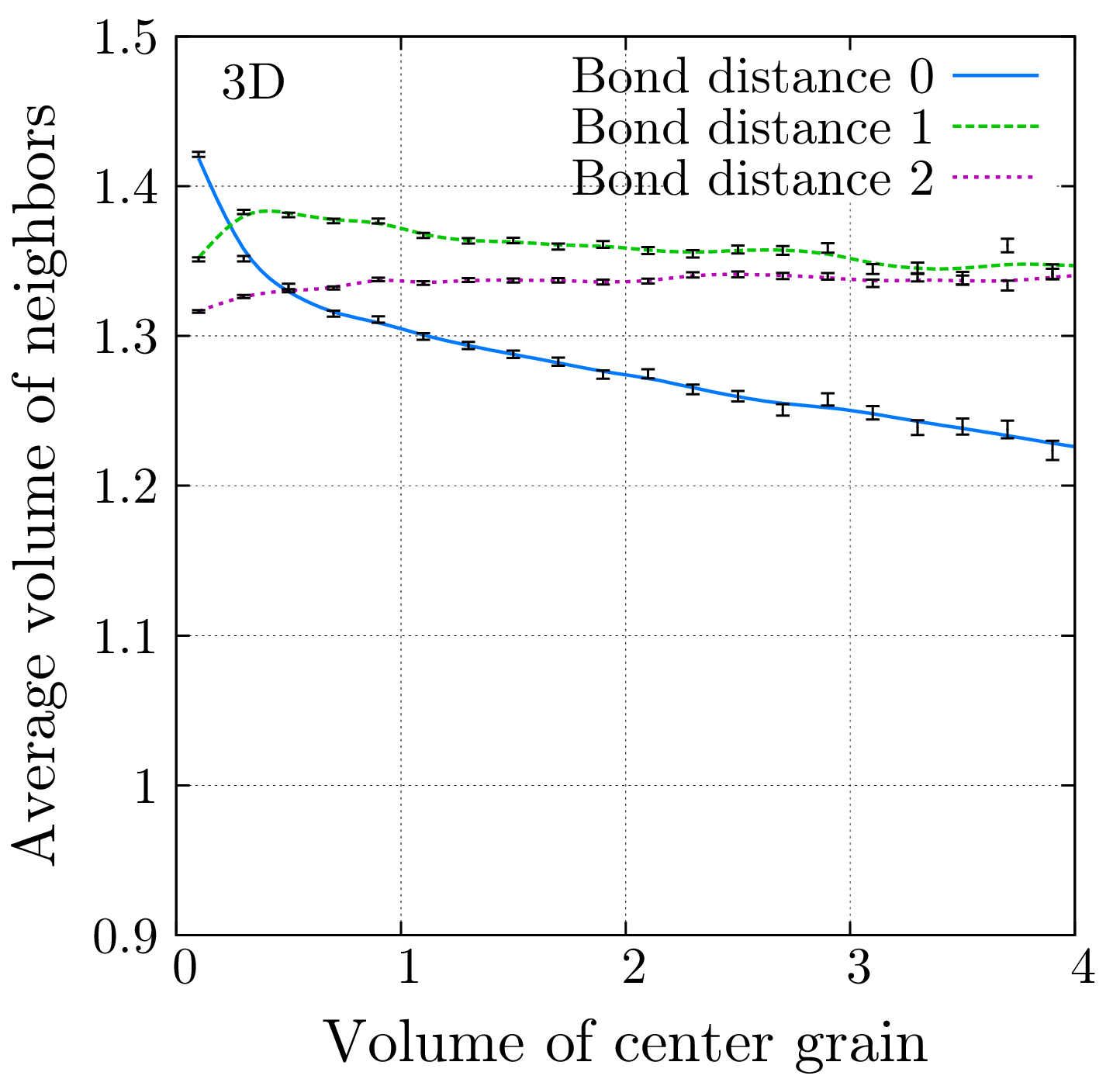}}}
\end{tabular}
\caption{Average size of neighboring grains as a function of size of a central grain (in normalized units) and bond distance.  Spline curves have been added to provide a guide for the eye.}
\label{topological_area}
\end{center}
\end{figure*}

A different mechanism is required to explain the observation that grains at bond distance $1$ have slightly more than the average number of sides when the central grain has many.  By reasoning analogous to that for the case of a central grain with few sides, a central grain with many sides will generally grow and acquire more sides.  This suggests that event (b) will occur frequently and event (a) will occur infrequently.  Event (c) should occur at an intermediate frequency since this edge-flip does not interact directly with the central grain.  Notice that the most frequent events (b) and (c) move grains from bond distance $2$ into the population at bond distance $1$.  From Fig.~\ref{topological_sides}, grains at bond distance $2$ should have the same expected number of sides as the system average.  While event (b) moves these grains to bond distance $1$ without changing the number of sides, event (c) increases the number of sides of the relevant grain by one.  The average number of sides of a grain at bond distance $1$ from a central grain with many sides should therefore be expected to be slightly higher than the system average.

We now examine the sizes of grains separated by a particular bond distance, as indicated in Fig.~\ref{topological_area}.  A consistent feature of these results is that the average area or volume of neighboring grains at bond distance $0$ decreases with increasing area or volume of a central grain; this is similar to the behavior of the average number of sides of grains at bond distance $0$ as a function of the number of sides of a central grain in Fig.~\ref{topological_sides}.  This similarity should be expected given the strong correlation between the number of sides and the size of a grain as seen in Figs.~\ref{edges_perim_area} and \ref{volumes_and_surface_areas}.  More striking, however, is the observation that for almost any bond distance and any size of central grain, the average size of surrounding grains is higher than the corresponding system average (i.e., unity).  This phenomenon is a consequence of the fact that grains with large areas and many sides are neighbors to many grains and are included in the averages of Fig.~\ref{topological_area} more frequently than grains with small areas and few sides. This shifts all of the curves upward; this effect increases with increasing connectivity of the network as occurs in going from two to three dimensions.

\begin{figure}
\center
\includegraphics[height=2.6cm]{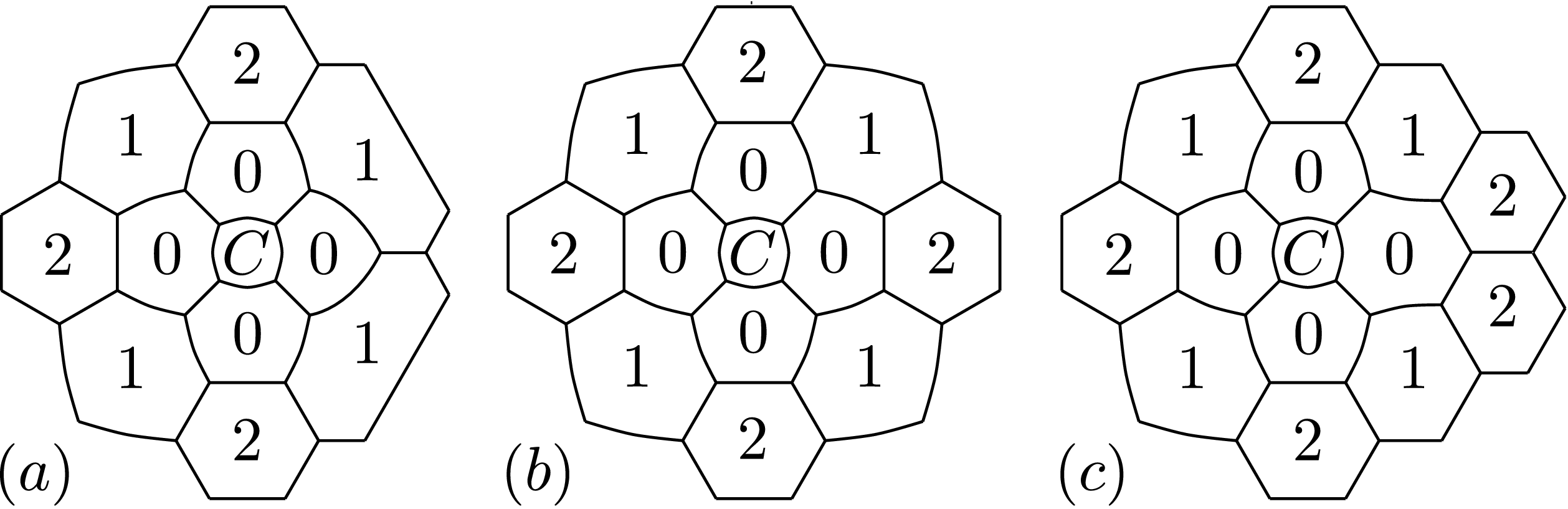}
\caption{A central grain $C$ with neighboring grains at various bond distances.  Illustrations highlight the effect of the number of sides of grains at bond distance $0$ on the total number of grains at bond distance $2$, as described in the text.}
\label{sampling}
\end{figure}

However, this does not explain why the average size of neighboring grains at bond distance $1$ is consistently higher than the average size of grains at bond distance $2$.  For this purpose, consider the grains in the vicinity of the central grain $C$ in Fig.~\ref{sampling}.  Suppose that the number of sides of the central grain is $n$.  When the central grain is surrounded by grains all with six sides (e.g., see (b)), the number of grains at bond distance $1$ is $n$ and the number of grains at bond distance $2$ is $n$.  A five-sided grain at bond distance $0$, as in (a), reduces the number of grains at bond distance $2$ by one.  Conversely, a grain with seven or more sides at bond distance $0$, as in (c), increases the number of grains at bond distance $2$ by one.  Now, grains of small area generally have few sides (see Fig.~\ref{edges_perim_area}) and grains with few sides generally have neighbors with many sides (shown in Fig.~\ref{topological_sides}).  A small central grain will therefore have more neighbors at bond distance $2$ than at bond distance $1$.  By analogous reasoning, a large central grain will have fewer neighbors at bond distance $2$ than at bond distance $1$.  Finally, notice that bond distance is a symmetric function.  Therefore, a large grain will be observed more often at bond distance $1$ from a central grain than at bond distance $2$, while a small grain will be observed more often at bond distance $2$ from a central grain than at bond distance $1$.  This difference in sampling will drive the curves in Fig.~\ref{topological_area} for bond distance $1$ higher than the curves for bond distance $2$.  The same reasoning pertains to the three-dimensional case.

\begin{figure*}
\begin{center}
\begin{tabular}{ccc}
\subfigure[]{\label{perim-corr-a}\resizebox{0.31\linewidth}{!}{\includegraphics{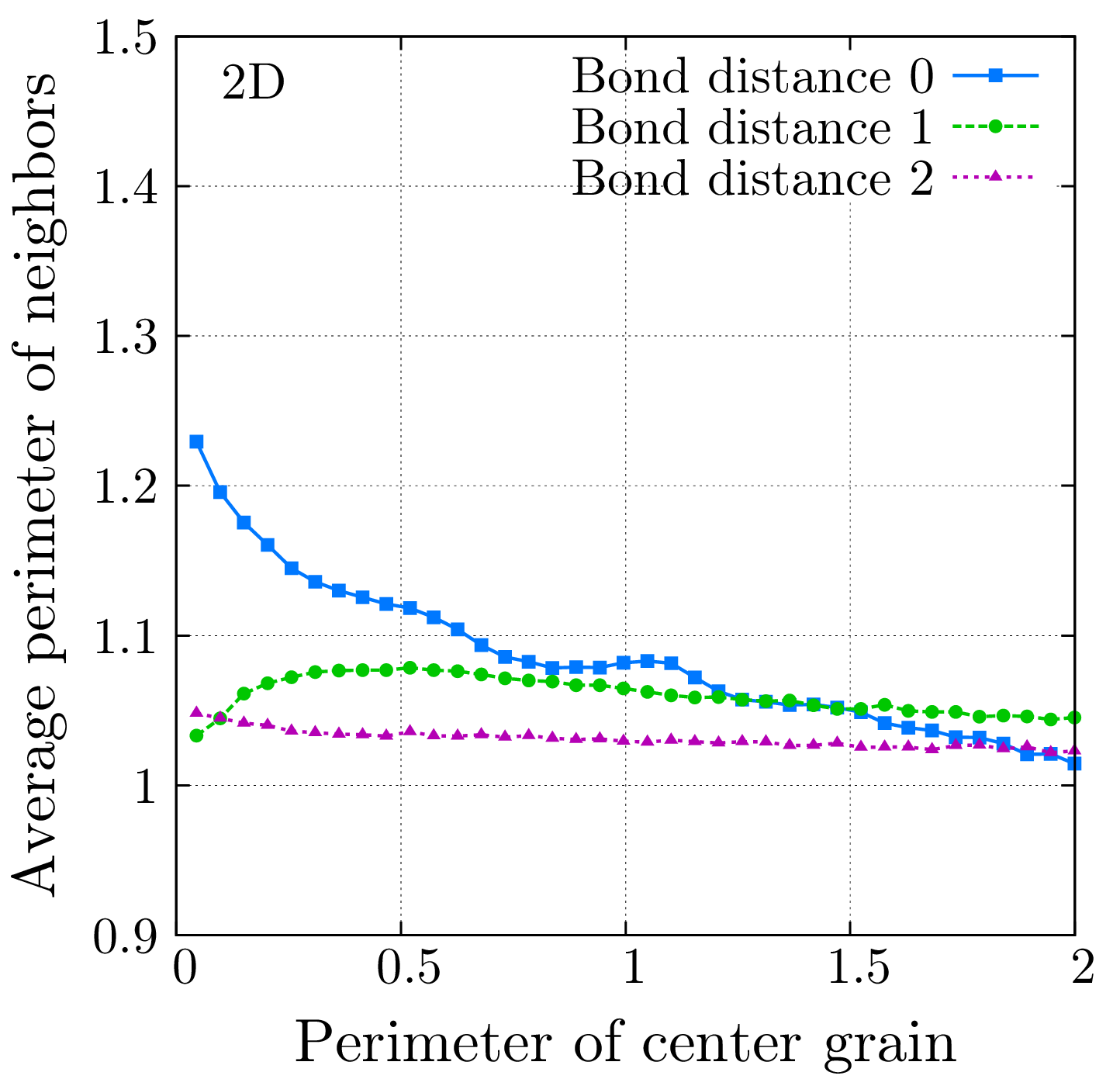}}} &
\subfigure[]{\label{perim-corr-b}\resizebox{0.31\linewidth}{!}{\includegraphics{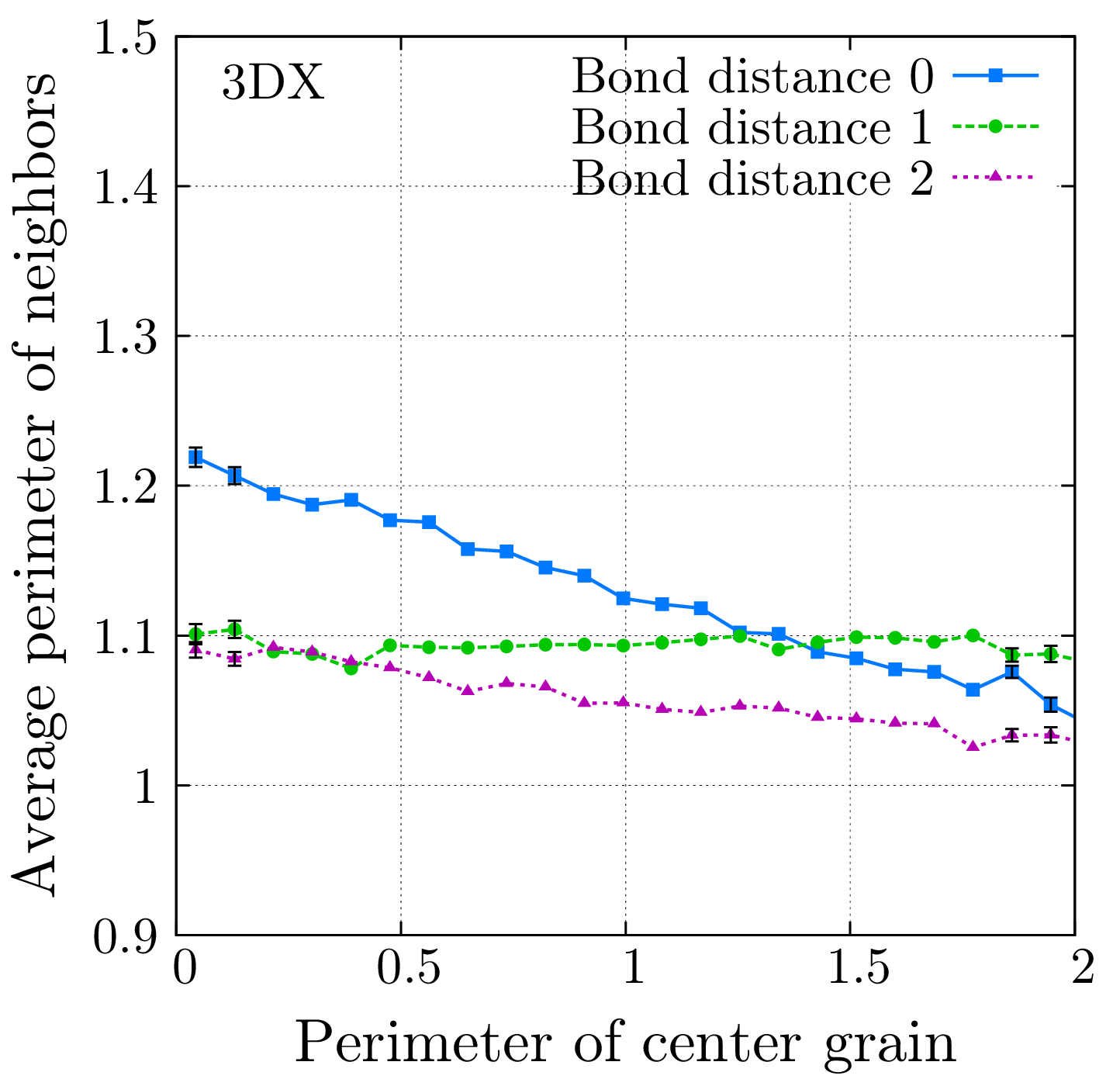}}} &
\subfigure[]{\label{perim-corr-c}\resizebox{0.31\linewidth}{!}{\includegraphics{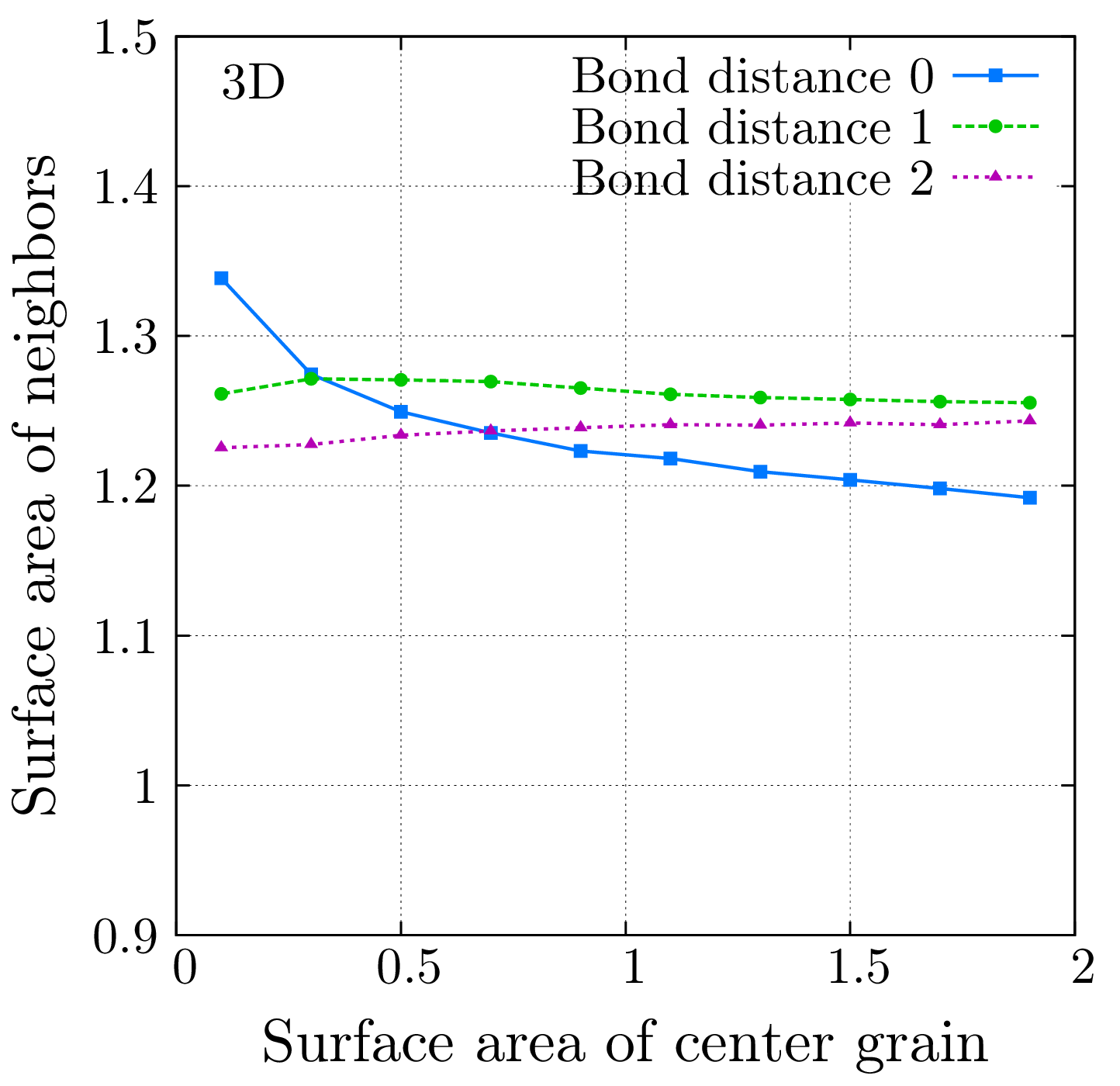}}}
\end{tabular}
\caption{Average boundary size of neighboring grains as a function of boundary size of a central grain (in normalized units).  Some error bars are smaller than the marker size.}
\label{topological_perim}
\end{center}
\end{figure*}

The final property of interest considered is the perimeter or surface area of a grain; results are presented in Fig.~\ref{topological_perim}.  Overall, the behavior is similar to that in Fig.~\ref{topological_area} and analogous explanations for this behavior apply.  One significant difference, however, is the appearance of distinct peaks in the average perimeter of grains at bond distance $0$ from a central grain in the 2D system.  The resemblance of these peaks to the ones in Figs.~\ref{dist_areas} and \ref{dist_perim} suggests that they arise from the same source, i.e., from the separate contributions of grains with different number of sides.

The results in this section suggest that  analysis of a microstructure by shell distance is generally not sufficient to distinguish the arrangement or the characteristics of grains around a central grain, and that a more complete description of topological relative position is required.  While bond distance provides one such alternative, it is clear that it remains an incomplete measure of topological relative position.

\subsection{Poisson--Voronoi microstructure}
\label{}

While we consider the steady-state, isotropic grain-growth (curvature flow) microstructure to be the prototype or canonical form of a polycrystalline microstructure, other relatively simples types of microstructures are also discussed in the literature.  In general, different evolution or microstructure formation laws yield microstructures with very different statistical (geometric and topological) properties. Probably the most widely discussed ``other'' microstructure is the Poisson-Voronoi (PV) construction.  In a materials science context, the PV microstructure is often described as arising from the simultaneous nucleation of grains at random positions in space, and the isotropic growth of these grains such that the velocity of the growth front is constant in time.  When two growth fronts impinge, they are replaced by stationary grain boundaries, resulting in a static PV microstructure.  This microstructure may apply in the extreme limits of phase transformations or recrystallization \cite{1953meijering, johnson1939reaction}. Other similar nucleation and growth laws have also been proposed \cite{christian2002theory, johnson1939reaction}.

\begin{figure}
\begin{center}
\subfigure[]{\label{gg_vol_tops_a}\resizebox{0.85\linewidth}{!}{\includegraphics{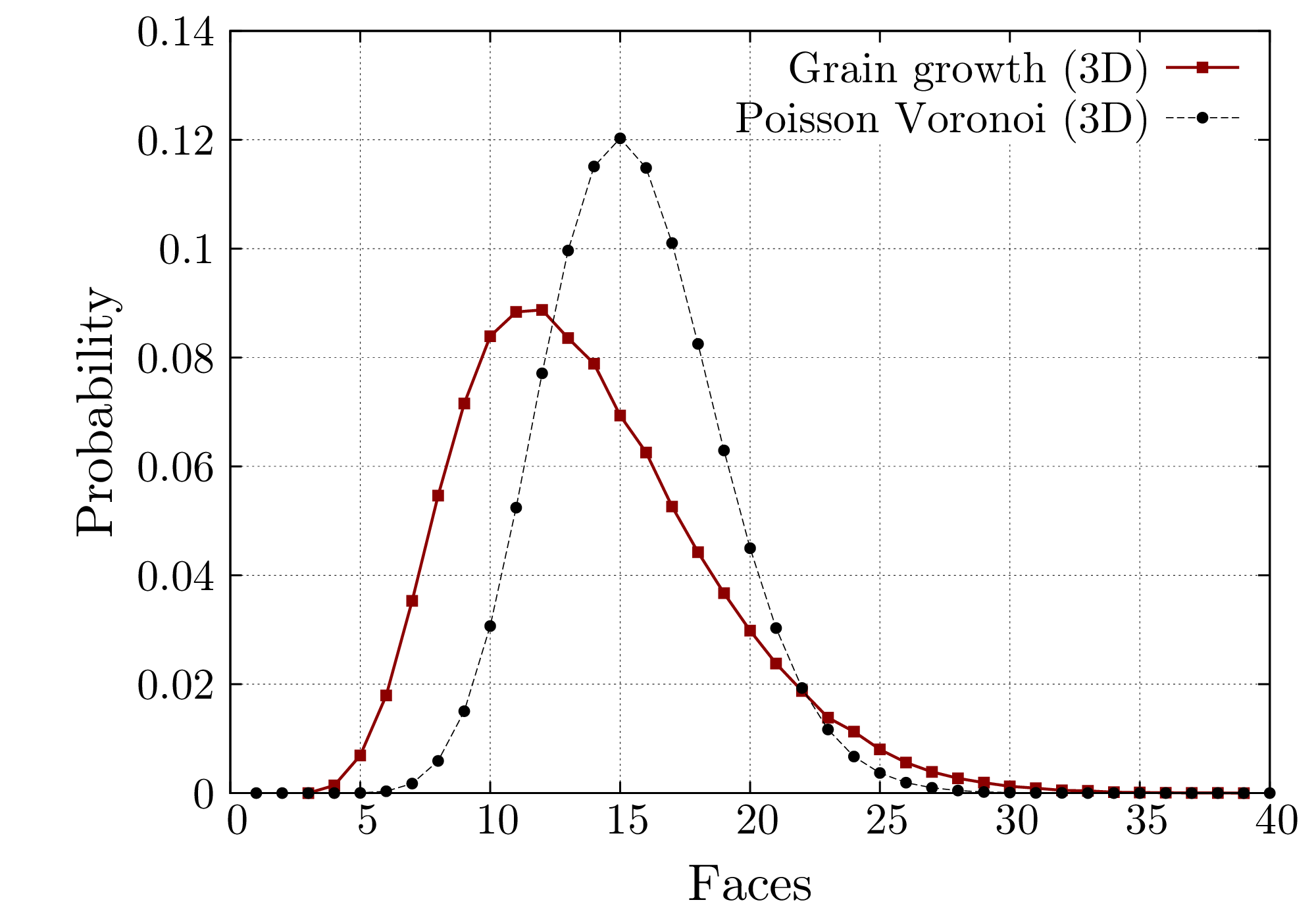}}} \\
\subfigure[]{\label{gg_vol_tops_b}\resizebox{0.85\linewidth}{!}{\includegraphics{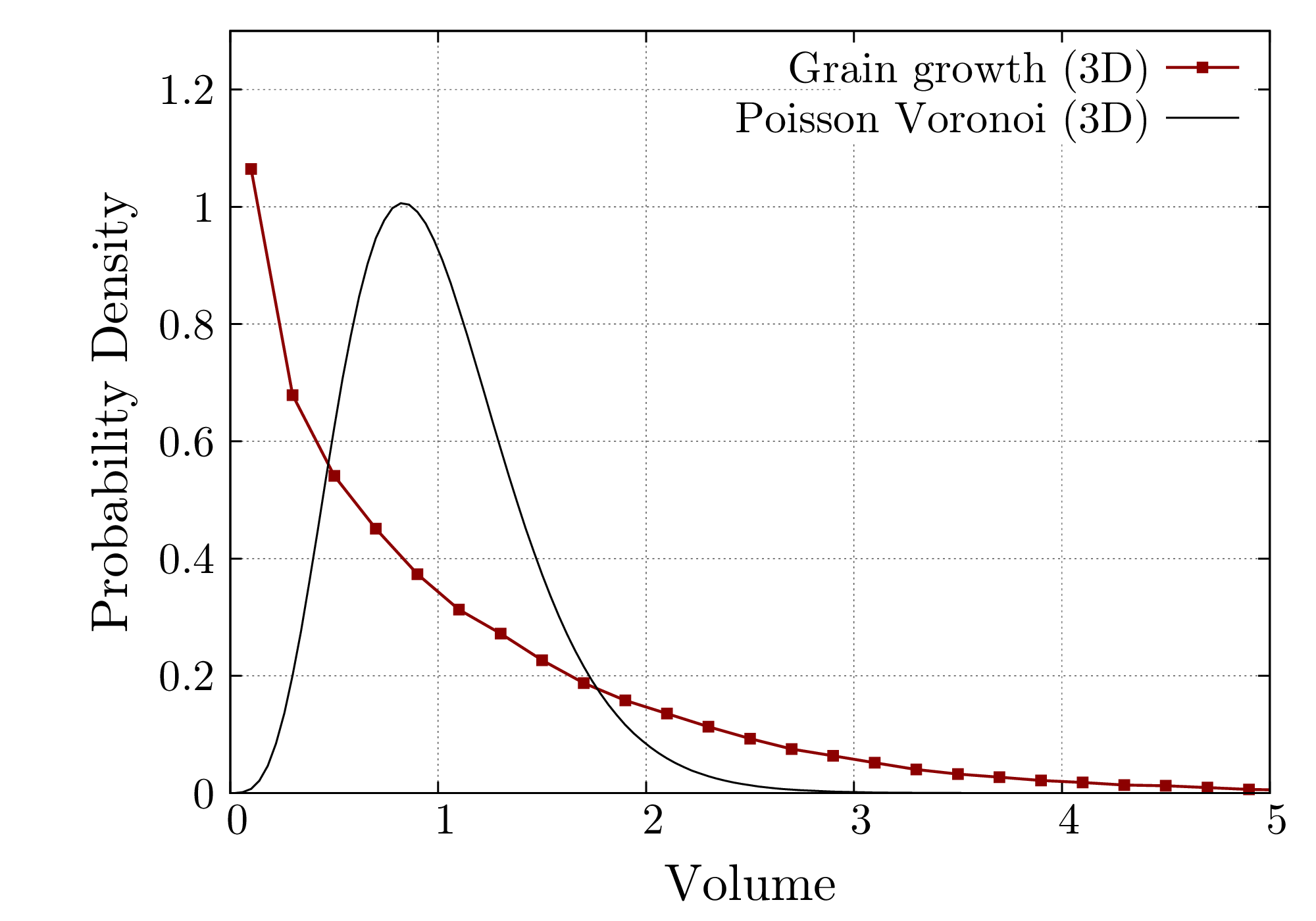}}}
\caption{Distribution of (a) faces per grain and (b) normalized grain volumes from grain-growth and Poisson-Voronoi microstructures.}
\label{gg_vol_tops}
\end{center}
\end{figure}

To illustrate several simple examples of the differences between the PV and the steady-state, isotropic grain-growth microstructures, we compare the distribution of faces per grain and grain volume in the two microstructures in Fig.~\ref{gg_vol_tops_a}.  Not only is the average number of faces per grain noticeably smaller in the grain-growth microstructure, the distribution is also considerably wider.  The distribution of grain volumes in the two systems is also significantly different.  Whereas cell volumes follow a near-exponential distribution in the grain growth microstructure, small grains in PV microstructures are vanishingly rare and the grain volume distribution is peaked at finite grain size.  Much more extensive sets of PV statistics to compare with the other measurements in this paper can be found elsewhere \cite{lazar2013statistical}.



\section{Discussion and Conclusion}
\label{sec:discussion}

The existence of a statistically steady-state grain-growth microstructure has been commonly conjectured in studies of grain growth.  To achieve such a statistical steady-state, the microstructure must undergo grain growth for sufficient time for the system to ``forget'' its initial state.  In simulations, the initial state is typically a Voronoi tessellation of a random distribution of points, while in experiment it is inherited from prior processing.  Substantial evolution requires growth growth simulations to begin from a large number of initial grains such that when statistics are measured, most of the initial grains have been consumed.  In the present study, we report statistical analysis of some of the largest two- and three-dimensional normal grain growth simulation data sets produced to date.  These simulations are accurate in that they are consistent with the von Neumann--Mullins relation in two dimensions \cite{1952vonneumann, 1956mullins} and the MacPherson-Srolovitz relation in three dimensions \cite{2007macpherson}.  A detailed description of the simulation methods and their numerical errors can be found elsewhere \cite{2010lazar, 2011lazar, lazar2011evolution}.

The  simulation results presented here vividly illustrate a number of qualitative differences between two-dimensional systems and cross-sections of three-dimensional ones.  The distribution of edges in the cross-sections is noticeably wider than in the genuine two-dimensional system, as is the distribution of edge lengths.  Cross-sections thus contain a larger fraction of grains with many sides, as well as a higher fraction of very long edges.  Moreover, grains in the genuine two-dimensional system are significantly more circular than grains in cross-sections.  These three differences can be accounted for by considering that two-dimensional systems evolve through a steepest descent path in energy, which in two-dimensional systems is proportional to the total length of the grain boundaries.  This inhibits the development of very long edges, grains with many edges, and grains with low isoperimetric ratios.  Because the evolution of cross-sections is not governed by the same energy-minimization considerations, grains tend to be less circular, and long edges and grains with many sides arise more frequently.

The more prominent feature distinguishing genuine two-dimensional systems from cross-sections of three-dimensional ones can be observed in the distributions of grain areas and perimeters.  Two-dimensional systems exhibit bimodal distributions of these quantities, whereas cross-sections exhibit unimodal ones.  Although these features have been observed previously, insufficient data have been available to draw definitive conclusions.  Given the accuracy in the data reported here, these two trends cannot be attributed to noise.  They appear to result from the role that topology plays in the evolution of genuine two-dimensional systems, which it does not play in the evolution of cross-sections.

We also observe important qualitative differences between the 2D and 3D systems.  In particular, we note that unlike the bimodal distribution of grain areas and perimeters in two-dimensional systems, the distribution of grain volumes and surfaces areas in three dimensions is unimodal.  This may similarly be attributed to the reduced role of topology in three dimensions: whereas the von Neumann-Mullins relationship relates the growth rate of a grain to its topology, the analogous MacPherson-Srolovitz relation does not. 

Another significant finding of our simulations is the near-exponential distribution of grain volumes in three-dimensional systems.  The probability density function of finding a grain with a given normalized volume $x$ is accurately described by the simple exponential $e^{-x}$.  The reason for this near-exponential distribution of the grain volumes is not known and requires further inquiry.

We also find that while neither Lewis' law or Feltham's law accurately describe the relationship between the number of edges of a grain in two dimensions and its size, an similar rule seems to hold in three dimensions.  That is, the volume of a grain appears to grow linearly with its number of faces $n$ for $n \ge 20$.  It should however be noted that the fraction of grains with 20 or more faces is rather small.  We do not currently understand the source of this relationship.

The last important set of results concerns pairwise correlations between grains separated by some metric or topological distance.  Our results illustrate a qualitative similarity between two-dimensional systems and cross-sections of three-dimensional ones.  The pairwise correlations between grains separated by some metric distance appear very similar in both systems despite the substantial difference in their origins.  Both of these systems exhibit correlations noticeably stronger than those that appear in the three-dimensional system.  It appears that the dimension of the systems plays a larger role in determining these correlations than the particular method of generating them.  In all three systems, we find that pairwise correlations of sizes and shapes are roughly zero at distances greater than three grain diameters.  This indicates that although the rules of evolution impose some local ordering on a structure, this ordering vanishes rather quickly.  

Pairwise correlations between grains at different topological separations provide additional insight into grain growth structures and methods of measuring their properties.  Neighboring grains at bond distance 0 tend to have fewer sides when a central grain has more sides and {\it vice versa}; this is consistent with the classical Aboav--Weaire law.  When considering generalizations of the law to grains that are not immediate neighbors, we observe that data for bond distance 1 is almost always distinguished from data for bond distance 2.  This suggests that the traditional notion of next-nearest neighbors -- which fails to distinguish between these two distances -- is inadequate in analyzing grain-growth microstructures.  In particular, we find that grains at bond distance 1 tend to have few sides when the central grain has few sides and {\it vice versa}; this correlation does not hold for bond distance 2.  All correlations between a central grain and its neighbors vanish for grains at bond distances 3 or 4, which illustrates that the ordering generated by grain-growth evolution is fairly local.

We also consider generalizations of the Aboav--Weaire law to correlations between metric properties of neighboring grains.  We note that for bond distance 0, the average area, perimeter, volume, and surface area of a neighboring grain decreases with increasing area, perimeter, volume, and surface area of a central grain.  This can be explained by the strong correlation between the topology of a grain and its metric size as investigated earlier in the paper.  This data also illustrates the role which topology plays in the genuine two-dimensional structure, which exhibits multiple peaks in the data for correlations of areas and perimeters between neighboring grains; these peaks do not appear in the data from the cross-sections.

Finally, since the steady-state grain-growth microstructure is reasonably well-defined, a comparison with the Poisson--Voronoi microstructure is meaningful.  The Poisson--Voronoi microstructure is found to be drastically different, suggesting that it is a poor representation of polycrystalline microstructures that have experienced extensive grain growth.  Nonetheless, the ease of constructing PV microstructures has led many researchers to use them to represent polycrystalline materials independent of the physical context \cite{hasnaoui2002non, latapie2004molecular, wu2013anatomy}. We respectfully argue that this practice should be discontinued.

\begin{acknowledgments}
The authors all acknowledge the support of the Institute for Advanced Study.  JKM acknowledges support from the Bo\u{g}azi\c{c}i University BAP Commission under Grant No.~8920.  EAL and DJS acknowledge support from the Division of Materials Research, National Science Foundation through CMMT award DMR-1507013.  JKM and EAL further acknowledge support from grant HR0011-08-1-0093.
\end{acknowledgments}

\bibliographystyle{apsrev4-1.bst}
\bibliography{canon_grain_growth.bbl}

\end{document}